\documentclass[11pt]{article}

\usepackage{amsfonts,amssymb,chet,dsfont,graphicx,mathrsfs,pgfplots,tikz,tensor,bm}
\allowdisplaybreaks

\newcommand{\Op}[2]{\mathcal{O}_{#1}(\eta_{#2})}

\newcommand{\ee}[3]{(\eta_{#1}\cdot\eta_{#2})^{#3}}
\newcommand{\e}[3]{\eta_{#1}^{#2_{#3}}}
\newcommand{\bbe}[1]{\bar{\bar{\eta}}_{#1}}

\newcommand{\A}{\mathcal{A}}

\newcommand{\aCF}[4]{{}_{#1}\alpha_{#2#3#4}}

\newcommand{\tOPE}[6]{{}_{#1}t_{#2#3}^{#5#6#4}}
\newcommand{\cCF}[4]{{}_{#1}c_{#2#3#4}}

\newcommand{\FCF}[6]{{}_{#1}F_{#2#3#4}^{#5#6}}

\title{Conformal Blocks in Three Dimensions}

\author{Jean-Fran\c{c}ois Fortin$^{\ast,}$\email{jean-francois.fortin@phy.ulaval.ca}, Jingping Li$^{\dagger,}$\email{jingpingl@andrew.cmu.edu}, Alex Sandomirsky$^{\ddag,}$\email{alex.sandomirsky@yale.edu}, and Witold Skiba$^{\ddag,}$\email{witold.skiba@yale.edu}}
\affiliation{
$^\ast$D\'epartement de Physique, de G\'enie Physique et d'Optique\\Universit\'e Laval, Qu\'ebec, QC G1V 0A6, Canada\\
$^\dagger$Department of Physics, Carnegie Mellon University, Pittsburgh, PA 15213, USA \\
$^\ddag$Department of Physics, Yale University, New Haven, CT 06520, USA
}

\abstract{We derive expressions for conformal blocks involving operators with arbitrary spins in 3-dimensional CFTs. We use previous results on the action of the OPE in the embedding space to derive the conformal blocks. The blocks are given as explicit power series in terms of the conformal cross ratios.}

\date{Decmber 2022} 

\begin{document}

\maketitle

\toc


\section{Introduction}\label{SecIntro}

Conformal field theories (CFTs) due to their enhanced symmetries tend to be more tractable at strong coupling compared to relativistic theories that do not have any additional space-time symmetry beyond Poincare invariance. The conformal  bootstrap program~\cite{Ferrara:1973yt,Polyakov:1974gs} is a widely known method for constraining the dynamics of CFTs based on consistency conditions of correlation functions. For a review of modern bootstrap literature see~\cite{Poland:2018epd}. A crucial ingredient of the bootstrap are four-point correlation functions, which can be expressed in terms of numbers, the so-called conformal data, that depend on the dynamics and in terms of conformal blocks that are restricted by symmetries. Bootstrap could be formulated in terms of correlators with more than four points as well, but this approach is not yet well developed. 

In this article we focus on conformal blocks in 3-dimensional CFTs. Such CFTs describe condensed-matter systems at criticality. We provide explicit expressions for blocks without any restrictions on the spins of either exchange or external operators. There is no simpler non-trivial case as the 3-dimensional Lorentz group is the simplest non-Abelian Lie group. The operator spin representations are all traceless symmetric tensors, so there are only two different classes of representations that need to be considered: either bosonic or fermionic.

Our method of deriving conformal blocks  is based on using the operator product expansion (OPE) and utilizing the embedding space. Since the OPE has a finite radius of convergence in CFTs~\cite{Mack:1976pa} an $M$-point correlation function can be expressed in terms of derivative operators acting on an $(M-1)$-point function and so on until the derivatives act on either a two-point or a three-point function, both of which are easy to construct. The embedding space for CFTs makes all expressions manifestly covariant since the conformal symmetry acts linearly on the embedding space~\cite{Dirac:1936fq,Mack:1969rr}. This path has been setup in detail in \cite{Fortin:2019fvx,Fortin:2019dnq} and further refined in \cite{Fortin:2020ncr}. For earlier work on use of the OPE and embedding space in CFTs see~\cite{Ferrara:1971vh,Ferrara:1971zy,Ferrara:1972cq,Ferrara:1973eg,Ferrara:1973vz,Ferrara:1974nf,Dobrev:1975ru}. In particular, a convenient choice of a derivative operator on the right-hand side of the OPE allowed an explicit calculation of the action of the OPE on any correlation function in a CFT. The action of the OPE is then given in terms of a specific function of the conformal cross ratios, which is given as a power series of the cross ratios. 

Armed with the results in \cite{Fortin:2019dnq,Fortin:2020ncr} the derivation of the blocks is reduced to a few steps. First is finding group-theoretic tensor structures that appear in the OPE and three-point functions. These structures intertwine the representations of the operators appearing on either both sides of the OPE or in a three-point function. For practical reasons, we find it simpler to use the OPE only once in a reduction of a  four-point conformal block; that is, we act with the derivative operator on a three-point function. This leads to what we call ``mixed basis" conformal blocks. The last step of making the derivation complete is finding a transformation that relates the two bases such that the blocks can be converted from the mixed basis to a pure basis because this is most useful for the bootstrap. 

It is worth noting that there are several other approaches to deriving conformal blocks. These involve solving the Casimir equations \cite{Dolan:2003hv,Dolan:2011dv,Kravchuk:2017dzd}, the shadow formalism \cite{Ferrara:1972xe,Ferrara:1972uq,SimmonsDuffin:2012uy}, the weight-shifting formalism \cite{Karateev:2017jgd,Costa:2018mcg}, integrability \cite{Isachenkov:2016gim,Schomerus:2016epl,Schomerus:2017eny,Isachenkov:2017qgn,Buric:2019dfk}, 
as well as AdS/CFT \cite{Hijano:2015zsa,Nishida:2016vds,Castro:2017hpx,Dyer:2017zef,Chen:2017yia,Sleight:2017fpc}. 

In the next section we briefly review the formalism we use. Readers familiar with Refs.~\cite{Fortin:2019fvx} and \cite{Fortin:2020ncr} can safely skip this section. In Section~\ref{sec:tensors}, we discuss a basis for the tensor structures and show how to construct all independent tensors. There are two sets of  tensors we consider. One of the two sets is applicable to writing the OPE and can only involve two coordinates that appear in the OPE. Another set leads directly to three-point functions and therefore involves three coordinates. As we already mentioned, it turns out that deriving conformal blocks is most convenient when using one tensor structure of each kind. The two different bases are related by a linear transformation that we present in Section~\ref{sec:3pt}. We derive conformal blocks in mixed basis in  Section~\ref{sec:4pt} and conclude in Section~\ref{sec:conclusions}. Certain technical details are relegated to Appendix~\ref{sec:ldep}. There, we discuss removing dependence on the spin of exchange operators such that all coefficients are given in terms of finite sums even if the spin of an exchange operator is arbitrarily large.

\section{Formalism and Relevant Results}\label{sec:formalism}

This section presents a brief summary of key results for conformal correlation functions in general dimensions obtained via the embedding space OPE method and customized here to the case at hand; that is to $(3+2)$-dimensional embedding space. These results were derived in~\cite{Fortin:2019fvx} and \cite{Fortin:2020ncr}, but an abbreviated version is useful here to establish notation. Any details omitted here can be found in those references. 

\subsection{OPE in the Embedding Space}
The central tool for us is the OPE of operators ${\mathcal{O}}_{i}$ and ${\mathcal{O}}_{j}$
\begin{equation}
\begin{aligned}\Op{i}{1}\Op{j}{2}= & (\mathcal{T}_{12}^{\boldsymbol{N}_{i}}\Gamma)(\mathcal{T}_{21}^{\boldsymbol{N}_{j}}\Gamma)\cdot\sum_{k}\sum_{a=1}^{N_{ijk}}\frac{\tensor*[_{a}]{c}{_{ij}^{\ \ k}}{}\tensor*[_{a}]{t}{_{ij}^{12k}}}{(\eta_{1}\cdot\eta_{2})^{p_{ijk}}}\cdot\mathcal{D}_{12}^{(d,h_{ijk}-n_{a}/2,n_{a})}(\mathcal{T}_{12\boldsymbol{N}_{k}}\Gamma)*\Op{k}{2},\\
 & p_{ijk}=\frac{1}{2}(\tau_{i}+\tau_{j}-\tau_{k}),\qquad h_{ijk}=-\frac{1}{2}(\chi_{i}-\chi_{j}+\chi_{k}),\\
\tau_{\mathcal{O}} & =\Delta_{\mathcal{O}}-S_{\mathcal{O}},\qquad\chi_{\mathcal{O}}=\Delta_{\mathcal{O}}-\xi_{\mathcal{O}},\qquad\xi_{\mathcal{O}}=S_{\mathcal{O}}-\lfloor S_{\mathcal{O}}\rfloor,
\end{aligned}
\label{eq:ope}
\end{equation}
where $\Delta_{\mathcal{O}}$ denotes the dimension and $S_{\mathcal{O}}$ the spin of a primary operator. Meanwhile, $\eta$ denote coordinates in the embedding space which is restricted to the light cone $\eta \cdot \eta=0$, $\mathcal{D}_{12}$ is a differential operator, and $\tensor*[_{a}]{c}{_{ij}^{\ \ k}}{}$ are the OPE coefficients. We have suppressed explicit indices of Lorentz representations here to make the formula more readable, but the operators carry only spinor that is half-integer representation indices. All operators are assumed to transform in irreducible representations denoted $\boldsymbol{N}$ which means in particular that the spinor indices are fully symmetrized. In $d=3$, $\boldsymbol{N}$ is simply a non-negative half integer so we can use $S$ and $\boldsymbol{N}$ interchangeably. The tensor structures that depend on $\boldsymbol{N}_i$,  $\boldsymbol{N}_j$, $\boldsymbol{N}_k$ are called  $\tensor*[_{a}]{t}{_{ij}^{12k}}$. Detailed definition of the differential operator $\mathcal{D}_{12}$  is not needed here because we will use previously derived results on how the operator acts on combinations of coordinates that appear in correlation functions.

Since the coordinates are vectors it is more convenient to realize representations in terms of vector indices and at most one spinor index. This is accomplished by ``half projectors," $(\mathcal{T}_{12}^{\boldsymbol{N}}\Gamma)$, that convert $2 S$ spinor indices to $\lfloor S \rfloor$ symmetrized vector indices and, when $S$ is not integer, one spinor index. 
On the right-hand side of the OPE in (\ref{eq:ope}) above, the $*$ product indicates that all spinor indices in the half projector are contracted with those in $\Op{k}{2}$. 

In position space, the half projectors are simply products of $\gamma$ matrices and projection operators 
\begin{equation}
(\mathcal{T}^S)_{\alpha_1 \ldots \alpha_{2 S}}^{\mu_1 \ldots \mu_{\lfloor S \rfloor}\delta}= \left( \frac{1}{\sqrt{2}}(\gamma^{\mu_1} C^{-1})_{\alpha_1\alpha_2} \ldots\frac{1}{\sqrt{2}}(\gamma^{\mu_{\lfloor S \rfloor}}C^{-1})_{\alpha_{2\lfloor S\rfloor-1 }\alpha_{2 \lfloor S\rfloor}}  \delta_{\alpha_{2S}}^{\delta'} \right)
(\hat{P}^S)_{\delta' \mu'_{\lfloor S \rfloor} \ldots \mu'_1}^{\phantom{\hat{P}^S_{\delta' \mu'_{\lfloor S \rfloor} \ldots \mu'_1}}\!\!\!\!  \mu_1 \ldots  \mu_{\lfloor S \rfloor}  \delta},
\end{equation}
where the indices $\delta$ and $\delta'$ are only present for half-integer $S$. The matrices $C$ relate the $\gamma$ matrices to their transposes as $\gamma_\mu^T=-\, C \gamma_\mu C^{-1}$. In three dimensions, there are only two types of representations and the corresponding projectors are
\begin{equation}
\begin{aligned}
& (\hat{\mathcal{P}}^{\ell})_{\mu_\ell\cdots\mu_1}^{\phantom{\mu_\ell\cdots\mu_1}\mu'_1\cdots\mu'_\ell}=
   \sum_{i=0}^{\lfloor\ell/2\rfloor}a_i(d,\ell)g_{(\mu_1\mu_2}g^{(\mu'_1\mu'_2}\cdots 
      g_{\mu_{2i-1}\mu_{2i}}g^{\mu'_{2i-1}\mu'_{2i}}g_{\mu_{2i+1}}^{\phantom{\mu_{2i+1}}\mu'_{2i+1}}\cdots g_{\mu_\ell)}^{\phantom{\mu_\ell)}\mu'_\ell)}, \\ 
& (\hat{\mathcal{P}}^{\ell+\frac{1}{2}})_{\alpha\mu_\ell\cdots\mu_1}^{\phantom{\alpha\mu_\ell\cdots\mu_1}\mu'_1\cdots\mu'_\ell\alpha'}
  =\sum_{i=0}^{\lfloor\ell/2\rfloor}a_i(d+2,\ell)g_{(\mu_1\mu_2}g^{(\mu'_1\mu'_2}\cdots
     g_{\mu_{2i-1}\mu_{2i}}g^{\mu'_{2i-1}\mu'_{2i}}g_{\mu_{2i+1}}^{\phantom{\mu_{2i+1}}\mu'_{2i+1}}\cdots g_{\mu_\ell)}^{\phantom{\mu_\ell)}\mu'_\ell)}\delta_\alpha^{\phantom{\alpha}\alpha'}\\
\quad &+\sum_{i=0}^{\lfloor(\ell-1)/2\rfloor}\frac{\ell a_i(d+2,\ell-1)}{2(-\ell+1-d/2)}g_{(\mu_1\mu_2}g^{(\mu'_1\mu'_2}\cdots g_{\mu_{2i-1}\mu_{2i}}g^{\mu'_{2i-1}\mu'_{2i}}g_{\mu_{2i+1}}^{\phantom{\mu_{2i+1}}\mu'_{2i+1}}\cdots g_{\mu_{\ell-1}}^{\phantom{\mu_{\ell-1}}\mu'_{\ell-1}}(\gamma_{\mu_\ell)}\gamma^{\mu'_\ell)})_\alpha^{\phantom{\alpha}\alpha'},
\end{aligned}
\label{eq:projectorsposition}
\end{equation}
where $a_i(d,\ell)=\frac{(-\ell)_{2i}}{2^{2i}i!(-\ell+2-d/2)_i}$ and $d=3$.  We keep an adjustable number of dimensions $d$ in the projectors because it will be useful a little bit later when we discuss a ``shifted" form of the projectors, for which $d\neq3$. 

In the embedding space, the dimensions of irreducible representations cannot change despite the indices spanning the larger embedding space. This is achieved by contracting all embedding space indices with the ``transverse metric"
\begin{equation}
\label{eq:metric}
 \A^{AB}_{ij}=g^{AB} - \frac{\eta_i^A \eta_j^B}{\ee{i}{j}{}}- \frac{\eta_i^B \eta_j^A}{\ee{i}{j}{}},
\end{equation}
which is symmetric and transverse to both $\eta_i$ and $\eta_j$ that is $ \A^{AB}\eta_{iB} =  \A^{AB}\eta_{jB} = 0$. When we turn to constructing tensor structures in the next section we will also need the $\epsilon$ tensor that is uplifted from three to five dimensions. This is implemented by defining
\begin{equation}
\label{eq:epsilon}
 \epsilon^{ABC}_{ij}=\frac{1}{\ee{i}{j}{}} \eta_{i Z'}\, \epsilon^{Z'  A' B' C' D'} \eta_{j D'}   \A_{ijA'}^{\phantom{ijA'}A}   \A_{ijB'}^{\phantom{ijB'}B}    \A_{ijC'}^{\phantom{ijC'}C}, 
 \end{equation}
 which is also manifestly transverse to both coordinates $\eta_i$ and $\eta_j$. 
 
 Using the transverse metric we can easily translate the projectors in (\ref{eq:projectorsposition}) to the embedding space. For the integer spin representations, we have
 \begin{equation}
 (\hat{\mathcal{P}}^{\ell}_{ij})_{A_\ell\cdots A_1}^{\phantom{A_\ell\cdots\A_1}A'_1\cdots A'_\ell}=
   \sum_{i=0}^{\lfloor\ell/2\rfloor}a_i(d,\ell) \A_{ij(A_1 A_2} \A^{(A'_1 A'_2}_{ij}\cdots     \A_{ij A_{2i+1}}^{\phantom{A_{2i+1}}A'_{2i+1}}\cdots \A_{ij A_\ell)}^{\phantom{A_\ell)}A'_\ell)},
   \label{eq:projeembeddingboson}
\end{equation}
where all the position-space metrics  $g$ in  (\ref{eq:projectorsposition}) were simply replaced by the embedding-space transverse metrics $\A_{ij}$ and we still set $d=3$  in the coefficients $a_i(d,\ell)$. For the half-integer representations, the projector is 
\begin{equation}
\begin{aligned}
  (\hat{\mathcal{P}}^{\ell + \frac{1}{2}}_{ij})_{a A_\ell\cdots A_1}^{\phantom{A_\ell\cdots\A_1}A'_1\cdots A'_\ell a'}
  &=\frac{(\eta_i \cdot \Gamma)_a^{\phantom{a} b}}{ 2 \, \eta_i \cdot \eta_j} \left( \sum_{i=0}^{\lfloor\ell/2\rfloor}a_i(d+2,\ell) \A_{ij} \ldots \A_{ij} \delta_b^{\phantom{b} b'} \right. \\
&\left. +\sum_{i=0}^{\lfloor(\ell-1)/2\rfloor}\frac{\ell a_i(d+2,\ell-1)}{2(-\ell+1-d/2)}  \A_{ij} \ldots \A_{ij} 
    (\Gamma_{ij \A_\ell)}\Gamma_{ij}^{A'_\ell)})_b^{\phantom{a}b'} \right) (\eta_j \cdot \Gamma)_{b'}^{\phantom{b'} a'},
\end{aligned}
\label{eq:projeembeddingfermion}
\end{equation}
where for readability reasons we only displayed spinor indices and omitted all vector indices on the metrics $\A_{ij}$. Those missing  indices are directly ported from the position-space projector by replacing $\mu_i$ with $A_i$, as we did in the integer-spin case. The interesting parts here are the fermionic parts, where we introduced $\eta\cdot \Gamma=\eta_A \Gamma^A$ and  $\Gamma_{ij}^A=\Gamma^{A'}  \A_{ij A'}^A$.

The half-projectors can also be uplifted to the embedding space. This is accomplished by
\begin{equation}
(\mathcal{T}^S_{ij} \Gamma)=\left( \left( \frac{\sqrt{2}}{\sqrt{\eta_i\cdot \eta_j}} \Gamma^{A'B'}  \eta_{iA'} \A_{ijB'}^{\phantom{ijB}A} C^{-1} \right)^{\lfloor S \rfloor} 
   \left( \frac{\eta_i \cdot \Gamma\, \eta_j \cdot \Gamma}{ 2 \eta_i \cdot \eta_j} \right)^{2(S-\lfloor S \rfloor)}  \right) \hat{\mathcal{P}}_{ij}^S. 
\label{eq:halfprojectorembed}
\end{equation}    
Since the half-projectors appear in both the OPE and correlation functions we summarize some of their properties below. The half-projectors carry two sets of indices  $(\mathcal{T}^S_{ij} \Gamma)_{\, a_1 \ldots a_{2S}}^{ A_1 \ldots A_{\lfloor S \rfloor} a}$ and they are transverse with respect to both sets of indices 
\begin{equation} 
  (\eta_i \cdot \Gamma)_x^{\phantom{x} a_k} (\mathcal{T}^S_{ij} \Gamma)_{  \ldots  a_k \ldots }^{A_1 \ldots A_{\lfloor S \rfloor} a } =0, \quad 
    \eta_{i A}  (\mathcal{T}^S_{ij} \Gamma)_{ \, a_1 \ldots a_{2S}}^{\, \ldots  A \ldots } 
       = \eta_{j A}  (\mathcal{T}^S_{ij} \Gamma)_{a_1 \ldots a_{2S}}^{\, \ldots  A \ldots  }  =0.
 \label{eq:halfptransverse}
 \end{equation}
The following properties  are apparent by noting that  there is an explicit projector $ \hat{\mathcal{P}}_{ij}^S$ in (\ref{eq:halfprojectorembed}). Projectors to different irreducible representations are orthogonal: 
\begin{equation}
  (\mathcal{T}^S_{ij})  \hat{\mathcal{P}}_{ij}^{S'} =  (\mathcal{T}^S_{ij} \Gamma) \,  \delta_{S,S'},
\end{equation}
and due to their irreducibility the half-projectors, and the projectors, must be traceless:
\begin{equation} 
  \A_{ij A A'} (\mathcal{T}^S_{ij} \Gamma)_{ \, a_1 \ldots a_{2S}}^{\,\ldots  A  \ldots A' \ldots }=0, \quad   (\mathcal{T}^S_{ij} \Gamma)_{ a_1 \ldots a_{2S}}^{\, \ldots  A  \ldots a } (\Gamma_A)_a^{\phantom{a} x} =0.
\end{equation}
   
\subsection{Two-Point Functions}

From the OPE formula (\ref{eq:ope}), it is not difficult to derive
the two-point correlation function \cite{Fortin:2019fvx} 

\begin{align}
\langle\mathcal{O}_{i}(\eta_{1})\mathcal{O}_{j}(\eta_{2})\rangle & =(\mathcal{T}_{12}^{\boldsymbol{N}_{i}}\Gamma)(\mathcal{T}_{21}^{\boldsymbol{N}_{j}}\Gamma)\cdot\frac{\lambda_{\boldsymbol{N}_{i}}\tensor*[_{a}]{c}{_{ij}^{k}}\hat{\mathcal{P}}_{12}^{\boldsymbol{N}_{i}}}{(\eta_{1}\cdot\eta_{2})^{\tau_{i}}}=(\mathcal{T}_{12}^{\boldsymbol{N}_{i}}\Gamma)\cdot(\mathcal{T}_{21}^{\boldsymbol{N}_{j}}\Gamma)\frac{\lambda_{\boldsymbol{N}_{i}}\tensor*[_{a}]{c}{_{ij}^{\mathds{1}}}}{(\eta_{1}\cdot\eta_{2})^{\tau_{i}}},\label{eq:2pt}
\end{align}
which will not play any role in later derivations, but it shows how half-projectors appear in correlation functions. There needs to be a half projector on the right-hand size for every operator, and the representations of the operators and the corresponding half-projectors match. This will be the case for higher-point functions as well. Going from the middle to the right in (\ref{eq:2pt}), projector $\hat{\mathcal{P}}_{12}^{\boldsymbol{N}_{i}}$ is dropped because the half-projectors already contain the same $\hat{\mathcal{P}}_{12}^{\boldsymbol{N}_{i}}$ and projectors are idempotent. The overall constant $\lambda_{\boldsymbol{N}_{i}}$ is usually absorbed into operator normalization. 

\subsection{Three-Point Functions from the OPE}
\label{sec:3ptOPE}

Applying the OPE (\ref{eq:ope}) to the first two operators in $\langle\mathcal{O}_{i}(\eta_{1})\mathcal{O}_{j}(\eta_{2})\mathcal{O}_{m}(\eta_{3})\rangle$
and using (\ref{eq:2pt}) for the leftover two-point function produces
\begin{equation}
\begin{aligned}\langle\mathcal{O}_{i}(\eta_{1})\mathcal{O}_{j}(\eta_{2})\mathcal{O}_{m}(\eta_{3})\rangle & =\frac{(\mathcal{T}_{12}^{\boldsymbol{N}_{i}}\Gamma)(\mathcal{T}_{21}^{\boldsymbol{N}_{j}}\Gamma)(\mathcal{T}_{31}^{\boldsymbol{N}_{m}}\Gamma)}{(\eta_{1}\cdot\eta_{2})^{\frac{1}{2}(\tau_{i}+\tau_{j}-\chi_{m})}(\eta_{1}\cdot\eta_{3})^{\frac{1}{2}(\chi_{i}-\chi_{j}+\tau_{m})}(\eta_{2}\cdot\eta_{3})^{\frac{1}{2}(-\chi_{i}+\chi_{j}+\chi_{m})}}\\
 & \qquad\qquad\cdot\sum_{a=1}^{N_{ijm}}\tensor*[_{a}]{c}{_{ijm}}\mathscr{G}_{(a|}^{ij|m},
\end{aligned}
\label{eq:3ptOPE}
\end{equation}
where $\mathscr{G}_{(a|}^{ij|m}$ is the three-point function in the OPE basis. The OPE coefficients $\tensor*[_{a}]{c}{_{ijm}}$ were introduced in (\ref{eq:ope}), while the tensor structures $\tensor*[_{a}]{t}{_{ij}^{12m}}$ are absorbed into $\mathscr{G}_{(a|}^{ij|m}$. 

When we later turn to four-point functions, operator $\mathcal{O}_{m}$ will not be an external operator, but instead be an exchange one. It is convenient to consider at once a tower of exchange operators with increasing spin. Consider specific representations $\boldsymbol{N}_{i}$ and $\boldsymbol{N}_{j}$ of operators $\mathcal{O}_{i}$ and $\mathcal{O}_{j}$, respectively. If a representation with spin $\boldsymbol{N}_{m}+i_a$ is allowed by symmetry for $\mathcal{O}_{m}$, so are representations with spin $\boldsymbol{N}_{m}+\ell$, for any $\ell \geq i_a$. In three dimensions, the smallest representations that do not contain any positive integer spins are $\boldsymbol{N}_{m}=0,\frac{1}{2}.$ We can separate the tensor structures 
\begin{equation}
    \tensor*[_{a}]{t}{_{ij}^{12m+\ell}} =   \tensor*[_{a}]{t}{_{ij}^{12m+i_a}} (\A_{12})^{\ell-i_a},
 \label{eq:tensorspecial}
\end{equation}
where $m+i_a$ is the smallest spin allowed by symmetry and $\ell-i_a$ label the increasing spins in the tower of operators.  

Using the tensor structure decomposition (\ref{eq:tensorspecial}), the three-point function can be simplified as
\begin{equation}
\begin{aligned}(\mathscr{G}_{(a|}^{ij|m+\ell})_{\{aA\}\{bB\}\{eE\}} & =\lambda_{\boldsymbol{N}_{m+\ell}}(\tensor*[_{a}]{t}{_{ij}^{12,m+i_{a}}})_{\{aA\}\{bB\}\{e'E'\}\{F\}}\\ 
 & \times\left((\tensor{\mathcal{A}}{_{321E}^{E'}})^{n_{v}^{m}+i_{a}}(\bbe{3}\cdot\Gamma\, \bbe{2}\cdot\Gamma\tensor{)}{_{e}^{e'}}(\tensor{\mathcal{A}}{_{321E}^{E'}})^{\ell-i_{a}}\right)_{cs_{3}}(-\bbe{2E'}\bbe{1F})^{\ell-i_{a}},
\end{aligned}
\label{eq:3ptfromOPE}
\end{equation}
where we specified the indices explicitly and introduced $\A_{321}=\A_{32}\cdot \A_{21}$. Upper-case indices are vectorial, but each label stands for a collective set of symmetrized indices. Lower-case indices $a$, $b$, $e$, and $e'$ are spinorial and each appears at most once. Note that lower-case $a$ is used as both a spinor index and a counting label enumerating diffrent tensor structures, but the two would be difficult to confuse. The index sets $\{aA\}$, $\{bB\}$ and $\{eE\}$ are, respectively, associated with the operators $\mathcal{O}_{i}$, $\mathcal{O}_{j}$, and $\mathcal{O}_{m}$. Indices $F$ are contracted with implicit indices inside $(\ldots)_{cs_3}$ that originate from the derivative operator in the OPE\@. The $\ell-i_a$ power of the transverse metric $\A_{12}$ in (\ref{eq:tensorspecial}) were reduced to $(-\bbe{2E'}\bbe{1F})$ in (\ref{eq:3ptfromOPE}) because the other terms in  $\A_{12}$ vanish under contraction given their index assignments.  Moreover in (\ref{eq:3ptfromOPE}), $n_v^m$ is the number of vector indices in the representation $m$, which turns out to always be 0 in three dimensions.  The double-barred coordinates $\eta$ are homogenized
\eqn{\bbe{i}^A=\frac{\ee{j}{m}{\frac{1}{2}}}{\ee{i}{j}{\frac{1}{2}}\ee{i}{m}{\frac{1}{2}}}\e{i}{A}{}.}[Eqetab3]
The double-bar notation might appear odd. We introduced it because later on we will use coordinates homogenized in a different way and those coordinates will be denoted with single bars.

We refer to $(\ldots)_{cs_{3}}$ as the ``conformal substitution" for a 3-point function. It is given by \cite{Fortin:2019pep} 
\begin{equation}
 (\ldots)_{cs_{3}} = (\ldots)_{(g)^{s_0}(\bbe{1})^{s_1}(\bbe{2})^{s_2}(\bbe{3})^{s_3} \to (g)^{s_0}(\bbe{1})^{s_1}(\bbe{3})^{s_3} 
     \bar{I}_{12}^{(d,h-n/2-s_2,n+s_2;\chi-s_1/2+s_2/2+s_3/2)}},
\label{eq:cs3}
\end{equation}
where the substitution implies collecting powers of $g$ and the homogenized coordinates $\bbe{}$ and substituting them with the expression involving
the three-point tensorial function derived in \cite{Fortin:2019fvx,Fortin:2019dnq} 
\eqn{\bar{I}_{12}^{(d,h,n;p)}=\rho^{(d,h;p)}\sum_{\substack{q_0,q_1,q_2,q_3\geq0\\\bar{q}=2q_0+q_1+q_2+q_3=n}}S_{(q_0,q_1,q_2,q_3)}K^{(d,h;p;q_0,q_1,q_2,q_3)},}[EqIb3]
where the totally symmetric $S$-tensor, the $\rho$-function, and the $K$-function are
\eqna{
S_{(q_0,q_1,q_2,q_3)}^{A_1\cdots A_{\bar{q}}}&=g^{(A_1A_2}\cdots g^{A_{2q_0-1}A_{2q_0}}\bbe{1}^{A_{2q_0+1}}\cdots\bbe{1}^{A_{2q_0+q_1}}\\
&\phantom{=}\qquad\times\bbe{2}^{A_{2q_0+q_1+1}}\cdots\bbe{2}^{A_{2q_0+q_1+q_2}}\bbe{3}^{A_{2q_0+q_1+q_2+1}}\cdots\bbe{3}^{A_{\bar{q}})},\\
\rho^{(d,h;p)}&=(-2)^h(p)_h(p+1-d/2)_h,\\
K^{(d,h;p;q_0,q_1,q_2,q_3)}&=\frac{(-1)^{\bar{q}-q_0-q_1-q_2}(-2)^{\bar{q}-q_0}\bar{q}!}{q_0!q_1!q_2!q_3!}\frac{(-h-\bar{q})_{\bar{q}-q_0-q_2}(p+h)_{\bar{q}-q_0-q_1}}{(p+1-d/2)_{-q_0-q_1-q_2}}.
}[EqK3]
In \eqref{EqK3}, $\bar{q}=2q_0+q_1+q_2+q_3=n$ and is the total number of indices on $\bar{I}_{12}^{(d,h,n;p)}$. 

The three-point tensorial function is totally symmetric and satisfies several convenient contiguous relations \cite{Fortin:2019fvx,Fortin:2019dnq}, given by
\eqna{
g\cdot\bar{I}_{12}^{(d,h,n;p)}&=0,\\
\bbe{1}\cdot\bar{I}_{12}^{(d,h,n;p)}&=\bar{I}_{12}^{(d,h+1,n-1;p)},\\
\bbe{2}\cdot\bar{I}_{12}^{(d,h,n;p)}&=\rho^{(d,1;-h-n)}\bar{I}_{12}^{(d,h,n-1;p)},\\
\bbe{3}\cdot\bar{I}_{12}^{(d,h,n;p)}&=\bar{I}_{12}^{(d,h+1,n-1;p-1)}.
}[EqCont3]
These will be of great utility in the determination of rotation matrices.  For future convenience, we also introduce $\widetilde{K}^{(d,h;p;q_0,q_1,q_2,q_3)}=\rho^{(d,h;p)}K^{(d,h;p;q_0,q_1,q_2,q_3)}$ to simplify computations.

Performing the conformal substitution and contracting the $\ell$-dependent indices in (\ref{eq:3ptfromOPE}), we obtain
\begin{equation}
\begin{aligned}( & \mathscr{G}_{(a|}^{ij|m+\ell})_{\{aA\}\{bB\}\{eE\}}\\
= & \lambda_{\boldsymbol{N}_{m+\ell}}({}_{a}t_{ij,m+i_{a}}^{12})_{\{aA\}\{bB\}\{e'E'\}\{F\}}(\bbe{3}\cdot\Gamma\Gamma_{F})_{e}^{\ e'}(-1)^{\ell_{a}}\\
 & \times\sum_{\sigma}\sum_{r_{0},r_{3},t_{0},t_{3}\geq0}\binom{n_{v}^{m+i_{a}}}{r_{0}+r_{3}}\binom{r_{0}+r_{3}}{r_{3}}\frac{(-1)^{r_{3}}}{n_{v}^{m+i_{a}}!}\binom{\ell_{a}}{t_{0}+t_{3}}\binom{t_{0}+t_{3}}{t_{3}}(-1)^{t_{3}}\\
 & \times\rho^{(d,\ell_{a}-t_{0}-t_{3};-h-n-\ell_{a})}\delta_{E_{\sigma(1)}}^{E'_{\sigma(1)}}\cdots\delta_{E_{\sigma(r_{0})}}^{E'_{\sigma(r_{0})}}\bbe{3}^{E'_{\sigma(r_{0}+1)}}\cdots\bbe{3}^{E'_{\sigma(r_{0}+r_{3})}}\\
 & \times\bar{I}_{12}^{(d,h'+2t_{0}+t_{3},n'-t_{0};p'-t_{0})}{}_{E_{\sigma(r_{0}+1)}\cdots E_{\sigma(n_{v}^{m+i_{a}})}E^{\ell_{a}-t_{0}}}^{E'_{\sigma(r_{0}+r_{3}+1)}\cdots E'_{\sigma(n_{v}^{m+i_{a}})}F^{n_{a}-\ell_{a}+2\xi_{m}}}(\bbe{2E})^{t_{0}}.
\end{aligned}
\label{eq:3ptblk}
\end{equation}
This results can be further contracted in the general case, but we find this form to be the most convenient to work with in three dimensions. 

\subsection{Three-Point Functions and Rotation Matrices}

Three-point functions can be written directly, without the use of the OPE, simply by using the representations of the operators 
\begin{equation}
\begin{aligned}\langle\mathcal{O}_{i}(\eta_{1})\mathcal{O}_{j}(\eta_{2})\mathcal{O}_{m}(\eta_{3})\rangle & =\frac{(\mathcal{T}_{12}^{\boldsymbol{N}_{i}}\Gamma)(\mathcal{T}_{21}^{\boldsymbol{N}_{j}}\Gamma)(\mathcal{T}_{31}^{\boldsymbol{N}_{m}}\Gamma)}{(\eta_{1}\cdot\eta_{2})^{\frac{1}{2}(\tau_{i}+\tau_{j}-\chi_{m})}(\eta_{1}\cdot\eta_{3})^{\frac{1}{2}(\chi_{i}-\chi_{j}+\tau_{m})}(\eta_{2}\cdot\eta_{3})^{\frac{1}{2}(-\chi_{i}+\chi_{j}+\chi_{m})}}\\
 & \qquad\qquad\cdot\sum_{a=1}^{N_{ijm}}\tensor*[_{a}]{\alpha}{_{ijm}}\mathscr{G}_{[a|}^{ij|m},
\end{aligned}
\label{eq:3ptdirect}
\end{equation}
where 
\eqn{\mathscr{G}_{[a|}^{ij|m}=\bbe{3}\cdot\Gamma\,\FCF{a}{i}{j}{m}{1}{2}(\A_{12},\Gamma_{12},\epsilon_{12};\A_{12}\cdot\bbe{3}),}[Eq3ptTS]
is referred to as the three-point basis. Factor $\bbe{3}\cdot\Gamma$  in the equation above only appears if $\mathcal{O}_{m}$ is fermionic. $\FCF{a}{i}{j}{m}{1}{2}$ are tensor structures constructed from $\A_{12}$, $\Gamma_{12}$, $\epsilon_{12}$, and $\A_{12}\cdot\bbe{3}$. In general, the set of all such structures is redundant and in Section \ref{sec:tensors} we will discuss how to find a basis of independent structures. The same can be said about the OPE tensor structures. When we discussed three point functions computed using the OPE we considered that the spin of $\mathcal{O}_{m}$  is $\boldsymbol{N}_{m}+\ell$. Just like we did in (\ref{eq:tensorspecial}), we can separate three-point tensor structures into the special part and the symmetric parts
\eqn{
\FCF{a}{i}{j}{,m+\ell}{1}{2}=\FCF{a}{i}{j}{,m+i_a}{1}{2}(\A_{12}\cdot\bar{\eta}_3)^{\ell-i_a}.
 }[Eq3ptTS-sp]

 Since (\ref{eq:3ptOPE}) and (\ref{eq:3ptdirect}) must be identical we have
 \eqn{\sum_{a=1}^{N_{ijm}}\cCF{a}{i}{j}{m}\mathscr{G}_{(a|}^{ij|m}=\sum_{a=1}^{N_{ijm}}\aCF{a}{i}{j}{m}\mathscr{G}_{[a|}^{ij|m}.}[EqRcoeff]
 The OPE and three-point calculations are related by a change of basis and must be related by linear transformations
 \eqn{\mathscr{G}_{(a|}^{ij|m}=\sum_{a'=1}^{N_{ijm}}(R_{ijm}^{-1})_{aa'}\mathscr{G}_{[a'|}^{ij|m},\qquad\qquad\cCF{a}{i}{j}{m}=\sum_{a'=1}^{N_{ijm}}\aCF{a'}{i}{j}{m}(R_{ijm})_{a'a},}[EqRbases]
 which are determined by comparing  (\ref{eq:3ptOPE}) to (\ref{eq:3ptdirect}). If both the OPE and the 3-point tensors are restricted to their minimal basis 
 we refer to the matrices $(R_{ijm})_{a'a}$ as the rotation matrices. As we will sometimes do, we can also work with non-minimal sets of tensors structures, in which case  $(R_{ijm})_{a'a}$ will not be invertible and we then call these transformation matrices. We will derive these objects in Section \ref{sec:3pt}.

\subsection{Shifted Projection Operators}

For the four-point calculations, it is useful to decompose the projection operators in the following fashion
\begin{equation}
\hat{\mathcal{P}}^{\bm{N}_{m}+\ell}=\sum_{t}\mathscr{A}_{t}(d,t)\hat{\mathcal{Q}}_{t}^{\bm{N}_{m}+\ell_{t}}\hat{\mathcal{P}}_{d+d_{t}}^{(\ell-\ell_{t})},\label{eq:projdec}
\end{equation}
where we refer to $\hat{\mathcal{Q}}$ as the special part and $\hat{\mathcal{P}}_{d+d_{t}}$ as a shifted projector when $d=3$, but $d_{t}\neq 0$. The shifted projectors are not actually projectors, at least not at $d=3$. For example, they are not traceless. Nevertheless, when we discuss towers of exchange operators such objects are handy because we can isolate $\ell$ dependence more easily. 

In (\ref{eq:projectorsposition}), we listed the projectors into integer and half-integer representations. The integer representations require no decomposition, the sum over $t$ in (\ref{eq:projdec}) contains only one term and $\hat{\mathcal{Q}}=1$. Furthermore, $\ell_t=d_t=0$. 

The half-integer case is more interesting,
\eqn{
    (\hat{\mathcal{P}}^{\ell+\frac{1}{2}})_{\alpha\mu_\ell\cdots\mu_1}^{\phantom{\alpha\mu_\ell\cdots\mu_1}\mu'_1\cdots\mu'_\ell\alpha'}    
    =\delta_\alpha^{\phantom{\alpha}\alpha'}(\hat{\mathcal{P}}_{d+2}^{\ell})_{\mu^\ell}^{\phantom{\mu^\ell}\mu'^\ell}
        +\frac{\ell}{2(-\ell+1-d/2)}(\gamma_{(\mu}\gamma^{(\mu'})_\alpha^{\phantom{\alpha}\alpha'}(\hat{\mathcal{P}}_{d+2}^{(\ell-1)})_{\mu^{\ell-1})}^{\phantom{\mu^{\ell-1})}\mu'^{\ell-1})}.
 }[EqPerple1odd]
All vector indices are symmetrized. There are two terms, so $t=1,2$ and the shifts, coefficients, and the special parts can be summarized as
\eqn{
\begin{array}{|cccc|}\hline
t & (d_t,\ell_t) & \mathscr{A}_t(d,\ell) & \hat{\mathcal{Q}}_t\\\hline
1 & (2,0) & 1 & \delta_\alpha^{\phantom{\alpha}\alpha'}\\
2 & (2,1) & \frac{\ell}{2(-\ell+1-d/2)} & (\gamma_\mu\gamma^{\mu'})_\alpha^{\phantom{\alpha}\alpha'}\\
\hline
\end{array}
}
This is a special case of a slightly more general result presented in \cite{Fortin:2020ncr}.

\subsection{Four-Point Conformal Blocks}

Finding four-point functions is most convenient by applying the OPE to the first two operators in a correlator and evaluating the remaining part in the three-point basis. One then finds \cite{Fortin:2019gck}
\begin{equation}
\begin{aligned} & \langle\mathcal{O}_{i}(\eta_{1})\mathcal{O}_{j}(\eta_{2})\mathcal{O}_{k}(\eta_{3})\mathcal{O}_{l}(\eta_{4})\rangle\\
= & \frac{(\mathcal{T}_{12}^{\boldsymbol{N}_{i}}\Gamma)(\mathcal{T}_{21}^{\boldsymbol{N}_{j}}\Gamma)(\mathcal{T}_{34}^{\boldsymbol{N}_{k}}\Gamma)(\mathcal{T}_{43}^{\boldsymbol{N}_{l}}\Gamma)}{(\eta_{1}\cdot\eta_{2})^{\frac{1}{2}(\tau_{i}-\chi_{i}+\tau_{j}+\chi_{j})}(\eta_{1}\cdot\eta_{3})^{\frac{1}{2}(\chi_{i}-\chi_{j}+\chi_{k}-\chi_{l})}(\eta_{1}\cdot\eta_{4})^{\frac{1}{2}(\chi_{i}-\chi_{j}-\chi_{k}+\chi_{l})}(\eta_{3}\cdot\eta_{4})^{\frac{1}{2}(-\chi_{i}+\chi_{j}+\tau_{k}+\tau_{l})}}\\
 & \qquad\cdot\sum_{m}\sum_{a=1}^{N_{ijm}}\sum_{b=1}^{N_{klm}}\tensor*[_{a}]{c}{_{ij}^{m}}\tensor*[_{b}]{\alpha}{_{klm}}\mathscr{G}_{(a|b]}^{ij|m|kl},
\end{aligned}
\end{equation}
where the mixed-basis four-point conformal blocks is explicitly given by
\eqna{
(\mathscr{G}_{(a|b]}^{ij|m|kl})_{\{aA\}\{bB\}\{cC\}\{dD\}}&=(\tOPE{a}{i}{j}{m}{1}{2})_{\{aA\}\{bB\}}^{\phantom{\{aA\}\{bB\}}\{Ee\}\{F\}}\left((-x_3)^{2\xi_m}(\A_{123E}^{\phantom{123E}E'})^{n_v^m}(\bbe{2}\cdot\Gamma)_e^{\phantom{e}e'}\right.\\
&\phantom{=}\qquad\left.\times(\bar{\eta}_3\cdot\Gamma\,\hat{\mathcal{P}}_{13}^{\boldsymbol{N}_m}\,\bar{\eta}_2\cdot\Gamma)_{e'E'}^{\phantom{e'E'}E''e''}(\FCF{b}{k}{l}{m}{3}{4})_{\{cC\}\{dD\}\{e''E''\}}\right)_{cs_4},
}[Eq4ptCBind]
and $\A_{123}=\A_{12}\cdot \A_{23}$. As before, the round bracket subscript $(a|$ refers to the OPE basis, while the square bracket $|b]$ to the three-point basis. In parallel to the discussion below (\ref{eq:3ptfromOPE}), we restored the Lorentz indices. The sets $\{aA\}$, $\{bB\}$, $\{cC\}$ and $\{dD\}$ correspond to the external operators at points 1, 2, 3, and 4, respectively, and are contracted with the matching half projectors. Meanwhile all terms inside $(\ldots)_{cs_4}$ are subject to a 4-point conformal substitution.

We need a few more definitions before we can completely unpack this result. First, the single-barred coordinates are
\eqn{
\begin{gathered}
\bar{\eta}_1^A=\frac{\ee{3}{4}{\frac{1}{2}}}{\ee{1}{3}{\frac{1}{2}}\ee{1}{4}{\frac{1}{2}}}\eta_1^A,\qquad\qquad\bar{\eta}_2^A=\frac{\ee{1}{3}{\frac{1}{2}}\ee{1}{4}{\frac{1}{2}}}{\ee{1}{2}{}\ee{3}{4}{\frac{1}{2}}}\eta_2^A,\\
\bar{\eta}_3^A=\frac{\ee{1}{4}{\frac{1}{2}}}{\ee{3}{4}{\frac{1}{2}}\ee{1}{3}{\frac{1}{2}}}\eta_3^A,\qquad\qquad\bar{\eta}_4^A=\frac{\ee{1}{3}{\frac{1}{2}}}{\ee{3}{4}{\frac{1}{2}}\ee{1}{4}{\frac{1}{2}}}\eta_4^A,\\
\end{gathered}
}[Eqetab4]
which are homogenized using all four coordinates in order to avoid the appearance of $\eta_2$ in all the barred coordinates except  $\bar{\eta}_2$. This is because the differential operator in the OPE acts on $\eta_2$, but does not act on any other coordinate. We define conformal cross-ratios as
\eqn{
x_3=\frac{\ee{1}{2}{}\ee{3}{4}{}}{\ee{1}{4}{}\ee{2}{3}{}}=\frac{u}{v},\qquad\qquad x_4=\frac{\ee{1}{2}{}\ee{3}{4}{}}{\ee{1}{3}{}\ee{2}{4}{}}=u.}
The conformal substitution is quite simple 
\eqn{ (\ldots)_{cs_4}=(\ldots)_{(\bar{\eta}_2)^{s_2}x_3^{r_3}x_4^{r_4}\to\bar{I}_{12;34}^{(d,h_1-n_1/2-s_2,n_1+s_2;-h_2+r_3,\chi+h_2+r_4)}},}
which only requires keeping track of the powers of $\bar{\eta}_2$ and of the cross ratios $x_3$ and $x_4$.

Not surprisingly, the four-point $\bar{I}$-function is complicated \cite{Fortin:2019fvx,Fortin:2019dnq}:
\eqn{\bar{I}_{12;34}^{(d,h,n;p_3,p_4)}=\sum_{\substack{q_0,q_1,q_2,q_3,q_4\geq0\\\bar{q}=2q_0+q_1+q_2+q_3+q_4=n}}S_{(\boldsymbol{q})}\rho^{(d,h;p_3+p_4)}x_3^{p_3+p_4+h+q_0+q_2+q_3+q_4}K_{12;34;3}^{(d,h;p_3,p_4;q_0,q_1,q_2,q_3,q_4)}(x_3;y_4),}[EqIb4]
with the totally symmetric object $S_{(\boldsymbol{q})}$ given by
\eqn{S_{(\boldsymbol{q})}^{A_1\cdots A_{\bar{q}}}=g^{(A_1A_2}\cdots g^{A_{2q_0-1}A_{2q_0}}\bar{\eta}_1^{A_{2q_0+1}}\cdots\bar{\eta}_1^{A_{2q_0+q_1}}\cdots\bar{\eta}_4^{A_{\bar{q}-q_4+1}}\cdots\bar{\eta}_4^{A_{\bar{q}})},}[EqS4]
where $\bar{q}=2q_0+q_1+q_2+q_3+q_4$ and $y_4=1-x_3/x_4$.  The function \eqref{EqS4} is an extension of \eqref{EqK3} to four points.

The $K$-function in (\ref{EqIb4}) is prescribed  by 
\eqna{
K_{12;34;3}^{(d,h;\boldsymbol{p};\boldsymbol{q})}(x_3;y_4)&=\frac{(-1)^{q_0+q_3+q_4}(-2)^{\bar{q}-q_0}\bar{q}!}{q_0!q_1!q_2!q_3!q_4!}\frac{(-h-\bar{q})_{\bar{q}-q_0-q_2}(p_3)_{q_3}(p_3+p_4+h)_{\bar{q}-q_0-q_1}}{(p_3+p_4)_{q_3+q_4}(p_3+p_4+1-d/2)_{-q_0-q_1-q_2}}(p_4)_{q_4}\\
&\phantom{=}\qquad\times K_{12;34;3}^{(d+2\bar{q}-2q_0,h+q_0+q_2;p_3+q_3,p_4+q_4)}(x_3;y_4),
}[EqK4]
where
\eqna{
K_{12;34;3}^{(d,h;p_3,p_4)}(x_3;y_4)&=\sum_{n_4,n_{34}\geq0}\frac{(-h)_{n_{34}}(p_3)_{n_{34}}(p_3+p_4+h)_{n_4}}{(p_3+p_4)_{n_4+n_{34}}(p_3+p_4+1-d/2)_{n_{34}}}\frac{(p_4)_{n_4}}{n_{34}!(n_4-n_{34})!}y_4^{n_4}\left(\frac{x_3}{y_4}\right)^{n_{34}}\\
&=G(p_4,p_3+p_4+h,p_3+p_4+1-d/2,p_3+p_4;u/v,1-1/v),
}[EqK0]
is the usual Exton $G$-function $G(\alpha,\beta,\gamma,\delta;x,y)$ with appropriately shifted parameters \cite{Exton_1995}, which is related to a conformal block for scalar external operators and  scalar exchange operator. Hence, the tensorial $\bar{I}$-function is constructed from linear combinations of the Exton $G$-function, or alternatively scalar exchange blocks.

Like the three-point tensorial function, the four-point $\bar{I}$-function \eqref{EqIb4} satisfies a set of contiguous relations \cite{Fortin:2019fvx,Fortin:2019dnq},
\eqna{
g\cdot\bar{I}_{12;34}^{(d,h,n;p)}&=0,\\
\bar{\eta}_1\cdot\bar{I}_{12;34}^{(d,h,n;p_3,p_4)}&=\bar{I}_{12;34}^{(d,h+1,n-1;p_3,p_4)},\\
\bar{\eta}_2\cdot\bar{I}_{12;34}^{(d,h,n;p_3,p_4)}&=\rho^{(d,1;-h-n)}\bar{I}_{12;34}^{(d,h,n-1;p_3,p_4)},\\
\bar{\eta}_3\cdot\bar{I}_{12;34}^{(d,h,n;p_3,p_4)}&=\bar{I}_{12;34}^{(d,h+1,n-1;p_3-1,p_4)},\\
\bar{\eta}_4\cdot\bar{I}_{12;34}^{(d,h,n;p_3,p_4)}&=\bar{I}_{12;34}^{(d,h+1,n-1;p_3,p_4-1)},
}[EqCont4]

Inserting the shifted form of the projectors in (\ref{eq:projdec}) as well as the separated forms of the tensor structures in (\ref{eq:tensorspecial}) and (\ref{Eq3ptTS-sp}) into (\ref{Eq4ptCBind}) leads to 
\eqna{
&(\mathscr{G}_{(a|b]}^{ij|m+\ell|kl})_{\{aA\}\{bB\}\{cC\}\{dD\}}\\
&\qquad=\sum_t\mathscr{A}_t(d,\ell)(\tOPE{a}{i}{j}{,m+i_a}{1}{2})_{\{aA\}\{bB\}}^{\phantom{\{aA\}\{bB\}}\{Ee\}\{F\}}(-\bar{\eta}_2^E\bar{\eta}_1^F)^{\ell-i_a}\left((-x_3)^{2\xi_m}(\A_{123E}^{\phantom{123E}E'})^{n_v^m+\ell}(\bbe{2}\cdot\Gamma)_e^{\phantom{e}e'}\right.\\
&\qquad\phantom{=}\left.\times(\bar{\eta}_3\cdot\Gamma\,\hat{\mathcal{Q}}_{13|t}^{\boldsymbol{N}_m+\ell_t\boldsymbol{e}_1}\hat{\mathcal{P}}_{13|d+d_t}^{(\ell-\ell_t)\boldsymbol{e}_1}\,\bar{\eta}_2\cdot\Gamma)_{e'E'}^{\phantom{e'E'}E''e''}(\FCF{b}{k}{l}{,m+i_b}{3}{4})_{\{cC\}\{dD\}\{e''E''\}}[(\A_{34}\cdot\bar{\bar{\eta}}_2)_{E''}]^{\ell-i_b}\right)_{cs_4}.
}
Inside the conformal substitution above, the terms $ \A_{123E}^{\ell'} \hat{\mathcal{P}}_{13|d'}^{\ell'}(\A_{34}\cdot\bbe{2})^{\ell'}$ can be combined into \cite{Fortin:2019gck,Fortin:2020ncr} Gegenbauer polynomials $C_{\ell'}^{(d'/2-1)}$. It is useful to regard these as polynomials in the variable 
\eqn{X=\frac{(\alpha_4-\alpha_2)x_4-(\alpha_3-\alpha_2)x_3}{2},}[EqX]
where $\alpha_i$ are counting parameters whose powers determine the details of the eventual substitution 
\eqn{\alpha_2^{s_2}\alpha_3^{s_3}\alpha_4^{s_4}x_3^{r_3}x_4^{r_4}\to G_{(\ell'-\ell,n_a-\ell,n_a+\ell'-2\ell+2i_a,0,0)F^{n_a-\ell+i_a}}^{ij|m+\ell|kl},}[EqSpp]
and $ G^{ij|m+\ell|kl}$ is given by
\eqna{
&G_{(n_1,n_2,n_3,n_4,n_5)A_1\cdots A_n}^{ij|m+\ell|kl}\\
&\qquad=\rho^{(d,(\ell+s_2-s_3-s_4+n_1)/2;-h_{ij,m+\ell}-(\ell+n_2)/2)}x_3^{-s_3}x_4^{-s_4}\\
&\qquad\phantom{=}\times\bar{I}_{12;34}^{(d,h_{ij,m+\ell}-(s_2-s_3-s_4+n_3)/2,n;-h_{kl,m+\ell}+(r_3-r_4+n_4)/2,\chi_{m+\ell}+h_{kl,m+\ell}-(r_3-r_4+n_5)/2)}{}_{A_1\cdots A_n}.
}[EqG]

We now have collected all the ingredients needed to give the final result, which we will use in Section \ref{sec:4pt}
\begin{equation}
\begin{aligned}( & \mathscr{G}_{(a|b]}^{ij|m+\ell|kl})_{\{aA\}\{bB\}\{cC\}\{dD\}}\\
= & \sum_{t}\mathscr{A}_{t}(d,\ell)\sum_{j_{a},j_{b}\geq0}\binom{i_{a}}{j_{a}}\binom{i_{b}}{j_{b}}\frac{(-\ell_{t})_{i_{a}-j_{a}}(-\ell_{t})_{i_{b}-j_{b}}(-\ell+\ell_{t})_{j_{a}}(-\ell+\ell_{t})_{j_{b}}}{(-\ell)_{i_{a}}(-\ell)_{i_{b}}}\\
 & \times\sum_{\substack{r,\boldsymbol{r}',\boldsymbol{r}''\geq0\\
r+2r'_{0}+r'_{1}+r'_{2}=j_{a}\\
r+2r''_{0}+r''_{1}+r''_{2}=j_{b}\\
r'_{0}+r'_{1}+r'_{3}=r''_{0}+r''_{1}+r''_{3}
}
}(-1)^{\ell-\ell'-i_{a}+r'_{1}+r'_{2}}\frac{(-2)^{r'_{3}+r''_{3}}\ell'!}{(d'/2-1)_{\ell'}}\mathscr{C}_{j_{a},j_{b}}^{(d+d_{t},\ell-\ell_{t})}(r,\boldsymbol{r}',\boldsymbol{r}'')\\
 & \times\left(C_{\ell'}^{(d'/2-1)}(X)\right)_{s_{(a|b]}^{ij|m+\ell|kl}(t,j_{a},j_{b},r,\boldsymbol{r}',\boldsymbol{r}'')},
\end{aligned}
\label{eq:4ptblk}
\end{equation}
where the $s$-substitution is
\begin{equation}
\begin{aligned} & s_{(a|b]}^{ij|m+\ell|kl}(t,j_{a},j_{b},r,\boldsymbol{r}',\boldsymbol{r}''):\alpha_{2}^{s_{2}}\alpha_{3}^{s_{3}}\alpha_{4}^{s_{4}}x_{3}^{r_{3}}x_{4}^{r_{4}}\\
 & \to(-1)^{2\xi_{m}}(\tensor*[_{a}]{t}{_{ij}^{12,m+i_{a}}}\tensor{)}{_{\{aA\}\{bB\}}^{\{Ee\}\{F\}}}(g_{E_{s}E_{s}})^{r'_{0}}(\tensor{\mathcal{S}}{_{E_{s}}^{E_{s}''}})^{r}[(\mathcal{S}\cdot\bbe{4})_{E_{s}}]^{r'_{2}}\\
 & \times\left(G_{(\ell'-\ell+2r'_{3},n_{a}-\ell,n'_{3},n'_{4},n'_{5})}^{ij|m+\ell|kl}\right)_{F^{n_{a}-\ell+i_{a}}E_{s}^{r'_{1}}}^{{E''_{s}}^{r''_{1}}{F''}^{4\xi_{m}}{F''}^{n_{b}-\ell+i_{b}}{F''}^{\ell_{t}-i_{b}+j_{b}}}(\bbe{2}^{E})^{\ell_{t}-i_{a}+j_{a}}\\
 & \times\tensor{(\Gamma_{F''}\,\bbe{3}\cdot\Gamma\,\mathcal{S}^{n_{v}^{m}+\ell_{t}}\hat{\mathcal{Q}}_{13|t}^{\boldsymbol{N}_{m}+\ell_{t}\boldsymbol{e}_{1}}\,\Gamma_{F''})}{_{eE^{n_{v}^{m}}E^{\ell_{t}-i_{a}+j_{a}}E_{s}^{i_{a}-j_{a}}}^{{E''_{s}}^{i_{b}-j_{b}}{E''}^{\ell_{t}-i_{b}+j_{b}}{E''}^{n_{v}^{m}}e''}}\\
 & \times[(\bbe{2}\cdot\mathcal{S})^{E''_{s}}]^{r''_{2}}(g^{E''_{s}E''_{s}})^{r''_{0}}(_{b}t_{kl}^{34,m+i_{b}})_{\{cC\}\{dD\}\{e''E''\}\{F''\}}(\mathcal{A}_{34E''F''})^{\ell_{t}-i_{b}+j_{b}}.
\end{aligned}
\label{eq:4ptsub}
\end{equation}
The coefficients $\mathscr{C}(r,\boldsymbol{r}',\boldsymbol{r}'')$ were introduced and computed in \cite{Fortin:2020ncr}:
\eqna{
\mathscr{C}_{j_{a},j_{b}}^{(d,\ell)}(r,\boldsymbol{r}',\boldsymbol{r}'')&=\frac{(-1)^{r+r'_2+r'_3+j_{a}+r''_2+r''_3+j_{b}}j_{a}!j_{b}!}{2^{r'_0+r'_3+r''_0+r''_3}r!r'_0!r'_1!r'_2!r'_3!r''_0!r''_1!r''_2!r''_3!}\frac{[(r'_0+r'_1-r'_3+r''_0+r''_1-r''_3)/2]!}{(-\ell)_{j_{a}}(-\ell)_{j_{b}}}\\
&\phantom{=}\qquad\times(-r'_0-r'_1)_{r''_3}(-r''_0-r''_1)_{r'_3}\frac{(-\ell)_{r+r'_0+r'_1+r'_2+r'_3+r''_0+r''_1+r''_2+r''_3}}{(-\ell+2-d/2)_{(r'_0+r'_1+r'_3+r''_0+r''_1+r''_3)/2}}.
}[EqCC]

\section{Tensor Structures}\label{sec:tensors}

In this section we enumerate independent tensor structures that appear in three-point functions and in the OPE. We start with the three-point function and build up the results from the simpler cases to more complicated ones. We will first consider three integer spin operators and afterwards  discuss operators with half integer spins as well. Finally, we will turn our attention to tensor structures that appear in the OPE. 

We start with the most general three-point function. Instead of the usual operator positions $\eta_1$, $\eta_2$, and $\eta_3$, that we used in Section \ref{sec:formalism}, we choose to place the operators at  $\eta_2$, $\eta_3$, and $\eta_4$ since these expressions will be later inserted into four-point functions. Then, 
\begin{equation}
 \langle\Op{m}{2}\Op{k}{3}\Op{l}{4}\rangle =
 \frac{(\mathcal{T}_{23}^{S_{m}}\Gamma)^{Ee} (\mathcal{T}_{34}^{S_{k}}\Gamma)^{Cc}(\mathcal{T}_{43}^{S_{l}}\Gamma)^{Dd}[(\bar{\bar{\eta}}_{2}\cdot\Gamma)^{2\xi_{m}}\tensor*[_{b}]{F}{_{klm}^{34}}]_{CcDdEe}}{\ee{3}{4}{\frac{1}{2}(\tau_{k}+\tau_{l}-\chi_{m})}\ee{3}{2}{\frac{1}{2}(\chi_{k}-\chi_{l}+\tau_{m})}\ee{4}{2}{\frac{1}{2}(-\chi_{k}+\chi_{l}+\chi_{m})}},
 \label{eq:3pt3pt}
\end{equation}
where we have restored indices but written them in a shortened form. For example, index $E$ stands for $l_m=\lfloor S_m \rfloor$ symmetrized indices while index $e$ appears $2 \xi_m=2(S_m-\lfloor S_m \rfloor)$ times, with analogous numbers for the indices $C$, $c$, $D$, $d$. Because these indices are all completely symmetrized, we need not distinguish the different indices and use the same collective notation for every index in any given symmetrized subset. We will need the explicit indices when we enumerate the three-point structures $\tensor*[_{b}]{F}{_{klm}^{34}}$. The subscript $b$ in  $\tensor*[_{b}]{F}{^{34}}$ enumerates different tensor structures and is not a spacetime index. 

 $\tensor*[_{b}]{F}{_{klm}^{34}}$ must saturate all spacetime indices of the half projectors and is a function of $\A_{34}$, $\Gamma_{34}$, $\epsilon_{34}$, and  $\A_{34}\cdot \bbe{2}$~\cite{Fortin:2019pep}. Because of the transversality properties of the half projectors, see (\ref{eq:halfptransverse}) coordinates with some indices cannot appear in $F_{34}$, as $\bbe{2E}$, $\bbe{3C}$, $\bbe{3D}$, $\bbe{3E}$, $\bbe{4C}$, $\bbe{4D}$ all vanish when contracted with the half projectors in (\ref{eq:3pt3pt}). For the same reason, many expressions are simplified when contracted with the half projectors, e.g. $\A_{34CD}\simeq g_{CD}$. We will use the $\simeq$ symbol to indicate expressions that are equivalent when contracted in (\ref{eq:3pt3pt}).

 \subsection{Integer spins only}

The most general expression for the three-point tensor can be written as 
\begin{equation}
\begin{aligned} & (\tensor[_{b}]{F}{}_{klm}^{34})_{C^{\ell_{k}}D^{\ell_{l}}E^{\ell_{m}}}=(\A_{34CD})^{b_{kl}}(\A_{34CE})^{b_{km}}(\A_{34DE})^{b_{lm}} (\epsilon_{34CDE})^{x} \\
 & \times(\epsilon_{34}\cdot\bbe{2})_{CD}^{\epsilon_{kl}}
  (\epsilon_{34}\cdot\bbe{2})_{CE}^{\epsilon_{km}} 
  (\epsilon_{34}\cdot\bbe{2})_{DE}^{\epsilon_{lm}}
 (\A_{34}\cdot\bbe{2})_{C}^{\ell_{k}-b_{kl}-b_{km}-\epsilon_{kl}-\epsilon_{km}-x}\\
 & \times(\A_{34}\cdot\bbe{2})_{D}^{\ell_{l}-b_{kl}-b_{lm}-\epsilon_{kl}-\epsilon_{lm}-x}
   (\A_{34}\cdot\bbe{2})_{E}^{\ell_{m}-b_{km}-b_{lm}-\epsilon_{km}-\epsilon_{lm}-x}\\
\end{aligned}
\label{eq:tenint}
\end{equation}
The powers indicate how many times each term appears. Due to index symmetry the powers uniquely determine a tensor structure. For generic powers,  the expressions form an over-complete set that is the tensors are linearly dependent. We will demonstrate how to choose a set of linearly independent structures. Clearly, the powers of $(\mathcal{A}_{34}\cdot\bbe{2})$ are determined once all the other powers are known as these objects simply saturate all the remaining indices. 

Since products of two Levi-Civita tensors can be always re-written in term of the metric tensors we only consider $x+ \epsilon_{kl}+\epsilon_{km}+\epsilon_{lm}\leq1$. First, we show that we can always set $x=0$. Since 6 indices cannot be anti-symmetrized in a 5-$d$ space we have
\begin{displaymath}
\begin{aligned} 
& 0=\epsilon_{[CDEGH}\bar{\bar{\eta}}_{3I]}\bar{\bar{\eta}}_{2}^{I}\bar{\bar{\eta}}_{4}^{G}\bar{\bar{\eta}}_{3}^{H}
    \simeq \frac{1}{6}(\epsilon_{34CDE}+\epsilon_{23CDE}),\\
 & 0=\epsilon_{[CDEGH}\bar{\bar{\eta}}_{2I]}\bar{\bar{\eta}}_{3}^{I}\bar{\bar{\eta}}_{2}^{G}\bar{\bar{\eta}}_{4}^{H}
      \simeq  \frac{1}{6}((\epsilon_{42CDE}+\epsilon_{23CDE}+\bar{\bar{\eta}}_{2D}(\epsilon_{34}\cdot\bar{\bar{\eta}}_{2})_{CE}-\bar{\bar{\eta}}_{2C}(\epsilon_{34}\cdot\bar{\bar{\eta}}_{2})_{DE}),\\
 & 0=\epsilon_{[CDEGH}\bar{\bar{\eta}}_{4I]}\bar{\bar{\eta}}_{3}^{I}\bar{\bar{\eta}}_{2}^{G}\bar{\bar{\eta}}_{4}^{H}\simeq 5!(\epsilon_{42CDE}+\epsilon_{34CDE}-\bar{\bar{\eta}}_{4E}(\epsilon_{34}\cdot\bar{\bar{\eta}}_{2})_{CD}),
\end{aligned}
\end{displaymath}
which we can solve for 
\begin{equation}
\epsilon_{34CDE}\simeq\frac{1}{2}[\bbe{4E}(\epsilon_{34}\cdot\bbe{2})_{CD}-\bbe{2C}(\epsilon_{34}\cdot\bbe{2})_{DE}+\bbe{2D}(\epsilon_{34}\cdot\bbe{2})_{CE}].
\label{eq:epsconv}
\end{equation}
This indicates that $x=1$ can be traded for non-zero powers of $\epsilon_{kl}$, $\epsilon_{km}$, and $\epsilon_{lm}$. 

Second, we use this result to show that we may set $b_{kl} b_{km} b_{lm}=0$. The product of three different metrics $\A_{34CD}\mathcal{A}_{34CE}\mathcal{A}_{34DE}$ can be traded for terms with only two $\A$'s and two $\A\cdot \bbe{2}$. We examine $g_{CD}g_{DE}g_{EC}\simeq \A_{34CD}\mathcal{A}_{34CE}\mathcal{A}_{34DE}$. Again, we do not need to distinguish different indices  in the same symmetrized index group, so index $D$ appearing twice stands for symmetrized indices $D_1$ and $D_2$, and so on. In the following equation we indicate anti-symmetrized indices by underscoring them while indices that are not underlined  in between the square brackets  are left out of the anti-symmetrization.
\begin{equation}
\begin{aligned} & g_{CD}g_{DE}g_{EC}
     \simeq \frac{3!}{2} g_{[\underline{C} D} g_{\underline{D} E} g_{\underline{E}]C}
     \simeq \frac{5!}{2}g_{[\underline{C} D} g_{\underline{D} E} g_{\underline{E}C}\bbe{3\underline{G}} \bbe{4\underline{H}]}\bbe{4}^{G}\bbe{3}^{H}\\
 & \simeq \frac{1}{2} g_{C'D}g_{D'E}g_{E'C}\bar{\bar{\eta}}_{3G'}\bar{\bar{\eta}}_{4H'}\epsilon^{C'D'E'G'H'}\epsilon_{CDEGH}\bbe{4}^{G}\bbe{3}^{H}\\
 &  \simeq  \frac{1}{2} \epsilon_{DECG'H'}\bbe{3}^{G'}\bbe{4}^{H'}\epsilon_{CDEGH}\bbe{4}^{G}\bbe{3}^{H}
  = - \frac{1}{2}  \epsilon_{34CDE}\epsilon_{34CDE}.
\end{aligned}
\label{eq:breduce}
\end{equation}
Next, we eliminate both $\epsilon_{34CDE}$'s in the last term above using (\ref{eq:epsconv}), yielding a square of the right-hand side of (\ref{eq:epsconv}). Expanding the square gives nine terms of the form $(\epsilon_{34}\cdot\bbe{2}) (\epsilon_{34}\cdot\bbe{2})$. Each of such terms can be re-written again as the determinant of the metric. Such a determinant is contracted with six coordinates: two each of $\bbe{2}$, $\bbe{3}$, and $\bbe{4}$ coming from $(\epsilon_{34}\cdot\bbe{2})$ squared. Thus, there can be at most two un-contracted metric tensors as there are only four free indices. This means that the product of three metrics at the beginning of (\ref{eq:breduce}) is a combination of terms with at most two metrics. Hence, every term has fewer $b_{xx}$'s compared to the original one and we can repeat this reasoning until one of the $b_{xx}$'s is $0$ that is until $b_{kl} b_{km} b_{lm}=0$.

Third, we turn to structures with $\epsilon_{kl}+\epsilon_{km}+\epsilon_{lm} = 1$. Choosing which of the three is non-zero does not lead to independent structures as long as all the powers $b_{xx}$ are kept fixed. For example, 
\begin{equation}
0=6! \epsilon_{[CDGHI}g_{E]C}\bbe{3}^{I}\bbe{4}^{H}\bbe{2}^{G}
   \simeq [(\epsilon_{34}\cdot\bbe{2})_{CD}g_{CE}-(\epsilon_{34}\cdot \bbe{2})_{CE}g_{CD}-\bbe{2C}\epsilon_{34CDE}], 
\label{eq:epsilonequiv}
\end{equation}
which means we can trade $\epsilon_{kl}$ for $\epsilon_{lm}$. This is because the last term on the right-hand side of (\ref{eq:epsilonequiv}) leads to terms with smaller values of $b_{xx}$, which are already included in the tensor structure basis. Therefore, we introduce $\epsilon_b=\epsilon_{kl}+\epsilon_{km}+\epsilon_{lm}$ and independent structures are obtained by choosing $\epsilon_b=0,1$. As an illustration, consider a structure with $b_{kl},b_{km}\neq 0$, but $b_{lm}=0$. In this case choosing either $\epsilon_{kl}=1$ (and all the other $\epsilon_{xx}=0$) is equivalent to choosing instead $\epsilon_{lm}=1$.

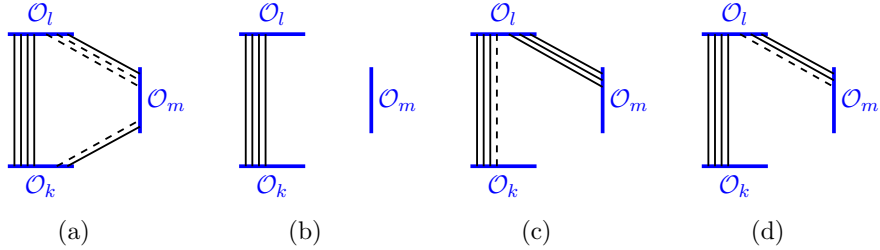
\begin{figure}[htb]
\centering
\resizebox{12cm}{!}{
\begin{tikzpicture}[thick]
\begin{scope}
\draw[-, ultra thick,blue] (0,0) -- (1,0);
\draw[- , ultra thick,blue] (0,2) -- (1,2);
\draw[-,  ultra thick,blue ] (2,0.5) -- (2,1.5);
\draw[-] (0.1,0) -- (0.1,2);
\draw[-] (0.2,0) -- (0.2,2);
\draw[-] (0.3,0) -- (0.3,2);
\draw[-] (0.4,0) -- (0.4,2);
\draw[-] (0.9,2) -- (2,1.4);
\draw[dashed] (0.74,2) -- (2,1.3);
\draw[dashed] (0.58,2) -- (2,1.2);
\draw[-] (0.9,0) -- (2,0.6);
\draw[dashed] (0.74,0) -- (2,0.7);
\draw[blue] (0.5,-0.3) node{$\mathcal{O}_{k}$};
\draw[blue] (0.5,2.3)  node{$\mathcal{O}_{l}$};
\draw[blue] (2.4,1)  node{$\mathcal{O}_{m}$};
\draw (1,-1) node{(a)};
\end{scope}
\begin{scope}[xshift=3.5cm]
\draw[-, ultra thick,blue] (0,0) -- (1,0);
\draw[- , ultra thick,blue] (0,2) -- (1,2);
\draw[-,  ultra thick,blue ] (2,0.5) -- (2,1.5);
\draw[-] (0.1,0) -- (0.1,2);
\draw[-] (0.2,0) -- (0.2,2);
\draw[-] (0.3,0) -- (0.3,2);
\draw[-] (0.4,0) -- (0.4,2);
\draw[blue] (0.5,-0.3) node{$\mathcal{O}_{k}$};
\draw[blue] (0.5,2.3)  node{$\mathcal{O}_{l}$};
\draw[blue] (2.4,1)  node{$\mathcal{O}_{m}$};
\draw (1,-1) node{(b)};
\end{scope}
\begin{scope}[xshift=7cm]
\draw[-, ultra thick,blue] (0,0) -- (1,0);
\draw[- , ultra thick,blue] (0,2) -- (1,2);
\draw[-,  ultra thick,blue ] (2,0.5) -- (2,1.5);
\draw[-] (0.1,0) -- (0.1,2);
\draw[-] (0.2,0) -- (0.2,2);
\draw[-] (0.3,0) -- (0.3,2);
\draw[dashed] (0.4,0) -- (0.4,2);
\draw[-] (0.9,2) -- (2,1.4);
\draw[-] (0.74,2) -- (2,1.3);
\draw[-] (0.58,2) -- (2,1.2);
\draw[blue] (0.5,-0.3) node{$\mathcal{O}_{k}$};
\draw[blue] (0.5,2.3)  node{$\mathcal{O}_{l}$};
\draw[blue] (2.4,1)  node{$\mathcal{O}_{m}$};
\draw (1,-1) node{(c)};
\end{scope}
\begin{scope}[xshift=10.5cm]
\draw[-, ultra thick,blue] (0,0) -- (1,0);
\draw[- , ultra thick,blue] (0,2) -- (1,2);
\draw[-,  ultra thick,blue ] (2,0.5) -- (2,1.5);
\draw[-] (0.1,0) -- (0.1,2);
\draw[-] (0.2,0) -- (0.2,2);
\draw[-] (0.3,0) -- (0.3,2);
\draw[-] (0.4,0) -- (0.4,2);
\draw[-] (0.9,2) -- (2,1.4);
\draw[-] (0.74,2) -- (2,1.3);
\draw[dashed] (0.58,2) -- (2,1.2);
\draw[blue] (0.5,-0.3) node{$\mathcal{O}_{k}$};
\draw[blue] (0.5,2.3)  node{$\mathcal{O}_{l}$};
\draw[blue] (2.4,1)  node{$\mathcal{O}_{m}$};
\draw (1,-1) node{(d)};
\end{scope}
\end{tikzpicture}
}
\caption{Pictorial representation of three-point tensor structures. Solid lines represent contractions with $\A_{34}$, while the dashed lines contractions with $\epsilon_{34} \cdot \bbe{2}$.  In the notation introduced in (\ref{eq:tenint}), diagram (a) corresponds to $b_{kl}=4$, $b_{lm}=b_{km}=1$, $\epsilon_{lm}=2$, and $\epsilon_{km}=1$. Diagram (a) does not depict an independent structure in our basis for two reasons. First, it contains more than one dashed line. Second, independent diagrams contain at most two sets of lines that is contractions among one pair of operators must vanish to satisfy $(b_{kl}+\epsilon_{kl})(b_{km}+\epsilon_{km})(b_{lm}+\epsilon_{lm})=0$. Diagrams (b) and (c) are examples of independent tensor structures in our basis. However, diagrams (c) and (d) are not independent and only one of the two belongs in the basis, as shown using (\ref{eq:epsilonequiv}). All the indices that are left over after lines are drawn correspond to contractions with $\A_{34} \cdot \bbe{2}$, for instance in diagram (b) all indices of $\mathcal{O}_{m}$ are saturated with $\A_{34} \cdot \bbe{2}$.}
\label{Fig3pt}
\end{figure}

Last, we observe that even when $\epsilon_b=1$ we can restrict to independent structures by satisfying  $(b_{kl}+\epsilon_{kl})(b_{km}+\epsilon_{km})(b_{lm}+\epsilon_{lm})=0$. We consider 
\begin{equation}
0=(\epsilon_{34}\cdot\bbe{2})_{[\underline{C} D} g_{\underline{D} E}g_{\underline{E} C}
   \bbe{2\underline{G}}  \bbe{3\underline{H}}  \bbe{4\underline{I}]} \bbe{2}^{H}\bbe{3}^{I}\bbe{4}^{G},
\end{equation}
which can be used to express tensor structures that include  $(\epsilon_{34}\cdot\bar{\bar{\eta}}_{2})_{CD}g_{DE}g_{EC}$ in terms of structures in which  that combination is absent, thus getting to $(b_{kl}+\epsilon_{kl})(b_{km}+\epsilon_{km})(b_{lm}+\epsilon_{lm})=0$.  Figure~\ref{Fig3pt} contains a graphic representation of tensor structures. The independent structures have simple geometric properties.

 \subsection{Two half integers and one integer spin}

None of the arguments in the fermion-free case in the previous section change at all in the presence of fermions. Therefore, from the beginning we set $x=0$ and consider the most general tensor structure with fermions
\begin{equation}
\begin{aligned} & ((\bbe{2}\cdot\Gamma)^{2\xi_{m}}\tensor[_{b}]{F}{}_{klm}^{34})_{C^{\ell_{k}}c^{2\xi_{k}}D^{\ell_{l}}d^{2\xi_{l}}E^{\ell_{m}}e^{2\xi_{m}}}=(\A_{34CD})^{b_{kl}}(\A_{34CE})^{b_{km}}(\A_{34DE})^{b_{lm}} \\
 & \times(\epsilon_{34}\cdot\bbe{2})_{CD}^{\epsilon_{kl}}
  (\epsilon_{34}\cdot\bbe{2})_{CE}^{\epsilon_{km}} 
  (\epsilon_{34}\cdot\bbe{2})_{DE}^{\epsilon_{lm}}
 (\A_{34}\cdot\bbe{2})_{C}^{\ell_{k}-b_{kl}-b_{km}-\epsilon_{kl}-\epsilon_{km}-4\delta_{b}(\xi_{l}\xi_{m})}\\
 & \times(\A_{34}\cdot\bbe{2})_{D}^{\ell_{l}-b_{kl}-b_{lm}-\epsilon_{kl}-\epsilon_{lm} -4\delta_{b}(\xi_{k}\xi_{m})}
  (\A_{34}\cdot\bbe{2})_{E}^{\ell_{m}-b_{km}-b_{lm}-\epsilon_{km}-\epsilon_{lm} -4\delta_{b}(\xi_{k}\xi_{l})}\\
  & \times[(\bbe{2}\cdot\Gamma)^{2\xi_{m}}(\Gamma_{34C^{4\xi_{l}\xi_{m}}D^{4\xi_{k}\xi_{m}}E^{4\xi_{k} \xi_{l}}})^{\delta_{b}}
  (\bbe{2}\cdot\Gamma_{34})^{|\epsilon'_{b}-\delta_{b}|} \, C_{\Gamma}^{-1}]_{c^{2\xi_{k}}d^{2\xi_{l}}e^{2\xi_{m}}}.
  \end{aligned}
\label{eq:tengen}
\end{equation}

As we did earlier, we eliminate redundant structures. We amend $\epsilon_b$ to include $\epsilon_{b}=\epsilon_{kl}+\epsilon_{km}+\epsilon_{lm}+\epsilon'_{b}$. In what follows we assume that $\xi_{k}=\xi_{l}=\frac{1}{2}$, but for the other two possibilities the arguments would have been completely analogous. 

First, we show that  structures with $\epsilon_{b}(1-\epsilon'_{b})=1$
are redundant. When
$\delta_{b}=0$, this is apparent from
\begin{equation}
\begin{aligned}0=6!\epsilon_{[CDGHI}\bbe{3}^{I}\bbe{4}^{H}\bbe{2}^{G}(\Gamma_{X]}\Gamma^{X}C_{\Gamma}^{-1})_{cd}
\simeq 5!(3(\epsilon_{34}\cdot\bbe{2})_{CD}(C_{\Gamma}^{-1})_{cd}+g_{CD}(\bbe{2}\cdot\Gamma_{34}C_{\Gamma}^{-1})_{cd}).\end{aligned}
\label{eq:epsgaconv}
\end{equation}
When $\delta_{b}=1$, the same rule follows from
\begin{equation}
\begin{aligned} & 0=6!\epsilon_{[CDGHI}\bbe{3}^{I}\bbe{4}^{H}\bbe{2}^{G}(\Gamma_{34E]}\Gamma_{34}\cdot\bar{\bar{\eta}}_{2}C_{\Gamma}^{-1})_{cd}\\
 & \simeq 5!((\epsilon_{34}\cdot\bbe{2})_{CD}(\Gamma_{34E}\bar{\bar{\eta}}_{2}\cdot\Gamma_{34}C_{\Gamma}^{-1})_{cd}+2\epsilon_{34CDE}(C_{\Gamma}^{-1})_{cd}-2\bbe{2D}(\epsilon_{34}\cdot\bbe{2})_{CE}(C_{\Gamma}^{-1})_{cd}),
\end{aligned}
\end{equation}
since we can use (\ref{eq:epsconv}) and (\ref{eq:epsgaconv}) in turn to reduce the last two terms above to structures that have not been shown to be redundant.

Second, we demonstrate that $\epsilon_b\leq1$;  that is, if $\epsilon_{kl}+\epsilon_{km}+\epsilon_{lm}=\epsilon'_{b}=1$, such a structure is redundant.  Indeed, when $\delta_{b}=0$
\begin{equation}
\begin{aligned} & (\epsilon_{34}\cdot\bbe{2})_{CD}(\Gamma_{34}\cdot\bar{\bar{\eta}}_{2}C_{\Gamma}^{-1})_{cd}=-\mathscr{K}(\epsilon_{34}\cdot\bbe{2})_{CD}\epsilon_{34}^{XYZ}\bar{\bar{\eta}}_{2Z}(\Gamma_{XY}C_{\Gamma}^{-1})_{cd}\\
 & \simeq -\mathscr{K}\delta_{CDGHI}^{XYZJK}\bar{\bar{\eta}}_{3}^{I}\bbe{4}^{H}\bbe{2}^{G}\bar{\bar{\eta}}_{2Z}\bbe{3K}\bbe{4J}(\Gamma_{X}\Gamma_{Y}C_{\Gamma}^{-1})_{cd},
\end{aligned}
\end{equation}
 only leads to non-redundant structures without any explicit $\epsilon$ tensors and whose spinor part is always $(C_{\Gamma}^{-1})_{cd}$. In other words, the original structure is shown to be redundant to structures having $\epsilon'_{b}=\epsilon_{b}=0$. Here, $\mathscr{K}$ is the proportionality constant in both $\gamma^{\mu_1\mu_2 \mu_3}=\mathscr{K}\epsilon^{\mu_1\mu_2\mu_3} $ and  $\Gamma^{A_1 \ldots  A_5}= \mathscr{K}\epsilon^{A_1 \ldots  A_5}$. A similar argument holds
when $\delta_{b}=1$
\begin{equation}
\begin{aligned} & (\epsilon_{34}\cdot\bbe{2})_{CD}(\Gamma_{34E}C_{\Gamma}^{-1})_{cd}=-\mathscr{K}(\epsilon_{34}\cdot\bbe{2})_{CD}\epsilon_{34}^{XYE'}g_{EE'}(\Gamma_{XY}C_{\Gamma}^{-1})_{cd}\\
 & \simeq -\mathscr{K}\delta_{CDGHI}^{XYE'JK}\bbe{3}^{I}\bbe{4}^{H}\bbe{2}^{G}g_{EE'}{}_{3K}\bbe{4J}(\Gamma_{X}\Gamma_{Y}C_{\Gamma}^{-1})_{cd}.
\end{aligned}
\end{equation}

The final observation is that if any pair of the $(k,l,m)$
operators $(p,q)$ are fermionic then $b_{pq}\delta_{b}=0$. To see this when $\epsilon_{b}=1$ note that
\begin{equation}
\begin{aligned}0=6!(g_{C[D \, }\epsilon_{XYEGH]}\bbe{3}^{H}\bbe{4}^{G})(\Gamma^{XY}C_{\Gamma}^{-1})_{cd}
   \simeq5!(\mathscr{K}g_{CD}(\Gamma_{E}C_{\Gamma}^{-1})_{cd}-2\epsilon_{34CDE}(C_{\Gamma}^{-1})_{cd}),\end{aligned}
\end{equation}
so, using (\ref{eq:epsconv}), the original redundant structure eventually
reduces to independent structures having $\delta_{b}=0$. When
$\epsilon_{b}=0$, the argument is not very different
\begin{equation}
\begin{aligned} & 0=6!(g_{C[D}\, \epsilon_{XYEGH]}\bbe{3}^{H}\bar{\bar{\eta}}_{4}^{G})(\Gamma_{34}^{XY}\bbe{2}\cdot\Gamma_{34}C_{\Gamma}^{-1})_{cd}\\
 & \simeq 5!(\mathscr{K}g_{CD}(\Gamma_{34E}\bbe{2}\cdot\Gamma_{34}C_{\Gamma}^{-1})_{cd}-2\epsilon_{34CDE}(\bbe{2}\cdot\Gamma_{34}C_{\Gamma}^{-1})_{cd}-2g_{CE}\bbe{2D}(C_{\Gamma}^{-1})_{cd}).
\end{aligned}
\end{equation}

\subsection{Tensor structure summary and counting}
\label{sec:counting}

To summarize,  the most general tensor structure is (\ref{eq:tengen}):
\begin{equation}
\begin{aligned} & ((\bbe{2}\cdot\Gamma)^{2\xi_{m}}\tensor[_{b}]{F}{}_{klm}^{34})_{C^{\ell_{k}}c^{2\xi_{k}}D^{\ell_{l}}d^{2\xi_{l}}E^{\ell_{m}}e^{2\xi_{m}}}=(\A_{34CD})^{b_{kl}}(\A_{34CE})^{b_{km}}(\A_{34DE})^{b_{lm}} \\
 & \times(\epsilon_{34}\cdot\bbe{2})_{CD}^{\epsilon_{kl}}
  (\epsilon_{34}\cdot\bbe{2})_{CE}^{\epsilon_{km}} 
  (\epsilon_{34}\cdot\bbe{2})_{DE}^{\epsilon_{lm}}
 (\A_{34}\cdot\bbe{2})_{C}^{i_{k}} (\A_{34}\cdot\bbe{2})_{D}^{i_{l}} (\A_{34}\cdot\bbe{2})_{E}^{i_m}\\
  & \times[(\bbe{2}\cdot\Gamma)^{2\xi_{m}}(\Gamma_{34C^{4\xi_{l}\xi_{m}}D^{4\xi_{k}\xi_{m}}E^{4\xi_{k} \xi_{l}}})^{\delta_{b}}
  (\bbe{2}\cdot\Gamma_{34})^{|\epsilon'_{b}-\delta_{b}|}C_{\Gamma}^{-1}]_{c^{2\xi_{k}}d^{2\xi_{l}}e^{2\xi_{m}}},
  \end{aligned}
\label{eq:tenfinal}
\end{equation}
where the term in the last line above is absent when there are no fermionic operators and  we introduced
\begin{equation}
\begin{aligned}
  i_{k} & =\ell_{k}-b_{kl}-b_{km}-\epsilon_{kl}-\epsilon_{km}-4\delta_{b}(\xi_{l}\xi_{m}),\\
  i_{l} & =\ell_{l}-b_{kl}-b_{lm}-\epsilon_{kl}-\epsilon_{lm}-4\delta_{b}(\xi_{k}\xi_{m}), \\
  i_{m} & =\ell-b_{km}-b_{lm}-\epsilon_{km}-\epsilon_{lm}-4\delta_{b}(\xi_{k}\xi_{l}).
\end{aligned}
\label{eq:tenis}
\end{equation}

Linearly independent structures obey the following constraints 
\begin{equation}
(b_{kl}+\epsilon_{kl})(b_{km}+\epsilon_{km})(b_{lm}+\epsilon_{lm})=0,\label{eq:tricon}
\end{equation}
(with $b_{pq}+\epsilon_{pq}\leq\min[\ell_{p},\ell_{q}]$) and 
\begin{equation}
\epsilon_{b}=\epsilon_{kl}+\epsilon_{km}+\epsilon_{lm}+\epsilon'_{b}=0\text{ or }1,\label{eq:epscons}
\end{equation}
while choosing any of the components to be $1$ is not independent from any other choices. For configurations with $(p,q)$ fermionic, we choose $\epsilon'_{b}=\epsilon_{b}$ and demand that $b_{pq}\delta_{b}=0$.

For bosonic cases, it is unambiguous to label the configurations by triplets of pairs
\begin{equation}
b=\{b_{km}\oplus\epsilon_{km},b_{lm}\oplus\epsilon_{lm},b_{kl}\oplus\epsilon_{kl}\}.
\end{equation}
When there are fermionic operators, it is necessary to include an additional label
for $\epsilon_{b}$. As an example, for $(k,l)$ fermionic, the unambiguous
labeling is given by
\begin{equation}
b=\{b_{km}\oplus\frac{\delta_{b}}{2},b_{lm}\oplus\frac{\delta_{b}}{2},b_{kl}\oplus\frac{1-\delta_{b}}{2};\epsilon_{b}\}.
\end{equation}
We note that  $\epsilon_{b}$ indicates the presence of the Levi-Civita tensor in the bosonic case and the number of $\Gamma$-matrix insertions in the fermionic case, and corresponds to the ``parity'' of the tensor structure.

We have thus reduced the set of all possible tensor structures to a subset. We eliminated tensors that linearly depend on other tensors. To ensure that the reduction is complete we need to verify that the number of the structures agrees with the expected number of independent tensor structures.  Without loss of generality, we assume that $\ell_{m}\geq\ell_{l}\geq\ell_{k}$.

We first consider the case of pure bosons with spins satisfying $\ell_{m}\geq\ell_{l}+\ell_{k}$. Firstly, there is always one configuration with $b=\{0,0,0\}$. Then, for contractions between a single pair of operators, there are $\ell_{k}$ possibilities each for $b=\{0,0,a_{kl}\}$ and $b=\{a_{km},0,0\}$, and $\ell_{l}$ possibilities for $b=\{0,a_{lm},0\}$. The counting is doubled by taking $\epsilon_{b}=1$, so the total number of single-pair configurations is $2(2\ell_{k}+\ell_{l})$. For contractions between two pairs, there are $\ell_{k}\ell_{l}$ possibilities for $b=\{a_{km},a_{lm},0\}$. Furthermore, $b=\{a_{km},0,a_{kl}\}$ would have $\sum_{n=1}^{\ell_{k}}(\ell_{k}-n)=\ell_{k}\ell_{k}-\frac{1}{2}\ell_{l}(\ell_{l}+1)$, while $b=\{0,a_{lm},a_{kl}\}$ have $\sum_{n=1}^{\ell_{k}}(\ell_{l}-n)=\ell_{k}\ell_{l}-\frac{1}{2}\ell_{l}(\ell_{l}+1)$. Doubling the count due to $\epsilon_{b}=1$, the total contribution from two-pair configuration is given by $2\ell_{l}(\ell_{k}+\ell_{l})-\ell_{l}(\ell_{l}+1)+\ell_{k}\ell_{l}$. There are no configurations with contractions between all pairs due to the constraint (\ref{eq:tricon}), so in total, the counting is
\begin{displaymath}
  1+2(2\ell_{k}+\ell_{l})+2\ell_{l}(\ell_{k}+\ell_{l})-2\ell_{l}(\ell_{l}+1)+2\ell_{k}\ell_{l}=(2\ell_{k}+1)(2\ell_{l}+1).
\end{displaymath}

On the other hand, when $\ell_{m}<\ell_{k}+\ell_{l}$, everything
else remains the same except the counting for $b=\{a_{km},a_{lm},0\}$
is no longer $2\ell_{k}\ell_{l}$. Instead, the counting is given
by $2(\ell_{m}-\ell_{l})\ell_{l}+2\sum_{n=\ell_{m}-\ell_{l}+1}^{\ell_{k}}(\ell_{m}-n)=2(\ell_{m}-\ell_{l})\ell_{l}+(\ell_{m}-\ell_{l}-\ell_{k})(\ell_{k}-\ell_{l}-\ell_{m}+1)$.
Therefore, the total count in this case is given by
\begin{align*}
(2\ell_{k} & +1)(2\ell_{l}+1)-2\ell_{k}\ell_{l}+2(\ell_{m}-\ell_{l})\ell_{l}+(\ell_{m}-\ell_{l}-\ell_{k})(\ell_{k}-\ell_{l}-\ell_{m}+1)\\
= & (2\ell_{k}+1)(2\ell_{l}+1)-(\ell_{k}+\ell_{l}-\ell_{m})(\ell_{k}+\ell_{l}-\ell_{m}+1).
\end{align*}

The two purely bosonic cases can be summarized by the formula
\begin{equation}
N(\ell_{k},\ell_{l},\ell_{m})=(2\ell_{k}+1)(2\ell_{l}+1)-p(p+1),
\end{equation}
where $p=\max[\ell_{k}+\ell_{l}-\ell_{m},0]$ which is consistent
with \cite{Costa:2011mg,Kravchuk:2016qvl}. The analysis goes through
along the same for the fermionic cases. Thus, we have proved that Eq.
(\ref{eq:tenfinal}) with the constraints described below that equation provides a basis of non-redundant tensor structures.  

\subsection{Obtaining the OPE Basis}

As pointed out in \cite{Fortin:2020ncr}, the OPE tensor structures 
can be obtained from the three-point basis by taking $\tensor[_{b}]{F}{}_{klm}^{34}$ and transforming
all $\A_{34}\cdot\bar{\bar{\eta}}_{2}\rightarrow\A_{34}$,
with the extra index, $F$, contracting with the OPE differential
operator. Hence, our convention is
\begin{equation}
\begin{aligned} & (\tensor*[_{b}]{t}{_{klm}^{34}})_{C^{\ell_{k}}c^{2\xi_{k}}D^{\ell_{l}}d^{2\xi_{l}}E^{\ell_{m}}e^{2\xi_{m}}F^{n_{b}}}\\
 & =(\A_{34CD})^{b_{kl}}(\A_{34CE})^{b_{km}}(\A_{34DE})^{b_{lm}}(\epsilon_{34CDF})^{\epsilon_{kl}}(\epsilon_{34CEF})^{\epsilon_{km}}(\epsilon_{34DEF})^{\epsilon_{lm}}\\
 & \times(\A_{34})_{CF}^{i_{k}}(\A_{34})_{DF}^{i_{l}}(\A_{34})_{EF}^{i_{m}}[(\Gamma_{34C^{4\xi_{l}\xi_{m}}D^{4\xi_{k}\xi_{m}}E^{4\xi_{k}\xi_{l}}})^{\delta_{b}}(\Gamma_{34F})^{|\epsilon'_{b}-\delta_{b}|}C_{\Gamma}^{-1}]_{c^{2\xi_{k}}d^{2\xi_{l}}e^{2\xi_{m}}},
\end{aligned}
\label{eq:klope}
\end{equation}
with the same rules governing the powers as outlined previously.

The basis for the other two operators are given by simple relabelings
\begin{equation}
\begin{aligned} & (\tensor*[_{a}]{t}{_{ijm}^{12}})_{A^{\ell_{i}}a^{2\xi_{i}}B^{\ell_{j}}b^{2\xi_{j}}E^{\ell_{m}}e^{2\xi_{m}}F^{n_{a}}}\\
 & =(\A_{12AB})^{a_{ij}}(\A_{12AE})^{a_{im}}(\A_{12BE})^{a_{jm}}(\epsilon_{12ABF})^{\epsilon_{ij}}(\epsilon_{12BEF})^{\epsilon_{im}}(\epsilon_{12BEF})^{\epsilon_{jm}}\\
 & \times(\A_{12})_{AF}^{i_{i}}(\A_{12})_{BF}^{i_{j}}(\A_{12})_{EF}^{i_{m}}[(\Gamma_{12A^{4\xi_{j}\xi_{m}}B^{4\xi_{i}\xi_{m}}E^{4\xi_{i}\xi_{j}}})^{\delta_{a}}(\Gamma_{12F})^{|\epsilon'_{a}-\delta_{a}|}C_{\Gamma}^{-1}]_{a^{2\xi_{i}}b^{2\xi_{j}}e^{2\xi_{m}}},
\end{aligned}
\end{equation}
where we have a similar set of shorthand notations
\begin{align}
i_{i} & =\ell_{i}-a_{ij}-a_{im}-\epsilon_{ij}-\epsilon_{im}-4\delta_{a}(\xi_{j}\xi_{m}),\\
i_{j} & =\ell_{j}-a_{ij}-a_{jm}-\epsilon_{ij}-\epsilon_{jm}-4\delta_{a}(\xi_{i}\xi_{m}),\\
i_{m} & =\ell-a_{im}-a_{jm}-\epsilon_{im}-\epsilon_{jm}-4\delta_{a}(\xi_{i}\xi_{j}).
\end{align}
Note that $i_m$ above and in (\ref{eq:tenis}) are not the same variable. Moreover, the structure $\tensor*[_{a}]{t}{_{ij}^{12,m+i_{a}}}$
($\tensor[_{b}]{t}{_{kl}^{34,m+i_{b}}}$) is obtained from the one above by omitting all factors of $\A_{12EF}$ ($\A_{34EF}$).

\section{Three-Point Correlators and Transformation Matrices}
\label{sec:3pt}

In this section we specialize the general recipe of three-point conformal blocks to three dimensions. Since the simplest way of computing four-point conformal blocks is in the mixed basis, as we mentioned in Section \ref{sec:formalism}, we need to relate the OPE and 3-point basis to one another. This is done by comparing the expressions for three-point functions obtained either directly or by the use of the OPE. Terms with the same dependence on spacetime coordinates and the metric tensor can differ by numerical coefficients only and such numerical differences simply reflect change of basis. 

Due to the redundancies among tensor structures shown in the previous section, we find that the most convenient way of relating different bases is using an over-complete set of tensor structures. Hence, the ``rotation matrices'' are more generalized transformations from the over-complete three-point bases to a complete OPE basis. If needed, it is not difficult to restrict the set of tensor structures to a minimal set. 

\subsection{General three-point function in 3D}

We start with the three-dimensional version of the general formula (\ref{eq:3ptblk})
\begin{equation}
\begin{aligned}( & \mathscr{G}_{\vert b)}^{m+\ell\vert kl})_{\{cC\}\{dD\}\{eE\}}\\
= & \lambda_{\ell}(_{b}t_{kl}^{34,m+i_{b}}\tensor{)}{_{\{cC\}\{dD\}}^{\{E'e'\}\{F\}}}[(\bar{\bar{\eta}}_{2}\cdot\Gamma\Gamma^{F}C_{\Gamma}^{-1})_{ee'}]^{2\xi_{m}}\\
\times & \sum_{r_{0},r_{2},t_{0},t_{2}\geq0}\left(\begin{array}{c}
i_{b}\\
r_{0}+r_{2}
\end{array}\right)\left(\begin{array}{c}
r_{0}+r_{2}\\
r_{2}
\end{array}\right)\left(\begin{array}{c}
\ell_{b}\\
t_{0}+t_{2}
\end{array}\right)\left(\begin{array}{c}
t_{0}+t_{2}\\
t_{2}
\end{array}\right)(-1)^{\ell_{b}+r_{2}+t_{2}}\\
\times & \rho^{(3,\ell_{b}-t_{0}-t_{2};-h-n_{b}-2i_{b}-\ell_{b}-2\xi_{m})}\sum_{\substack{q_0,q_2,q_3,q_4\geq0\\ \bar{q}=2 q_0+q_2+q_3+q_4}}\tilde{K}^{(3,h+2r_{0}+r_{2}+2t_{0}+t_{2};p-r_{0}-t_{0};q_{0},q_{1},q_{2},q_{3})}\\
\times & (g_{EE'})^{r_{0}}(\bar{\bar{\eta}}_{2E'})^{r_{2}}S_{(q_{0},q_{2},q_{3},q_{4})}{}_{E'^{i_{b}-r_{0}-r_{2}}F^{n_{b}-\ell_{b}+2\xi_{m}}E^{\ell-r_{0}-t_{0}}}(\bar{\bar{\eta}}_{4E})^{t_{0}},
\end{aligned}
\end{equation}
where $\bar{q}$ has the same value as in (\ref{eq:3ptOPEgeneral}) below. Any OPE tensor structures can be described by the expression
\begin{equation}
\begin{aligned} & (_{b}t_{kl}^{34,m+i_{b}}\tensor{)}{_{\{cC\}\{dD\}}^{\{E'e'\}\{F\}}}\\
\simeq & (\delta_{C}^{F})^{\ell_{k}-b_{kl}-b_{km}-a_{k}}(\delta_{D}^{F})^{\ell_{l}-b_{kl}-b_{lm}-a_{l}}(g_{CD})^{b_{kl}}(\delta_{C}^{E'})^{b_{km}}(\delta_{D}^{E'})^{b_{lm}}[T]_{c^{2\xi_{k}}C^{a_{k}}d^{2\xi_{l}}D^{a_{l}}}^{E'^{a_{m}}e'^{2\xi_{m}}F{}^{a_{s}}},
\end{aligned}
\end{equation}
where the ``special part'' $[T]_{c^{2\xi_{k}}C^{a_{k}}d^{2\xi_{l}}D^{a_{l}}}^{E'^{a_{m}}e'^{2\xi_{m}}F{}^{a_{s}}}$
denotes  either the $\epsilon$-tensor or the spinorial part of the tensor
structure. The special part is quite simple in three-dimensions because all representations are symmetric.  From the explicit formula (\ref{eq:klope}), it is straightforward
to read off the coefficients
\begin{align}
a_{k} & =\epsilon_{kl}+\epsilon_{km}+4\xi_{l}\xi_{m}\delta_{b},\\
a_{l} & =\epsilon_{kl}+\epsilon_{lm}+4\xi_{k}\xi_{m}\delta_{b},\\
a_{m} & =\epsilon_{km}+\epsilon_{lm}+4\xi_{k}\xi_{l}\delta_{b},\\
a_{s} & =\epsilon_{kl}+\epsilon_{km}+\epsilon_{lm}+|\epsilon'_{b}-\delta_{b}|.
\end{align}
Expressed in this form, the shorthands for the counting parameters are
\begin{align*}
i_{b} & =b_{km}+b_{lm}+a_{m},\\
\ell_{b} & =\ell-i_{b},\\
n_{b} & =\ell_{k}+\ell_{l}+\ell-2i_{kl}-2i_{km}-2i_{lm}-a_{k}-a_{l}-a_{m}+a_{s}.
\end{align*}

Substituting in the explicit expressions for the tensor structures,
it is then possible to contract the non-special vectorial indices
\begin{equation}
\begin{aligned}( & \mathscr{G}_{\vert b)}^{m+\ell\vert kl})_{\{cC\}\{dD\}\{eE\}}\\
= & \lambda_{\ell}\, (g_{CD})^{b_{kl}}\sum_{r_{0},r_{2},s_{0},s_{2},t_{0},t_{2}\geq0}\sum_{b_{0}+b_{2}+b_{s}=a_{m}}^{1}
[T]_{c^{2\xi_{k}}C^{a_{k}}d^{2\xi_{l}}D^{a_{l}}}^{E'^{a_{m}}e'^{2\xi_{m}}F{}^{a_{s}}}[(\bar{\bar{\eta}}_{2}\cdot\Gamma\Gamma^{F}C_{\Gamma}^{-1})_{ee'}]^{2\xi_{m}}(g_{EE'})^{b_{0}}(\bar{\bar{\eta}}_{2E'})^{b_{2}}\\
& \times 
 \left(\begin{array}{c}
   r_{0}\\
     s_{0},r_{0}-s_{0}-b_{0},b_{0}
\end{array}\right)
\left(\begin{array}{c}
   r_{2}\\
    s_{2},r_{2}-s_{2}-b_{2},b_{2}
\end{array}\right) \\
& \times 
\left(\begin{array}{c}
   i_{b}-r_{0}-r_{2}\\
   i_{km}-s_{0}-s_{2},i_{lm}-r_{0}-r_{2}+s_{0}+s_{2}+b_{0}+b_{2},b_{s}
\end{array}\right)
\left[ \left(\begin{array}{c}
   i_{b}\\
   b_{km},b_{lm},a_{m}
\end{array}\right)\right]^{-1} \\
& \times
 \left(\begin{array}{c}
    i_{b}\\
    r_{0},r_{2},i_{b}-r_{0}-r_{2}
\end{array}\right)\left(\begin{array}{c}
\ell_{b}\\
t_{0},t_{2},\ell_{b}-t_{0}-t_{2}
\end{array}\right)(-1)^{\ell_{b}+r_{2}+t_{2}}\rho^{(3,\ell_{b}-t_{0}-t_{2};-h-n_{b}-2i_{b}-\ell_{b}-2\xi_{m})}\\
& \times \sum_{\substack{q_0,q_2,q_3,q_4\geq0\\ \bar{q}= n-2r_0-r_2- t_0}}\tilde{K}^{(3,h+2r_{0}+r_{2}+2t_{0}+t_{2};p-r_{0}-t_{0};q_{0},q_{2},q_{3},q_{4})}(g_{CE})^{s_{0}}(g_{DE})^{r_{0}-s_{0}-b_{0}}(\bar{\bar{\eta}}_{2C})^{s_{2}}(\bar{\bar{\eta}}_{2D})^{r_{2}-s_{2}-b_{2}}\\
& \times  S_{(q_{0},q_{2},q_{3},q_{4})}{}_{C^{\ell_{k}-b_{kl}-s_{0}-s_{2}-a_{k}}D^{\ell_{l}-b_{kl}-r_{0}-r_{2}+s_{0}+s_{2}-a_{l}+b_{0}+b_{2}}E^{\ell-r_{0}-t_{0}}E'^{b_{s}}F^{a_{s}+2\xi_{m}}}(\bar{\bar{\eta}}_{4E})^{t_{0}},
\end{aligned}
\label{eq:3ptOPEgeneral}
\end{equation}
where $\bar{q}=2q_0+q_2+q_3+q_4$ and $n=2 n_v^m+2\xi_m+n_b+2 i_b$. Note that in three dimensions $n_v^m=0$ because the special part of the tensor has either spin 0 or $\frac{1}{2}$.  In the equation above, we used the trinomial symbols $\left(\begin{array}{c} N\\ a,b,c \end{array}\right)=\frac{N!}{a!b!c!}$, where $N=a+b+c$. Later on, we will also use the analogous multinomial coefficients.

\subsubsection{Disentangle the $S$ index structures}

Next, we need to disentangle the index structures of the special part $S$. To achieve this, first we note that there can be at most two of the special
$E'F$ contractions to $S$. This is equivalent to $a_m+a_s+2\xi_m \leq 2$, which follows from the choice of basis in Section \ref{sec:tensors}. In the rest of this section, we will refer collectively to the special indices $E'$ and $F$ as the $Z$ indices.  The term $\Gamma_{34}^{Z_{1}}\Gamma_{34}^{Z_{2}}$ appears in the  case of two contractions  (even the $\Gamma^{F}$ inside $\bar{\bar{\eta}}_{2}\cdot\Gamma\Gamma^{F}$ can be expressed as $\Gamma_{34}^{F}+\bar{\bar{\eta}}_{4}\cdot\Gamma\bar{\bar{\eta}}_{3}^{F}+\bar{\bar{\eta}}_{3}\cdot\Gamma\bar{\bar{\eta}}_{4}^{F}$) and since the indices are symmetrized the Clifford algebra implies equivalence of this term to $-2\bar{\bar{\eta}}_{3}^{Z_{1}}\bar{\bar{\eta}}_{4}^{Z_{2}}$. These two contractions of the coordinates are evaluated by the contiguous relations, after which there is at most a single special contraction to be considered. Therefore, we can express the general result as
\begin{equation}
\begin{aligned}( & \mathscr{G}_{\vert b)}^{m+\ell\vert kl})_{\{cC\}\{dD\}\{eE\}}\\
  = & \lambda_{\ell}(g_{CD})^{b_{kl}}\sum_{r_{0},r_{2},s_{0},s_{2},t_{0},t_{2}\geq0}\sum_{\substack{b_{0}+b_{2}+b_{s}=a_{m}\\
     c_{3}+c_{4}+c_{s}=2\xi_{m}
}
}^{1}[T]_{c^{2\xi_{k}}C^{a_{k}}d^{2\xi_{l}}D^{a_{l}}}^{E'^{a_{m}}e'^{2\xi_{m}}F{}^{a_{s}}}  (g_{Z_{1}Z_{2}}/3)^{e_{2}} \\
\times & [(\bbe{2}\cdot\Gamma\, \Gamma^{F}_{34})^{c_{s}}(\bar{\bar{\eta}}_{2}\cdot\Gamma \, \bbe{3}\cdot\Gamma)^{c_{3}}(\bar{\bar{\eta}}_{2}\cdot\Gamma \, \bbe{4}\cdot\Gamma)^{c_{4}} C^{-1}_{\Gamma}]_{e  e'}    \left(\begin{array}{c}
\ell_{b}\\
t_{0},t_{2},\ell_{b}-t_{0}-t_{2}
\end{array}\right) \\
\times & \frac{\left(\begin{array}{c}
i_{b}\\
s_{0},s_{2},b_{km}-s_{0}-s_{2},r_{0}-s_{0}-b_{0},r_{2}-s_{2}-b_{2},b_{lm}-r_{0}-r_{2}+s_{0}+s_{2}+b_{0}+b_{2},a_{m}
\end{array}\right)}{\left(\begin{array}{c}
i_{b}\\
b_{km},b_{lm},a_{m}
\end{array}\right)}\\
\times &(g_{EE'})^{b_{0}} (g_{CE})^{s_{0}}(g_{DE})^{r_{0}-s_{0}-b_{0}} (\bar{\bar{\eta}}_{2E'})^{b_{2}}(-1)^{\ell_{b}+r_{2}+t_{2}+e_{2}}\rho^{(3,\ell_{b}-t_{0}-t_{2}+c_{3}+e_{2};-h-n_{b}-2i_{b}-\ell_{b}-2\xi_{m})}\\
\times & \sum_{\substack{q_0,q_2,q_3,q_4\geq0\\ \bar{q}= n-2r_0-r_2- t_0-2 e_2-c_3-c_4}} \tilde{K}^{(3,h+2r_{0}+r_{2}+2t_{0}+t_{2}+c_{4}+e_{2};p-r_{0}-t_{0};q_{0},q_{2},q_{3},q_{4})}(\bar{\bar{\eta}}_{2C})^{s_{2}}(\bar{\bar{\eta}}_{2D})^{r_{2}-s_{2}-b_{2}}\\
\times & S_{(q_{0},q_{2},q_{3},q_{4})}{}_{C^{\ell_{k}-b_{kl}-s_{0}-s_{2}-a_{k}}D^{\ell_{l}-b_{kl}-r_{0}-r_{2}+s_{0}+s_{2}-a_{l}+b_{0}+b_{2}}E^{\ell-r_{0}-t_{0}}Z^{e_{1}}}(\bar{\bar{\eta}}_{4E})^{t_{0}},
\end{aligned}
\label{eq:3ptspecial}
\end{equation}
where the metric $g_{Z_{1}Z_{2}}$ in the first line above (divided by $d=3$) is a bookkeeping device to account for the double special contractions of the tensor structure $[T]$ and the $\Gamma$ matrices in the second line because $\Gamma_{34}^{Z_1} \Gamma_{34}^{Z_2} g_{Z_{1}Z_{2}}=3$. These are the contractions that were discussed just above (\ref{eq:3ptspecial}). The remaining  index of the single special contraction is denoted by $Z$. The two new powers are defined as $e_{1}=\delta_{a_{s}+b_{s}+c_{s},1}$ and $e_{2}=\delta_{a_{s}+b_{s}+c_{s},2}$. (We note that
this means $a_{s}+b_{s}+c_{s}=e_{1}+2e_{2}$.) 

In this way, the indices are disentangled as
\begin{equation}
\begin{aligned} & S_{(q_{0},q_{2},q_{3},q_{4})}{}_{C^{\ell_{k}-b_{kl}-s_{0}-s_{2}-a_{k}}D^{\ell_{l}-b_{kl}-r_{0}-r_{2}+s_{0}+s_{2}-a_{l}+b_{0}+b_{2}}E^{\ell-r_{0}-t_{0}}Z^{e_{1}}}\\
= & \sum_{p_{2}\geq0}\sum_{\substack{d_{C}+d_{D}+d_{E}\\
     +d_{2}=e_{1} }}^{1}
\left(\begin{array}{c}
   q_{2}\\
  p_{2},q_{2}-p_{2}-d_{2},d_{2}
\end{array}\right)
\left(\begin{array}{c}
   q_{0}\\
    E_{kl},E_{km},E_{lm},d_{C}+d_{D}+d_{E}
\end{array}\right) \\
& \times 
\left[
 \left(\begin{array}{c}
n_{b}+2i_{b}-2r_{0}-r_{2}-t_{0}-a_{m}-a_{s}+b_{0}+b_{2}+e_{1}\\
\ell-r_{0}-t_{0},\ell_{k}-b_{kl}-s_{0}-s_{2}-a_{k},\ell_{l}-b_{kl}-r_{0}-r_{2}+s_{0}+s_{2}-a_{l}+b_{0}+b_{2},e_{1}
\end{array}\right) \right]^{-1}\\
\times & 2^{q_{0}}(g_{CD})^{E_{km}}(g_{CE})^{E_{km}} (g_{DE})^{E_{lm}} \\
\times & (g_{CZ})^{d_{C}}(g_{DZ})^{d_{D}}(g_{EZ})^{d_{E}}(\bar{\bar{\eta}}_{2C})^{p_{2}}(\bar{\bar{\eta}}_{2D})^{q_{2}-p_{2}-d_{2}}(\bar{\bar{\eta}}_{2Z})^{d_{2}}(\bar{\bar{\eta}}_{4E})^{q_{4}},
\end{aligned}
\end{equation}
where to keep the expressions compact we denoted the exponents of the metrics as 
\begin{eqnarray}
  E_{kl}& =& q_{0}+q_{4}+r_{0}+t_{0}-\ell-d_{C}-d_{D}, \\
  E_{km}&=&\ell_{k}+\ell-b_{kl}-p_{2}-q_{0}-q_{4}-r_{0}-s_{0}-s_{2}-t_{0}-a_{k}+d_{C}+d_{D}, \\
  E_{lm}&=&b_{kl}-\ell_{k}+p_{2}+q_{0}+s_{0}+s_{2}+a_{k}-d_{C}-d_{D}-d_{E}. 
\end{eqnarray}
Substituting these into the full expression for the three-point function we obtain
\begin{equation}
\begin{aligned}( & \mathscr{G}_{\vert b)}^{m+\ell\vert kl})_{\{cC\}\{dD\}\{eE\}}\\
= & \lambda_{\ell}\sum_{r_{0},r_{2},s_{0},s_{2},t_{0},t_{2}\geq0}\sum_{\substack{b_{0}+b_{2}+b_{s}=a_{m}\\
c_{3}+c_{4}+c_{s}=2\xi_{m}\\
d_{C}+d_{D}+d_{E}+d_{2}=e_{1}}}^{1}
\sum_{\substack{p_{2},q_{0},q_{2},q_{4}\geq0 \\ 
     \bar{q}= n-2r_0-r_2- t_0-2 e_2-c_3-c_4 }}  
\frac{(-1)^{\ell_{b}+r_{2}+t_{2}+e_{2}}}{\left(\begin{array}{c}
i_{b}\\
b_{km},b_{lm},a_{m}
\end{array}\right)}\\
\times & \left(\begin{array}{c}
i_{b}\\
s_{0},s_{2},b_{km}-s_{0}-s_{2},r_{0}-s_{0}-b_{0},r_{2}-s_{2}-b_{2},b_{lm}-r_{0}-r_{2}+s_{0}+s_{2}+b_{0}+b_{2},a_{m}
\end{array}\right)\\
\times &\frac{\left(\begin{array}{c}
\ell_{b}\\
t_{0},t_{2},\ell_{b}-t_{0}-t_{2}
\end{array}\right)\left(\begin{array}{c}
q_{2}\\
p_{2},q_{2}-p_{2}-d_{2},d_{2}
\end{array}\right)\left(\begin{array}{c}
q_{0}\\
E_{kl},E_{km},E_{lm},d_{C}+d_{D}+d_{E}
\end{array}\right)}{\left(\begin{array}{c}
n_{b}+2i_{b}-2r_{0}-r_{2}-t_{0}-a_{m}-a_{s}+b_{0}+b_{2}+e_{1}\\
\ell-r_{0}-t_{0},\ell_{k}-b_{kl}-s_{0}-s_{2}-a_{k},\ell_{l}-b_{kl}-r_{0}-r_{2}+s_{0}+s_{2}-a_{l}+b_{0}+b_{2},e_{1}
\end{array}\right)}\\
\times & 2^{q_{0}}\rho^{(3,\ell_{b}-t_{0}-t_{2}+c_{3}+e_{2};-h-n_{b}-2i_{b}-\ell_{b}-2\xi_{m})}\tilde{K}^{(3,h+2r_{0}+r_{2}+2t_{0}+t_{2}+c_{4}+e_{2};p-r_{0}-t_{0};q_{0},q_{2},0,q_{4})}\\
\times & (g_{CE})^{\ell_{k}+\ell-b_{kl}-p_{2}-q_{0}-q_{4}-r_{0}-s_{2}-t_{0}-a_{k}+d_{C}+d_{D}}(g_{DE})^{b_{kl}-\ell_{k}+p_{2}+q_{0}+r_{0}+s_{2}+a_{k}-b_{0}-d_{C}-d_{D}-d_{E}}\\
\times & (g_{CD})^{b_{kl}+q_{0}+q_{4}+r_{0}+t_{0}-\ell-d_{C}-d_{D}}(\bar{\bar{\eta}}_{2C})^{p_{2}+s_{2}}(\bar{\bar{\eta}}_{2D})^{q_{2}-p_{2}+r_{2}-s_{2}-b_{2}-d_{2}}(\bar{\bar{\eta}}_{4E})^{q_{4}+t_{0}}\\
\times & [T]_{c^{2\xi_{k}}C^{a_{k}}d^{2\xi_{l}}D^{a_{l}}}^{E'^{a_{m}}e'^{2\xi_{m}}F{}^{a_{s}}}[(\bar{\bar{\eta}}_{2}\cdot\Gamma\, \Gamma^{F}_{34})^{c_{s}} 
      (\bar{\bar{\eta}}_{2}\cdot\Gamma\, \bbe{3}\cdot\Gamma)^{c_{3}}(\bar{\bar{\eta}}_{2}\cdot\Gamma\, \bbe{4}\cdot\Gamma)^{c_{4}} C^{-1}_{\Gamma}]_{e e'}\\
\times & (g_{EE'})^{b_{0}}(\bbe{2E'})^{b_{2}}(g_{CZ})^{d_{C}}(g_{DZ})^{d_{D}}(g_{EZ})^{d_{E}}(\bar{\bar{\eta}}_{2Z})^{d_{2}}(g_{Z_{1}Z_{2}}/3)^{e_{2}}.
\end{aligned}
\end{equation}

\subsubsection{The one coefficient that rules them all}

We can make the relabelings
\begin{align}
\tilde{b}_{kl} & =b_{kl}+q_{0}+q_{4}+r_{0}+t_{0}-\ell-d_{C}-d_{D},\\
\tilde{b}_{km} & =\ell_{k}+\ell-b_{kl}-p_{2}-q_{0}-q_{4}-r_{0}-s_{2}-t_{0}-a_{k}+d_{C}+d_{D},\\
\tilde{b}_{lm} & =b_{kl}-\ell_{k}+p_{2}+q_{0}+r_{0}+s_{2}+a_{k}-b_{0}-d_{C}-d_{D}-d_{E},
\end{align}
so that the summation indices can be relabeled as
\begin{align*}
p_{2} & =\ell_{k}-\tilde{b}_{km}-\tilde{b}_{kl}-s_{2}-a_{k}-d_{C},\\
q_{0} & =\tilde{b}_{kl}+\tilde{b}_{km}+\tilde{b}_{lm}-b_{kl}-r_{0}+b_{0}+d_{C}+d_{D}+d_{E}\\
 & =i_{\tilde{b}}+\tilde{b}_{kl}-b_{kl}-r_{0}+b_{0}+d_{C}+d_{D}+d_{E},\\
q_{2} & =\ell_{k}+\ell_{l}-2\tilde{b}_{kl}-\tilde{b}_{km}-\tilde{b}_{lm}-r_{2}-a_{k}-a_{l}+b_{2}-d_{C}-d_{D}+d_{2}\\
 & =n_{\tilde{b}}-\ell_{\tilde{b}}-r_{2}-a_{k}-a_{l}+b_{2}-d_{C}-d_{D}+d_{2},\\
q_{4} & =\ell-\tilde{b}_{km}-\tilde{b}_{lm}-t_{0}-b_{0}-d_{E} =\ell_{\tilde{b}}-t_{0}-b_{0}-d_{E},
\end{align*}
where we introduced the shorthand
\begin{align*}
i_{\tilde{b}} & =\tilde{b}_{km}+\tilde{b}_{lm},\\
\ell_{\tilde{b}} & =\ell-i_{\tilde{b}},\\
n_{\tilde{b}} & =\ell_{k}+\ell_{l}+\ell-2\tilde{b}_{kl}-2\tilde{b}_{km}-2\tilde{b}_{lm}.
\end{align*}
In this notation, we may rewrite the three-point functions as
\begin{equation}
\begin{aligned}( & \mathscr{G}_{\vert b)}^{m+\ell\vert kl})_{\{cC\}\{dD\}\{eE\}}\\
= & \lambda_{\ell}\sum_{\tilde{b}}\sum_{\substack{b_{0}+b_{2}+b_{s}=a_{m}\\
c_{3}+c_{4}+c_{s}=2\xi_{m}\\
d_{C}+d_{D}+d_{E}+d_{2}=e_{1}
}
}^{1}{}_{b\tilde{b}}\omega{}_{(b_{0},b_{3},b_{s};c_{3},c_{4},c_{s};d_{C},d_{D},d_{E},d_{2})}^{klm}(g_{CD})^{\tilde{b}_{kl}}(g_{CE})^{\tilde{b}_{km}}(g_{DE})^{\tilde{b}_{lm}}\\
\times & (\bar{\bar{\eta}}_{2C})^{\ell_{k}-\tilde{b}_{km}-\tilde{b}_{kl}-a_{k}-d_{C}}(\bar{\bar{\eta}}_{2D})^{\ell_{l}-\tilde{b}_{lm}-\tilde{b}_{kl}-a_{l}-d_{D}}(-\bar{\bar{\eta}}_{4E})^{\ell-\tilde{b}_{km}-\tilde{b}_{lm}-b_{0}-d_{E}}\\
\times & [T]_{c^{2\xi_{k}}C^{a_{k}}d^{2\xi_{l}}D^{a_{l}}}^{E'^{a_{m}}e'^{2\xi_{m}}F{}^{a_{s}}}[(\bbe{2}\cdot\Gamma \, \Gamma^{F}_{34})^{c_{s}}(\bbe{2}\cdot\Gamma\, \bbe {3}\cdot\Gamma)^{c_{3}}(\bbe{2}\cdot\Gamma\, \bbe{4}\cdot\Gamma)^{c_{4}}  C^{-1}_{\Gamma}] _{e e'}\\
\times & (g_{EE'})^{b_{0}}(\bar{\bar{\eta}}_{2E'})^{b_{2}}(g_{CZ})^{d_{C}}(g_{DZ})^{d_{D}}(g_{EZ})^{d_{E}}(\bar{\bar{\eta}}_{2Z})^{d_{2}}(g_{Z_{1}Z_{2}}/3)^{e_{2}},
\end{aligned}
\end{equation}
where the index $\tilde{b}=\{\tilde{b}_{kl},\tilde{b}_{km},\tilde{b}_{lm}\}$
sums over all non-vanishing $\tilde{b}$s, and the primitive coefficients
are defined as
\begin{equation}
\begin{aligned} & _{b\tilde{b}}\omega{}_{(b_{0},b_{3},b_{s};c_{3},c_{4},c_{s};d_{C},d_{D},d_{E},d_{2})}^{klm}=  \lambda_{\ell}\sum_{r_{0},r_{2},s_{0},s_{2},t_{0},t_{2}\geq0} \\
\times &\left(\begin{array}{c}
n_{\tilde{b}}-\ell_{\tilde{b}}-r_{2}-a_{k}-a_{l}+b_{2}-d_{C}-d_{D}+d_{2}\\
\ell_{k}-\tilde{b}_{km}-\tilde{b}_{kl}-s_{2}-a_{k}-d_{C},n_{\tilde{b}}-\ell_{k}-\ell_{\tilde{b}}+\tilde{b}_{km}+\tilde{b}_{kl}-r_{2}+s_{2}-a_{l}+b_{2}-d_{D},d_{2}
\end{array}\right)\\
\times & \frac{\left(\begin{array}{c}
i_{b}\\
s_{0},s_{2},b_{km}-s_{0}-s_{2},r_{0}-s_{0}-b_{0},r_{2}-s_{2}-b_{2},b_{lm}-r_{0}-r_{2}+s_{0}+s_{2}+b_{0}+b_{2},a_{m}
\end{array}\right)}{\left(\begin{array}{c}
i_{b}\\
b_{km},b_{lm},a_{m}
\end{array}\right)(-1)^{\ell_{b}+\ell-\tilde{b}_{km}-\tilde{b}_{lm}+r_{2}-b_{0}-d_{E}+e_{2}}}\\
\times & \frac{\left(\begin{array}{c}
i_{\tilde{b}}+\tilde{b}_{kl}-b_{kl}-r_{0}+b_{0}+d_{C}+d_{D}+d_{E}\\
\tilde{b}_{kl}-i_{kl},\tilde{b}_{km}-s_{0},\tilde{b}_{lm}-r_{0}+s_{0}+b_{0},d_{C}+d_{D}+d_{E}
\end{array}\right)\left(\begin{array}{c}
\ell_{b}\\
t_{0},t_{2},\ell_{b}-t_{0}-t_{2}
\end{array}\right)}{(-1)^{t_{2}}\left(\begin{array}{c}
n_{b}+2i_{b}-2r_{0}-r_{2}-t_{0}-a_{m}-a_{s}+b_{0}+b_{2}+e_{1}\\
\ell-r_{0}-t_{0},\ell_{k}-b_{kl}-s_{0}-s_{2}-a_{k},\ell_{l}-b_{kl}-r_{0}-r_{2}+s_{0}+s_{2}-a_{l}+b_{0}+b_{2},e_{1}
\end{array}\right)}\\
\times & 2^{i_{\tilde{b}}+\tilde{b}_{kl}-b_{kl}-r_{0}+b_{0}+d_{C}+d_{D}+d_{E}}\rho^{(3,\ell_{b}-t_{0}-t_{2}+c_{3}+e_{2};-h-n_{b}-2i_{b}-\ell_{b}-2\xi_{m})}\\
\times & \tilde{K}^{(3,h+2r_{0}+r_{2}+2t_{0}+t_{2}+c_{4}+e_{2};p-r_{0}-t_{0};i_{\tilde{b}}+\tilde{b}_{kl}-b_{kl}-r_{0}+b_{0}+d_{C}+d_{D}+d_{E},n_{\tilde{b}}-\ell_{\tilde{b}}-r_{2}-a_{k}-a_{l}+b_{2}-d_{C}-d_{D}+d_{2},0,\ell_{\tilde{b}}-t_{0}-b_{0}-d_{E})}.
\end{aligned}
\end{equation}
The expression above still has sums over $t_0$ and $t_2$ that are bounded by $\ell$. Since $\ell$ can be arbitrarily large it is best to avoid such sums. It is possible to remove the $\ell$-dependence using a trick similar to what was done in \cite{Fortin:2020ncr}---see Appendix \ref{sec:ldep} for details---and simplify the result  to
\begin{equation}
\begin{aligned} & _{b\tilde{b}}\omega{}_{(b_{0},b_{3},b_{s};c_{3},c_{4},c_{s};d_{C},d_{D},d_{E},d_{2})}^{klm}\\
= & \lambda_{\ell}\sum_{\substack{r_{0},r_{2},s_{0},s_{2},\\
t_{0},t_{2}\geq0
}
}(-1)^{\ell_{b}+r_{2}+t_{2}+e_{2}}(-2)^{h+n_{\tilde{b}}+2i_{\tilde{b}}+2\tilde{b}_{kl}-2b_{kl}-a_{k}-a_{l}+b_{0}+b_{2}+c_{3}+c_{4}+e_{1}+2e_{2}}\\
\times & \frac{(-i_{\tilde{b}}+r_{0}-b_{0}-d_{E})_{t_{0}}(-n_{\tilde{b}}+\ell_{\tilde{b}}+r_{2}+a_{k}+a_{l}+b_{0}-d_{2})_{t_{2}}(-\ell_{b})_{t_{0}+t_{2}}}{s_{0}!s_{2}!t_{0}!t_{2}!(b_{km}-s_{0}-s_{2})!(r_{0}-s_{0}-b_{0})!(r_{2}-s_{2}-b_{2})!}\\
\times & \frac{b_{km}!b_{lm}!(\ell-r_{0}-t_{0})!(\ell_{k}-b_{kl}-s_{0}-s_{2}-a_{k})!(\ell_{l}-b_{kl}-r_{0}-r_{2}+s_{0}+s_{2}-a_{l}+b_{0}+b_{2})!}{(b_{lm}-r_{0}-r_{2}+s_{0}+s_{2}+b_{0}+b_{2})!(\tilde{b}_{kl}-b_{kl})!(\ell_{\tilde{b}}-b_{0}-d_{E})!(\tilde{b}_{km}-s_{0})!(\tilde{b}_{lm}-r_{0}+s_{0}+b_{0})!}\\
\times & \frac{(-p+n_{b}+2i_{b}-r_{0}-r_{2}-a_{s}-b_{s}+e_{1}+2)_{\ell_{b}}(-p+r_{0}+1)_{t_{0}}(p-r_{0}-t_{0})_{h+2r_{0}+r_{2}+c_{4}+e_{2}}}{(\ell_{k}-\tilde{b}_{km}-\tilde{b}_{kl}-s_{2}-a_{k}-d_{C})!(n_{\tilde{b}}-\ell_{k}-\ell_{\tilde{b}}+\tilde{b}_{km}+\tilde{b}_{kl}-r_{2}+s_{2}-a_{l}+b_{2}-d_{D})!}\\
\times & \frac{\big(-h-n_{b}-2i_{b}-2\xi_{m}-\frac{1}{2}\big)_{c_{3}+e_{2}}(-h-n_{b}-\ell_{b}-2i_{b}-2\xi_{m})_{\ell_{b}+c_{3}+e_{2}}}{(-p+n_{b}+2i_{b}-r_{0}-r_{2}-a_{s}-b_{s}+e_{1}+2+\ell_{b})_{t_{0}+t_{2}}}\\
\times & \frac{(-h-n_{\tilde{b}}-2i_{\tilde{b}}+a_{k}+a_{l}-b_{0}-b_{2}-c_{4}-e_{1}-e_{2})_{n_{\tilde{b}}+i_{\tilde{b}}+\tilde{b}_{kl}-b_{kl}-r_{0}-r_{2}-a_{k}-a_{l}+b_{2}+d_{2}}}{(h+\ell-b_{kl}+\tilde{b}_{kl}+r_{0}+r_{2}+b_{0}+b_{2}+c_{4}+e_{2}+1)_{t_{0}+t_{2}}}\\
\times & \frac{\big(h+p+r_{0}+r_{2}+c_{4}+e_{2}-\frac{1}{2}\big)_{t_{0}+t_{2}}(h+p+r_{0}+r_{2}-t_{0}+c_{4}+e_{2})_{\ell_{\tilde{b}}+i_{\tilde{b}}+\tilde{b}_{kl}-b_{kl}-r_{0}+d_{C}+d_{D}}}{\big(h+p+r_{0}+r_{2}-t_{0}+c_{4}+e_{2}-\frac{1}{2}\big)_{-h-n_{\tilde{b}}+\ell_{\tilde{b}}-i_{\tilde{b}}-\tilde{b}_{kl}+b_{kl}-r_{0}+a_{k}+a_{l}-b_{0}-b_{2}-c_{4}-d_{E}-d_{2}-e_{2}}},
\end{aligned}
\end{equation}
where now the $t$-sums are bounded by  the $\ell$-independent terms $i_{\tilde{b}}-r_{0}+b_{0}+d_{E}$
and $n_{\tilde{b}}-\ell_{\tilde{b}}-r_{2}-a_{k}-a_{l}-b_{0}+d_{2}$.

The goal is then to rewrite the OPE basis expression in terms of the three-point basis as
\begin{equation}
(\mathscr{G}_{\vert b)}^{m+\ell\vert kl})=\sum_{b'}(R_{klm}^{-1})_{bb'}(\bar{\bar{\eta}}_{2}\cdot\Gamma){}_{b'}F_{kl,m}^{12,3}.\label{eq:3pt-1}
\end{equation}
We will apply this change of basis to mixed-basis four-point conformal blocks in order to bring them to a pure basis. The parity $\epsilon_{b}$ is always conserved for bosonic exchanges but always flipped for fermionic exchanges, so the $b'$-sum in (\ref{eq:3pt-1})
runs over all possible choices of $\{b'_{km},b'_{lm},b'_{kl}\}$ with the required parity label. In general, the matrix elements $(R_{klm}^{-1})_{bb'}$
are expressible as linear combinations of the coefficients 
\[
(R_{klm}^{-1})_{bb'}=\sum_{\substack{b_{0},b_{3},b_{s}\\
c_{3},c_{4},c_{s}\\
d_{C},d_{D},d_{E},d_{2}
}} 
\lambda \times {}_{b\tilde{b}}\omega{}_{(b_{0},b_{3},b_{s};c_{3},c_{4},c_{s};d_{C},d_{D},d_{E},d_{2})}^{klm},
\]
where the $\tilde{b}$ label depends on the $b'$ label. In fact, each $\omega$ comes with its own coefficient $\lambda$, but we suppressed the indices on the $\lambda$'s  for clarity. We note that this result involves an over-complete three-point basis. This does not present a problem because we use the same over-complete basis for the four-point tensor structures in mixed basis. In the following, we derive the explicit sums case by case.

\subsection{Boson-Boson OPE}

For purely bosonic operators, a complete set of independent OPE-basis
tensor structures is given by
\begin{align*}
(\tensor*[_{\{b_{km}\oplus\epsilon_{b},b_{lm},0\}}]{t}{_{kl}^{34,m+i_{b}}}\tensor{)}{_{\{C\}\{D\}}^{\{E\}\{F\}}} & \simeq(\delta_{C}^{F})^{\ell_{k}-b_{km}-\epsilon_{b}}(\delta_{D}^{F})^{\ell_{l}-b_{lm}}(\delta_{C}^{E})^{b_{km}}(\delta_{D}^{E})^{b_{lm}}(\tensor{\epsilon}{_{34C}^{EF}})^{\epsilon_{b}},\\
(\tensor*[_{\{b_{km},0,b_{kl}\oplus\epsilon_{b}\}}]{t}{_{kl}^{34,m+i_{b}}}\tensor{)}{_{\{C\}\{D\}}^{\{E\}\{F\}}} & \simeq(\delta_{C}^{F})^{\ell_{k}-b_{kl}-b_{km}-\epsilon_{b}}(\delta_{D}^{F})^{\ell_{l}-b_{kl}-\epsilon_{b}}(\delta_{C}^{E})^{b_{kl}}(\delta_{D}^{E})^{b_{km}}(\tensor{\epsilon}{_{34CD}^{F}})^{\epsilon_{b}},\\
(\tensor*[_{\{0,b_{lm}\oplus\epsilon_{b},b_{kl}\}}]{t}{_{kl}^{34,m+i_{b}}}\tensor{)}{_{\{C\}\{D\}}^{\{E\}\{F\}}} & \simeq(\delta_{C}^{F})^{\ell_{k}-b_{kl}}(\delta_{D}^{F})^{\ell_{l}-b_{kl}-b_{lm}-\epsilon_{b}}(\delta_{C}^{E})^{i_{kl}}(\delta_{D}^{E})^{i_{lm}}(\tensor{\epsilon}{_{34D}^{EF}})^{\epsilon_{b}}.
\end{align*}
In these cases, we never have any $c$ or $b_{s}$ labels. For the
natural three-point basis, we use the over-complete set labelled by
\[
\begin{aligned} & (\tensor*[_{\{b_{km}\oplus\epsilon_{km},b_{lm}\oplus\epsilon_{lm},b_{kl}\oplus\epsilon_{kl}\}}]{F}{_{klm}^{34}})_{\{C\}\{D\}\{E\}}\\
\simeq & (\bar{\bar{\eta}}_{2C})^{\ell_{k}-b_{km}-b_{kl}-\epsilon_{km}-\epsilon_{lm}}(\bar{\bar{\eta}}_{2D})^{\ell_{l}-b_{lm}-b_{kl}-\epsilon_{lm}-\epsilon_{kl}}(-\bar{\bar{\eta}}_{4E})^{\ell_{l}-b_{lm}-b_{kl}-\epsilon_{lm}-\epsilon_{kl}}\\
\times & (g_{CE})^{b_{km}}(g_{DE})^{b_{lm}}(g_{CD})^{b_{kl}}((\epsilon_{34}\cdot\bar{\bar{\eta}}_{2})_{CE})^{\epsilon_{km}}((\epsilon_{34}\cdot\bar{\bar{\eta}}_{2})_{DE})^{\epsilon_{lm}}((\epsilon_{34}\cdot\bar{\bar{\eta}}_{2})_{CD})^{\epsilon_{kl}},
\end{aligned}
\]
We note that in addition to this, it is also naively possible to have $\epsilon_{34CDE}$, but this structure can be re-expressed using \ref{eq:epsconv}).

\subsubsection{$\epsilon_{b}=0$}

For the simplest case of $b=\{b_{km},b_{lm},b_{kl}\}$, there is only
one matrix element depending on a single coefficient
\begin{equation}
(R_{klm}^{-1})_{b\{b'_{km},b'_{lm},b'_{kl}\}}={}_{b\{b'_{km},b'_{lm},b'_{kl}\}}\omega{}_{(0,0,0;0,0,0;0,0,0,0)}^{klm}.
\end{equation}

\subsubsection{$\epsilon_{b}=1$}

For the insertion of $\tensor{\epsilon}{_{34C}^{E'F'}}$, the non-vanishing special structure is given by
\[
\begin{aligned} & \tensor{\epsilon}{_{34C}^{E'F'}}(g_{EE'})^{b_{0}}(\bar{\bar{\eta}}_{2E'})^{b_{2}}(g_{DF'})^{d_{D}}(g_{EF'})^{d_{E}}(\bar{\bar{\eta}}_{2F'})^{d_{2}}\\
\simeq & \Big(\frac{1}{2}[(\epsilon_{34}\cdot\bar{\bar{\eta}}_{2})_{DE}\bar{\bar{\eta}}_{2C}-(\epsilon_{34}\cdot\bar{\bar{\eta}}_{2})_{CE}\bar{\bar{\eta}}_{2D}-(\epsilon_{34}\cdot\bar{\bar{\eta}}_{2})_{CD}\bar{\bar{\eta}}_{4E}]\Big)^{d_{D}}\\
\times & ((\epsilon_{234CE})^{d_{2}})^{b_{0}}(-(\epsilon_{234CD})^{d_{D}}(\epsilon_{234CE})^{d_{E}})^{b_{2}}.
\end{aligned}
\]
Therefore, for $b=\{b_{km}\oplus1,b_{lm},b_{kl}\}$, there are 3 parity odd matrix elements
\begin{align}
(R_{klm}^{-1}){}_{b\{b'_{km}\oplus1,b'_{lm},b'_{kl}\}} & =-\frac{1}{2}{}_{b\{b'_{km},b'_{lm},b'_{kl}\}}\omega{}_{(1,0,0;0,0,0;0,1,0,0)}^{klm}\nonumber \\
+ & _{b\{b'_{km},b'_{lm},b'_{kl}\}}\omega{}_{(1,0,0;0,0,0;0,0,0,1)}^{klm}+{}_{b\{b'_{km},b'_{lm},b'_{kl}\}}\omega{}_{(0,1,0;0,0,0;0,0,1,0)}^{klm},\\
(R_{klm}^{-1})_{b\{b'_{km},b'_{lm}\oplus1,b'_{kl}\}} & =\frac{1}{2}{}_{b\{b'_{km},b'_{lm},b'_{kl}\}}\omega{}_{(1,0,0;0,0,0;0,1,0,0)}^{klm},\\
(R_{klm}^{-1})_{b\{b'_{km},b'_{lm},b'_{kl}\oplus1\}} & =-\frac{1}{2}{}_{b\{b'_{km},b'_{lm},b'_{kl}\}}\omega{}_{(1,0,0;0,0,0;0,1,0,0)}^{klm}-{}_{b\{b'_{km},b'_{lm},b'_{kl}\}}\omega{}_{(0,1,0;0,0,0;0,1,0,0)}^{klm}.
\end{align}
When $\tensor{\epsilon}{_{34D}^{E'F'}}$ appears, the special structure
is
\[
\begin{aligned} & \tensor{\epsilon}{_{34D}^{E'F'}}(g_{EE'})^{b_{0}}(\bar{\bar{\eta}}_{2E'})^{b_{2}}(g_{CF'})^{d_{C}}(g_{EF'})^{d_{E}}(\bar{\bar{\eta}}_{2F'})^{d_{2}}\\
\simeq & \bigg(\Big(\frac{1}{2}[(\epsilon_{34}\cdot\bar{\bar{\eta}}_{2})_{CE}\bar{\bar{\eta}}_{2D}+(\epsilon_{34}\cdot\bar{\bar{\eta}}_{2})_{CD}\bar{\bar{\eta}}_{4E}-(\epsilon_{34}\cdot\bar{\bar{\eta}}_{2})_{DE}\bar{\bar{\eta}}_{2C}]\Big)^{d_{C}}\\
\times & ((\epsilon_{34}\cdot\bar{\bar{\eta}}_{2})_{DE})^{d_{2}}\bigg)^{b_{0}}\Big(((\epsilon_{34}\cdot\bar{\bar{\eta}}_{2})_{CD})^{d_{C}}(-(\epsilon_{34}\cdot\bar{\bar{\eta}}_{2})_{DE})^{d_{E}}\Big)^{b_{2}},
\end{aligned}
\]
so we have matrix elements with $b=\{b_{km},b_{lm}\oplus1,b_{kl}\}$
given by
\begin{align}
(R_{klm}^{-1}){}_{b\{b'_{km}\oplus1,b'_{lm},b'_{kl}\}} & =\frac{1}{2}{}_{b\{b'_{km},b'_{lm},b'_{kl}\}}\omega{}_{(1,0,0;0,0,0;1,0,0,0)}^{klm}\\
(R_{klm}^{-1})_{b\{b'_{km},b'_{lm}\oplus1,b'_{kl}\}} & =_{b\{b'_{km},b'_{lm},b'_{kl}\}}\omega{}_{(1,0,0;0,0,0;0,0,0,1)}^{klm}\nonumber \\
- & _{b\{b'_{km},b'_{lm},b'_{kl}\}}\omega{}_{(0,1,0;0,0,0;0,0,1,0)}^{klm}-\frac{1}{2}{}_{b\{b'_{km},b'_{lm},b'_{kl}\}}\omega{}_{(1,0,0;0,0,0;1,0,0,0)}^{klm},\\
(R_{klm}^{-1})_{b\{b'_{km},b'_{lm},b'_{kl}\oplus1\}} & =_{b\{b'_{km},b'_{lm},b'_{kl}\}}\omega{}_{(0,1,0;0,0,0;1,0,0,0)}^{klm}+\frac{1}{2}{}_{b\{b'_{km},b'_{lm},b'_{kl}\}}\omega{}_{(1,0,0;0,0,0;1,0,0,0)}^{klm}.
\end{align}
Lastly, for the $\tensor{\epsilon}{_{34CD}^{F'}}$ insertion, the
special structure is given by
\[
\tensor{\epsilon}{_{34CD}^{F'}}(g_{EF'})^{d_{E}}(\bar{\bar{\eta}}_{2F'})^{d_{2}}\simeq\Big(\frac{1}{2}[(\epsilon_{34}\cdot\bar{\bar{\eta}}_{2})_{CE}\bar{\bar{\eta}}_{2D}+(\epsilon_{34}\cdot\bar{\bar{\eta}}_{2})_{CD}\bar{\bar{\eta}}_{4E}-(\epsilon_{34}\cdot\bar{\bar{\eta}}_{2})_{DE}\bar{\bar{\eta}}_{2C}]\Big)^{d_{E}}((\epsilon_{34}\cdot\bar{\bar{\eta}}_{2})_{CD})^{d_{2}}.
\]
So the $b=\{b_{km},b_{lm},b_{kl}\oplus1\}$ matrix elements are defined
as
\begin{align}
(R_{klm}^{-1}){}_{b\{b'_{km}\oplus1,b'_{lm},b'_{kl}\}} & =\frac{1}{2}{}_{b\{b'_{km},b'_{lm},b'_{kl}\}}\omega{}_{(0,0,0;0,0,0;0,0,1,0)}^{klm},\\
(R_{klm}^{-1})_{b\{b'_{km},b'_{lm}\oplus1,b'_{kl}\}} & =-\frac{1}{2}{}_{b\{b'_{km},b'_{lm},b'_{kl}\}}\omega{}_{(0,0,0;0,0,0;0,0,1,0)}^{klm},\\
(R_{klm}^{-1})_{b\{b'_{km},b'_{lm},b'_{kl}\oplus1\}} & =\frac{1}{2}{}_{b\{b'_{km},b'_{lm},b'_{kl}\}}\omega{}_{(0,0,0;0,0,0;0,0,1,0)}^{klm}+{}_{b\{b'_{km},b'_{lm},b'_{kl}\}}\omega{}_{(0,0,0;0,0,0;0,0,0,1)}^{klm}.
\end{align}

\subsection{Fermion-Fermion OPE}

The fermion-fermion OPE tensor structures are given by
\[
\begin{aligned} & (\tensor*[_{\{b_{km}\oplus\frac{\delta_{b}}{2},b_{lm}\oplus\frac{\delta_{b}}{2},b_{kl}\oplus\frac{1-\delta_{b}}{2};\epsilon_{b}\}}]{t}{_{kl}^{34,m+i_{b}}}\tensor{)}{_{\{cC\}\{dD\}}^{\{E\}\{F\}}}\\
\simeq & (\delta_{C}^{F})^{\ell_{k}-b_{kl}-b_{km}}(\delta_{D}^{F})^{\ell_{l}-b_{kl}-b_{lm}}(g_{CD})^{b_{kl}}(\delta_{C}^{E})^{b_{km}}(\delta_{D}^{E})^{b_{lm}}[(\Gamma_{34}^{E'})^{\delta_{b}}(\Gamma_{34}^{F'})^{\vert\epsilon_{b}-\delta_{b}\vert}C_{\Gamma}^{-1}]_{cd}.
\end{aligned}
\]
For these tensor structures, we still do not need $c$ labels, but there may be non-vanishing $b_{s}$. The three-point basis is given by
\[
\begin{aligned} & (\tensor*[_{\{b_{km}\oplus\frac{\delta_{b}}{2},b_{lm}\oplus\frac{\delta_{b}}{2},b_{kl}\oplus\frac{1-\delta_{b}}{2};\epsilon_{b}\}}]{F}{_{klm}^{34}})_{\{cC\}\{D\}\{E\}}\\
\simeq & (\bar{\bar{\eta}}_{2C})^{\ell_{k}-b_{km}-b_{kl}}(\bar{\bar{\eta}}_{2D})^{\ell_{l}-b_{lm}-b_{kl}}(-\bar{\bar{\eta}}_{4E})^{\ell-b_{km}-b_{lm}-\delta_{b}}\\
\times & (g_{CD})^{b_{kl}}(g_{CE})^{b_{km}}(g_{DE})^{b_{lm}}[(\Gamma_{34E})^{\delta_{b}}(\bar{\bar{\eta}}_{2}\cdot\Gamma_{34})^{\vert\epsilon_{b}-\delta_{b}\vert}C_{\Gamma}^{-1}]_{cd}.
\end{aligned}
\]

\subsubsection{$\epsilon_{b}=0$}

In the $\delta_{b}=0$ case, the special structure is trivial. Thus for $b=\{b_{km},b_{lm},b_{kl}\oplus\frac{1}{2};0\}$, we have
\begin{align}
(R_{klm}^{-1}){}_{b\{b'_{km},b'_{lm},b'_{kl}\oplus\frac{1}{2};0\}} & =_{b\{b'_{km},b'_{lm},b'_{kl}\}}\omega{}_{(0,0,0;0,0,0;0,0,0,0)}^{klm},\\
(R_{klm}^{-1}){}_{b\{b'_{km}+\frac{1}{2},b'_{lm}\oplus\frac{1}{2},b'_{kl};0\}} & =0.
\end{align}
In the $\delta_{b}=1$ scenario, the special structure is given by
\[
\begin{aligned} & [\Gamma_{34}^{E'}\Gamma_{34}^{F'}C_{\Gamma}^{-1}]_{cd}(g_{EE'})^{b_{0}}(\bar{\bar{\eta}}_{2E'})^{b_{2}}(g_{E'F'}/3)^{b_{s}}(g_{CF'})^{d_{C}}(g_{EF'})^{d_{E}}(\bar{\bar{\eta}}_{2F'})^{d_{2}}\\
\simeq & \Big[\Big((2g_{CE})^{d_{C}}(\Gamma_{34E}\bar{\bar{\eta}}_{2}\cdot\Gamma_{34})^{d_{2}}\Big)^{b_{0}}\Big((2\bar{\bar{\eta}}_{2C})^{d_{C}}(-2\bar{\bar{\eta}}_{4E}-\Gamma_{34E}\bar{\bar{\eta}}_{2}\cdot\Gamma_{34})^{d_{E}}(-2)^{d_{2}}\Big)^{b_{2}}(1)^{b_{s}}C_{\Gamma}^{-1}\Big]_{cd}.
\end{aligned}
\]
Henc , for $b=\{b_{km}\oplus\frac{1}{2},b_{lm}\oplus\frac{1}{2},b_{kl};0\}$,
the matrix elements are
\begin{align}
(R_{klm}^{-1}){}_{b\{b'_{km},b'_{lm},b'_{kl}\oplus\frac{1}{2};0\}} & =_{b\{b'_{km},b'_{lm},b'_{kl}\}}\omega{}_{(0,0,1;0,0,0;0,0,0,0)}^{klm}\nonumber \\
+ & 2_{b\{b'_{km},b'_{lm},b'_{kl}\}}\omega{}_{(0,1,0;0,0,0;1,0,0,0)}^{klm}-2_{b\{b'_{km},b'_{lm},b'_{kl}\}}\omega{}_{(0,1,0;0,0,0;0,0,1,0)}^{kl,m}\\
- & 2_{b\{b'_{km},b'_{lm},b'_{kl}\}}\omega{}_{(0,1,0;0,0,0;0,0,0,1)}^{kl,m}+2{}_{b\{b'_{km}-1,b'_{lm},b'_{kl}\}}\omega{}_{(1,0,0;0,0,0;1,0,0,0)}^{klm},\\
(R_{klm}^{-1}){}_{b\{b'_{km}\oplus\frac{1}{2},b'_{lm}\oplus\frac{1}{2},b'_{kl};0\}} & =_{b\{b'_{km},b'_{lm},b'_{kl}\}}\omega{}_{(1,0,0;0,0,0;0,0,0,1)}^{klm}-{}_{b\{b'_{km},b'_{lm},b'_{kl}\}}\omega{}_{(0,1,0;0,0,0;0,0,1,0)}^{klm}.
\end{align}

\subsubsection{$\epsilon_{b}=1$}

In the $\delta_{b}=0$ case, the special structure is given by
\[
[\Gamma_{34}^{F'}C_{\Gamma}^{-1}]_{cd}(g_{EF'})^{d_{E}}(\bar{\bar{\eta}}_{2F'})^{d_{2}}\simeq[(\Gamma_{34E})^{d_{E}}(\bar{\bar{\eta}}_{2}\cdot\Gamma_{34})^{d_{2}}C_{\Gamma}^{-1}]_{cd}.
\]
So the matrix elements with with $b=\{b_{km},b_{lm},b_{kl}\oplus\frac{1}{2};1\}$
are
\begin{align}
(R_{klm}^{-1}){}_{b\{b'_{km},b'_{lm},b'_{kl}\oplus\frac{1}{2};1\}} & =_{b\{b'_{km},b'_{lm},b'_{kl}\}}\omega{}_{(0,0,0;0,0,0;0,0,0,1)}^{klm},\\
(R_{klm}^{-1})_{b\{b'_{km}\oplus\frac{1}{2},b'_{lm}\oplus\frac{1}{2},b'_{kl};1\}} & =_{b\{b'_{km},b'_{lm},b'_{kl}\}}\omega{}_{(0,0,0;0,0,0;0,0,1,0)}^{klm}.
\end{align}
On the other hand, when $\delta_{b}=1$, the corresponding special structure is given by
\[
[\Gamma_{34}^{E'}C_{\Gamma}^{-1}]_{cd}(g_{EE'})^{b_{0}}(\bar{\bar{\eta}}_{2E'})^{b_{2}}(g_{EE'})^{d_{E}}(\bar{\bar{\eta}}_{2E'})^{d_{2}}\simeq[(\Gamma_{34E})^{b_{0}+d_{E}}(\bar{\bar{\eta}}_{2}\cdot\Gamma_{34})^{b_{2}+d_{2}}C_{\Gamma}^{-1}]_{cd}.
\]
The $b=\{b_{km}\oplus\frac{1}{2},b_{lm}\oplus\frac{1}{2},b_{kl};1\}$ matrix elements are
\begin{align}
(R_{klm}^{-1}){}_{b\{b'_{km},b'_{lm},b'_{kl}\oplus\frac{1}{2};1\}} & =_{b\{b'_{km},b'_{lm},b'_{kl}\}}\omega{}_{(0,1,0;0,0,0;0,0,0,0)}^{klm}+_{b\{b'_{km},b'_{lm},b'_{kl}\}}\omega{}_{(0,0,1;0,0,0;0,0,0,1)}^{klm},\\
(R_{klm}^{-1})_{b\{b'_{km}\oplus\frac{1}{2},b'_{lm}\oplus\frac{1}{2},b'_{kl};1\}} & =_{b\{b'_{km},b'_{lm},b'_{kl}\}}\omega{}_{(1,0,0;0,0,0;0,0,0,0)}^{klm}+_{b\{b'_{km},b'_{lm},b'_{kl}\}}\omega{}_{(0,0,1;0,0,0;0,0,1,0)}^{klm}.
\end{align}

\subsection{Fermion-Boson OPE}

The fermion-boson OPE tensor structures with $(k,m)$ fermionic are given by
\[
\begin{aligned} & (\tensor*[_{\{b_{km}\oplus\frac{1-\delta_{b}}{2},b_{lm}\oplus\frac{\delta_{b}}{2},b_{kl}\oplus\frac{\delta_{b}}{2};\epsilon_{b}\}}]{t}{_{kl}^{34,m+i_{b}}}\tensor{)}{_{\{cC\}\{D\}}^{\{Ee\}\{F\}}}\\
\simeq & (\delta_{C}^{F})^{\ell_{k}-b_{ij}-b_{im}}(\delta_{D}^{F})^{\ell_{l}-b_{kl}-b_{lm}-\delta_{b}}(g_{CD})^{b_{kl}}(\delta_{C}^{E})^{b_{km}}(\delta_{D}^{E})^{b_{lm}}[(\Gamma_{D})^{\delta_{b}}(\Gamma_{34}^{F})^{\vert\epsilon_{b}-\delta_{b}\vert}]_{c}^{\ e}.
\end{aligned}
\]
In these cases, we will now have $c$ labels but $b$ will never be present. The three-point basis are
\[
\begin{aligned} & (\tensor*[_{\{b_{km}\oplus\frac{1-\delta_{b}}{2},b_{lm}\oplus\frac{\delta_{b}}{2},b_{kl}\oplus\frac{\delta_{b}}{2}/2;\epsilon_{b}\}}]{F}{_{klm}^{34}})_{\{cC\}\{D\}\{Ee\}}\\
\simeq & (\bar{\bar{\eta}}_{2C})^{\ell_{k}-b_{km}-b_{kl}}(\bar{\bar{\eta}}_{2D})^{\ell_{l}-b_{lm}-b_{kl}-\delta_{b}}(-\bar{\bar{\eta}}_{4E})^{\ell-b_{lm}-b_{kl}}\\
\times & (g_{CD})^{b_{kl}}(g_{CE})^{b_{km}}(g_{DE})^{b_{lm}}[(\Gamma_{D})^{\delta_{b}}(\bar{\bar{\eta}}_{2}\cdot\Gamma_{34})^{\vert\epsilon_{b}-\delta_{b}\vert}C_{\Gamma}^{-1}]_{ce}.
\end{aligned}
\]
When $(l,m)$ are fermions, the situation is completely mirrored. The OPE tensor structures are given by
\[
\begin{aligned} & (\tensor*[_{\{b_{km}\oplus\frac{\delta_{b}}{2},b_{lm}\oplus\frac{1-\delta_{b}}{2},b_{kl}\oplus\frac{\delta_{b}}{2};\epsilon_{b}\}}]{t}{_{kl}^{34,m+i_{b}}}\tensor{)}{_{\{C\}\{dD\}}^{\{Ee\}\{F\}}}\\
\simeq & (\delta_{A}^{F})^{\ell_{k}-b_{kl}-b_{km}-\delta_{b}}(\delta_{B}^{F})^{\ell_{l}-b_{km}-b_{lm}}(g_{CD})^{b_{kl}}(\delta_{C}^{E})^{b_{km}}(\delta_{D}^{E})^{b_{lm}}[(\Gamma_{C})^{\delta_{b}}(\Gamma_{34}^{F})^{\vert\epsilon_{b}-\delta_{b}\vert}]_{d}^{\ e},
\end{aligned}
\]
and the three-point ones are
\[
\begin{aligned} & (\tensor*[_{\{b_{km}\oplus\frac{\delta_{b}}{2},b_{lm}\oplus\frac{1-\delta_{b}}{2},b_{kl}\oplus\frac{\delta_{b}}{2};\epsilon_{b}\}}]{F}{_{klm}^{34}})_{\{C\}\{dD\}\{Ee\}}\\
\simeq & (\bar{\bar{\eta}}_{2C})^{\ell_{k}-b_{km}-b_{kl}-\delta_{b}}(\bar{\bar{\eta}}_{2D})^{\ell_{l}-b_{lm}-b_{kl}}(-\bar{\bar{\eta}}_{4E})^{\ell-b_{lm}-b_{kl}}\\
\times & (g_{CD})^{b_{kl}}(g_{CE})^{b_{km}}(g_{DE})^{b_{lm}}[(\Gamma_{A})^{\delta_{b}}(\bar{\bar{\eta}}_{2}\cdot\Gamma_{34})^{\vert\epsilon_{b}-\delta_{b}\vert}C_{\Gamma}^{-1}]_{be}.
\end{aligned}
\]

\subsubsection{$\epsilon_{b}=0$}

The $\delta_{b}=0$ case have the special structure
\[
\begin{aligned} & (C_{\Gamma}^{-1})_{ce'}[(\bar{\bar{\eta}}_{2}\cdot\Gamma\Gamma_{34}^{F})^{c_{s}}(\bar{\bar{\eta}}_{2}\cdot\Gamma\bar{\bar{\eta}}_{3}\cdot\Gamma)^{c_{3}}]_{e}^{\ e'}(g_{DF})^{d_{D}}(g_{EF})^{d_{E}}(\bar{\bar{\eta}}_{2F})^{d_{2}}\\
\simeq & [(\Gamma_{D}\bar{\bar{\eta}}_{2}\cdot\Gamma)^{d_{D}}(\bar{\bar{\eta}}_{4E}\bar{\bar{\eta}}_{2}\cdot\Gamma_{34}\bar{\bar{\eta}}_{2}\cdot\Gamma/2)^{d_{E}}(\bar{\bar{\eta}}_{2}\cdot\Gamma_{34}\bar{\bar{\eta}}_{2}\cdot\Gamma)^{d_{2}}(-\bar{\bar{\eta}}_{2}\cdot\Gamma_{34}\bar{\bar{\eta}}_{2}\cdot\Gamma)^{c_{3}}C_{\Gamma}^{-1}]_{ce}.
\end{aligned}
\]
The $b=\{b_{km}\oplus\frac{1}{2},b_{lm},b_{kl};0\}$ matrix elements
are
\begin{align}
(R_{klm}^{-1}){}_{b\{b'_{km}\oplus\frac{1}{2},b'_{lm},b'_{kl};1\}} & =\frac{1}{2}{}_{b\{b'_{km},b'_{lm},b'_{kl}\}}\omega{}_{(0,0,0;0,0,1;0,0,1,0)}^{klm}\nonumber \\
+ & _{b\{b'_{km},b'_{lm},b'_{kl}\}}\omega{}_{(0,0,0;0,0,1;0,0,0,1)}^{klm}-_{b\{b'_{km},b'_{lm},b'_{kl}\}}\omega{}_{(0,0,0;1,0,0;0,0,0,0)}^{klm},\\
(R_{klm}^{-1}){}_{b\{b'_{km},b'_{lm}\oplus\frac{1}{2},b'_{kl}\oplus\frac{1}{2};1\}} & =_{b\{b'_{km},b'_{lm},b'_{kl}\}}\omega{}_{(0,0,0;0,0,1;0,1,0,0)}^{klm}.
\end{align}
In the $\delta_{b}=1$ scenario, the special structure is
\[
\begin{aligned} & [\Gamma_{D}\Gamma_{34}^{F'}C_{\Gamma}^{-1}]_{ce'}[(\bar{\bar{\eta}}_{2}\cdot\Gamma\Gamma_{F'}/3)^{c_{s}}(\bar{\bar{\eta}}_{2}\cdot\Gamma\bar{\bar{\eta}}_{3}\cdot\Gamma)^{c_{3}}]_{e}^{\ e'}(g_{CF'})^{d_{C}}(\bar{\bar{\eta}}_{2F'})^{d_{2}}\\
 & \simeq[(\Gamma_{D}\bar{\bar{\eta}}_{2}\cdot\Gamma)^{c_{s}}(-2g_{CD}\bar{\bar{\eta}}_{2}\cdot\Gamma_{34}\bar{\bar{\eta}}_{2}\cdot\Gamma)^{d_{C}}(2\Gamma_{D}\bar{\bar{\eta}}_{2}\cdot\Gamma)^{d_{2}}C_{\Gamma}^{-1}]_{ce},
\end{aligned}
\]
and the corresponding $b=\{b_{km},b_{lm}\oplus\frac{1}{2},b_{kl}\oplus\frac{1}{2};0\}$ matrix elements are given by
\begin{align}
(R_{klm}^{-1}){}_{b\{b'_{km}\oplus\frac{1}{2},b'_{lm},b'_{kl};1\}} & =-2_{b\{b'_{km},b'_{lm},b'_{kl}-1\}}\omega{}_{(0,0,0;1,0,0;1,0,0,0)}^{klm},\\
(R_{klm}^{-1}){}_{b\{b'_{km},b'_{lm}\oplus\frac{1}{2},b'_{kl}\oplus\frac{1}{2};1\}} & =_{b\{b'_{km},b'_{lm},b'_{kl}\}}\omega{}_{(0,0,0;0,0,1;0,0,0,0)}^{klm}+2_{b\{b'_{km},b'_{lm},b'_{kl}\}}\omega{}_{(0,0,0;1,0,0;0,0,0,1)}^{klm}.
\end{align}

In the mirroring situation, the $\delta_{b}=0$ special structure is given by
\[
\begin{aligned} & (C_{\Gamma}^{-1})_{de'}[(\bar{\bar{\eta}}_{2}\cdot\Gamma\Gamma_{34}^{F})^{c_{s}}(\bar{\bar{\eta}}_{2}\cdot\Gamma\bar{\bar{\eta}}_{4}\cdot\Gamma)^{c_{4}}]_{e}^{\ e'}(g_{CF})^{d_{C}}(\bar{\bar{\eta}}_{2F})^{d_{2}}\\
 & \simeq[(\Gamma_{C}\bar{\bar{\eta}}_{2}\cdot\Gamma)^{d_{C}}(-\bar{\bar{\eta}}_{4}\cdot\Gamma\bar{\bar{\eta}}_{2}\cdot\Gamma)^{d_{2}}(\bar{\bar{\eta}}_{4}\cdot\Gamma\bar{\bar{\eta}}_{2}\cdot\Gamma)^{c_{2}}C_{\Gamma}^{-1}]_{de}.
\end{aligned}
\]
The $b=\{b_{km},b_{lm}\oplus\frac{1}{2},b_{kl};0\}$ matrix elements
are then
\begin{align}
(R_{klm}^{-1}){}_{b\{b'_{km},b'_{lm}\oplus\frac{1}{2},b'_{kl};1\}} & =_{b\{b'_{km},b'_{lm},b'_{kl}\}}\omega{}_{(0,0,0;0,1,0;0,0,0,0)}^{klm}-_{b\{b'_{km},b'_{lm},b'_{kl}\}}\omega{}_{(0,0,0;0,0,1;0,0,0,1)}^{klm},\\
(R_{klm}^{-1}){}_{b\{b'_{km}\oplus\frac{1}{2},b'_{lm},b'_{kl}\oplus\frac{1}{2};1\}} & =_{b\{b'_{km},b'_{lm},b'_{kl}\}}\omega{}_{(0,0,0;0,0,1;1,0,0,0)}^{klm}.
\end{align}
In the $\delta=1$ case, we have the special structure
\[
\begin{aligned} & [\Gamma_{C}\Gamma_{34}^{F'}C_{\Gamma}^{-1}]_{de'}[(\bar{\bar{\eta}}_{2}\cdot\Gamma\Gamma_{F'}/3)^{c_{s}}(\bar{\bar{\eta}}_{2}\cdot\Gamma\bar{\bar{\eta}}_{4}\cdot\Gamma)^{c_{4}}]_{e}^{\ e'}(g_{DF'})^{d_{D}}(g_{EF'})^{d_{E}}(\bar{\bar{\eta}}_{2F'})^{d_{2}}\\
\simeq & [(\Gamma_{C}\bar{\bar{\eta}}_{2}\cdot\Gamma)^{c_{s}}((-2g_{CD}\bar{\bar{\eta}}_{2}\cdot\Gamma_{34}\bar{\bar{\eta}}_{2}\cdot\Gamma)^{d_{D}}(2\bar{\bar{\eta}}_{2E}\Gamma_{C}\bar{\bar{\eta}}_{2}\cdot\Gamma)^{d_{E}}(2\Gamma_{C}\bar{\bar{\eta}}_{2}\cdot\Gamma)^{d_{2}})^{c_{4}}C_{\Gamma}^{-1}]_{de},
\end{aligned}
\]
and the matrix elements with $b=\{b_{km}\oplus\frac{1}{2},b_{lm},b_{kl}\oplus\frac{1}{2};0\}$ are given by
\begin{align}
(R_{klm}^{-1}){}_{b\{b'_{km},b'_{lm}\oplus\frac{1}{2},b'_{kl};1\}} & =-2_{b\{i'_{km},i'_{lm},i'_{kl}-1\}}\omega{}_{(0,0,0;0,1,0;0,1,0,0)}^{klm},\\
(R_{klm}^{-1}){}_{b\{b'_{km}\oplus\frac{1}{2},b'_{lm},b'_{kl}\oplus\frac{1}{2};1\}} & =_{b\{b'_{km},b'_{lm},b'_{kl}\}}\omega{}_{(0,0,0;0,0,1;0,0,0,0)}^{klm}\nonumber \\
+ & 2_{b\{b'_{km},b'_{lm},b'_{kl}\}}\omega{}_{(0,0,0;0,1,0;0,0,1,0)}^{klm}+2_{b\{b'_{km},b'_{lm},b'_{kl}\}}\omega{}_{(0,0,0;0,1,0;0,0,0,1)}^{klm}.
\end{align}

\subsubsection{$\epsilon_{b}=1$}

In the $\delta_{b}=0$ scenario, the special structure is given by
\[
\begin{aligned} & [\Gamma_{34}^{F'}C_{\Gamma}^{-1}]_{ce'}[(\bar{\bar{\eta}}_{2}\cdot\Gamma\Gamma_{34F'}/3)^{c_{s}}(\bar{\bar{\eta}}_{4}\cdot\Gamma\bar{\bar{\eta}}_{2}\cdot\Gamma)^{c_{3}}]_{e}^{\ e'}(g_{DF'})^{d_{D}}(\bar{\bar{\eta}}_{3F'})^{d_{3}}\\
 & \simeq[(\bar{\bar{\eta}}_{2}\cdot\Gamma)^{c_{s}}(-\Gamma_{D}\bar{\bar{\eta}}_{2}\cdot\Gamma_{34}\bar{\bar{\eta}}_{2}\cdot\Gamma)^{d_{D}}(2\bar{\bar{\eta}}_{2}\cdot\Gamma)^{d_{2}}C_{\Gamma}^{-1}]_{ce},
\end{aligned}
\]
and the $b=\{b_{km}\oplus\frac{1}{2},b_{lm},b_{kl};1\}$ matrix elements are given by
\begin{align}
(R_{klm}^{-1}){}_{b\{b'_{km}\oplus\frac{1}{2},b'_{lm},b'_{kl};0\}} & =_{b\{b'_{km},b'_{lm},b'_{kl}\}}\omega{}_{(0,0,0;0,0,1;0,0,0,0)}^{klm}+2_{b\{b'_{km},b'_{lm},b'_{kl}\}}\omega{}_{(0,0,0;1,0,0;0,0,0,1)}^{klm},\\
(R_{klm}^{-1}){}_{b\{b'_{km},b'_{lm}\oplus\frac{1}{2},b'_{kl}\oplus\frac{1}{2};0\}} & =-_{b\{b'_{km},b'_{lm},b'_{kl}\}}\omega{}_{(0,0,0;1,0,0;0,1,0,0)}^{klm}.
\end{align}
When $\delta_{b}=1$, the special structure is now
\[
\begin{aligned} & [\Gamma_{D}C_{\Gamma}^{-1}]_{ce'}[(\bar{\bar{\eta}}_{2}\cdot\Gamma\Gamma_{34}^{F})^{c_{s}}(\bar{\bar{\eta}}_{2}\cdot\Gamma\bar{\bar{\eta}}_{3}\cdot\Gamma)^{c_{3}}]_{e}^{\ e'}(g_{CF})^{d_{C}}(g_{EF})^{d_{E}}(\bar{\bar{\eta}}_{2F})^{d_{2}}\\
\simeq & [(2g_{CD}\bar{\bar{\eta}}_{2}\cdot\Gamma)^{d_{C}}(\bar{\bar{\eta}}_{4E}\Gamma_{D}\bar{\bar{\eta}}_{2}\cdot\Gamma_{34}\bar{\bar{\eta}}_{2}\cdot\Gamma)^{d_{E}}(\Gamma_{D}\bar{\bar{\eta}}_{2}\cdot\Gamma_{34}\bar{\bar{\eta}}_{2}\cdot\Gamma)^{d_{2}}(-\Gamma_{D}\bar{\bar{\eta}}_{2}\cdot\Gamma_{34}\bar{\bar{\eta}}_{2}\cdot\Gamma)^{c_{3}}C_{\Gamma}^{-1}]_{ce},
\end{aligned}
\]
so the $b=\{b_{km},b_{lm}\oplus\frac{1}{2},b_{kl}\oplus\frac{1}{2};1\}$ matrix elements are
\begin{align}
(R_{klm}^{-1}){}_{b\{b'_{km}\oplus\frac{1}{2},b'_{lm},b'_{kl};0\}} & =2_{b\{b'_{km},b'_{lm},b'_{kl}-1\}}\omega{}_{(0,0,0;0,0,1;1,0,0,0)}^{klm},\\
(R_{klm}^{-1}){}_{b\{b'_{km},b'_{lm}\oplus\frac{1}{2},b'_{kl}\oplus\frac{1}{2};0\}} & =_{b\{b'_{km},b'_{lm},b'_{kl}\}}\omega{}_{(0,0,0;0,0,1;0,0,1,0)}^{klm}\nonumber \\
+ & _{b\{b'_{km},b'_{lm},b'_{kl}\}}\omega{}_{(0,0,0;0,0,1;0,0,0,1)}^{klm}-_{b\{b'_{km},b'_{lm},b'_{kl}\}}\omega{}_{(0,0,0;1,0,0;0,0,0,0)}^{klm}.
\end{align}

For the mirroring setup, we have the special structure in the $\delta_{b}=0$ case given by
\[
\begin{aligned} & [\Gamma_{34}^{F'}C_{\Gamma}^{-1}]_{de'}[(\bar{\bar{\eta}}_{2}\cdot\Gamma\Gamma_{F'}/3)^{c_{s}}(\bar{\bar{\eta}}_{2}\cdot\Gamma\bar{\bar{\eta}}_{4}\cdot\Gamma)^{c_{4}}]_{e}^{\ e'}(g_{CF'})^{d_{C}}(g_{DF'})^{d_{D}}(g_{EF'})^{d_{E}}(\bar{\bar{\eta}}_{2F'})^{d_{2}}\\
 & \simeq[(\bar{\bar{\eta}}_{2}\cdot\Gamma)^{c_{s}}(-\Gamma_{C}\bar{\bar{\eta}}_{2}\cdot\Gamma_{34}\bar{\bar{\eta}}_{2}\cdot\Gamma)^{d_{C}}(2\bar{\bar{\eta}}_{4E}\bar{\bar{\eta}}_{2}\cdot\Gamma)^{d_{E}}(2\bar{\bar{\eta}}_{2}\cdot\Gamma)^{d_{2}}C_{\Gamma}^{-1}]_{de}.
\end{aligned}
\]
The $b=\{b_{km},b_{lm}\oplus\frac{1}{2},b_{kl};1\}$ matrix elements are
\begin{align}
(R_{klm}^{-1}){}_{b\{b'_{km}\oplus\frac{1}{2},b'_{lm},b'_{kl};0\}} & =_{b\{b'_{km},b'_{lm},b'_{kl}\}}\omega{}_{(0,0,0;0,0,1;0,0,0,0)}^{klm}\nonumber \\
+ & 2_{b\{b'_{km},b'_{lm},b'_{kl}\}}\omega{}_{(0,0,0;0,1,0;0,0,1,0)}^{klm}+2_{b\{b'_{km},b'_{lm},b'_{kl}\}}\omega{}_{(0,0,0;0,1,0;0,0,0,1)}^{klm},\\
(R_{klm}^{-1}){}_{b\{b'_{km},b'_{lm}\oplus\frac{1}{2},b'_{kl}\oplus\frac{1}{2};0\}} & =-_{b\{b'_{km},b'_{lm},b'_{kl}\}}\omega{}_{(0,0,0;0,1,0;1,0,0,0)}^{klm}.
\end{align}
In the $\delta_{b}=1$ case, the special structure is
\[
\begin{aligned} & [\Gamma_{C}C_{\Gamma}^{-1}]_{de'}[(\bar{\bar{\eta}}_{2}\cdot\Gamma\Gamma_{34}^{F})^{c_{s}}(\bar{\bar{\eta}}_{2}\cdot\Gamma\bar{\bar{\eta}}_{4}\cdot\Gamma)^{c_{4}}]_{e}^{\ e'}(g_{DF})^{d_{D}}(\bar{\bar{\eta}}_{2F})^{d_{2}}\\
\simeq & [(2g_{CD}\bar{\bar{\eta}}_{2}\cdot\Gamma)^{d_{D}}(\Gamma_{C}\bar{\bar{\eta}}_{2}\cdot\Gamma_{34}\bar{\bar{\eta}}_{2}\cdot\Gamma)^{d_{2}}(-\Gamma_{A}\bar{\bar{\eta}}_{2}\cdot\Gamma_{34}\bar{\bar{\eta}}_{2}\cdot\Gamma)^{c_{4}}]_{e}^{\ e'}C_{\Gamma}^{-1}]_{de}.
\end{aligned}
\]
The corresponding $b=\{b_{km}\oplus\frac{1}{2},b_{lm},b_{kl}\oplus\frac{1}{2};1\}$ matrix elements are then
\begin{align}
(R_{klm}^{-1}){}_{b\{b'_{km}\oplus\frac{1}{2},b'_{lm},b'_{kl};0\}} & =2_{b\{b'_{km},b'_{lm},b'_{kl}-1\}}\omega{}_{(0,0,0;0,0,1;0,1,0,0)}^{klm},\\
(R_{klm}^{-1}){}_{b\{b'_{km},b'_{lm}\oplus\frac{1}{2},b'_{kl}\oplus\frac{1}{2};0\}} & =_{b\{b'_{km},b'_{lm},b'_{kl}\}}\omega{}_{(0,0,0;0,0,1;0,0,0,1)}^{klm}-{}_{b\{b'_{km},b'_{lm},b'_{kl}\}}\omega{}_{(0,0,0;0,1,0;0,0,0,0)}^{klm}.
\end{align}

\section{Four-Point Conformal Blocks}\label{sec:4pt}

In this section, we specialize formulas for the mixed-basis four-point blocks obtained in~\cite{Fortin:2020ncr} for generic $d$ dimensions to $d=3$. The projector (\ref{eq:projdec}) is written as a sum of special parts and shifted projectors to the symmetric representation. Therefore, we separate the exchange operators into two types, one bosonic ($\xi_{m}=0$) with no special parts of projectors needed and one fermionic ($\xi_{m}=\frac{1}{2}$) with a spinorial special part.

We write down the most general formulas for all possible $(a\vert b]$ labels in each case. These formulas are applicable to any pair of tensor structures in over-complete basis of both three-point and OPE structures. When applying these formulas, the OPE basis must be reduced to the complete set while the three-point basis must be left over-complete so that we could apply
the transformation matrices derived in the previous section. This way one obtains the appropriate number of blocks in the pure OPE-OPE basis.

\subsection{Bosonic Exchange}

From (\ref{eq:4ptblk}), when $\xi_{m}=0$, the blocks take the form

\begin{equation}
\begin{aligned}\mathscr{G}_{(a\vert b]}^{ij\vert m+\ell\vert kl}= & \sum_{\substack{r+2r'_{0}+r'_{1}+r'2=i_{a}\\
r+2r''_{0}+r''_{1}+r''2=i_{b}\\
r'_{0}+r'_{1}+r'_{3}=r''_{0}+r''_{1}+r''_{3}
}
}(-1)^{\ell-\ell'-i_{a}+r'_{1}+r'_{2}}\frac{(-2)^{r'_{3}+r''{}_{3}}\ell'!}{(d'/2-1)_{\ell'}}\\
&\quad  \times \mathscr{C}_{i_{a},i_{b}}^{(d,\ell)}(r,r',r'')(C_{\ell'}^{(d'/2-1)}(X))_{s_{(a\vert b]}^{ij\vert m+\ell\vert kl}(0,0,0,r,\bm{r}',\bm{r}'')},
\end{aligned}
\end{equation}
because for bosonic operators, it is trivial to see that in (\ref{eq:projdec}), $t$ is only $0$, with $\mathscr{A}_{0}=1=\mathcal{\hat{Q}}_{0}$ and $\ell_{0}=d_{0}=0$. This implies also that $j_{a}=j_{b}=0$.

\subsubsection{All Bosons}

When all of the operators in a four-point function are bosonic, the substitution $s$ is obtained by inserting the tensor structures into (\ref{eq:4ptsub}),
\[
\begin{aligned} & s_{(a\vert b]}^{ij\vert m+\ell\vert kl}(0,0,0,r,\bm{r}',\bm{r}''):\alpha_{2}^{s_{2}}\alpha_{3}^{s_{3}}\alpha_{4}^{s_{4}}x_{3}^{r_{3}}x_{4}^{r_{4}}\\
 & \rightarrow(g_{AB})^{a_{ij}}(\delta_{A}^{E})^{a_{im}}(\delta_{B}^{E})^{a_{jm}}(\tensor{\epsilon}{_{12AB}^{F}})^{\epsilon_{ij}}(\tensor{\epsilon}{_{12A}^{EF}})^{\epsilon_{im}}(\tensor{\epsilon}{_{12B}^{EF}})^{\epsilon_{jm}}(\delta_{A}^{F})^{i_{i}}(\delta_{B}^{F})^{i_{j}}\\
 & \times(g_{EE})^{r'_{0}}(\tensor{\mathcal{S}}{_{E}^{E''}})^{r}[(\mathcal{S}\cdot\bar{\eta}_{4})_{E}]^{r'_{2}}(G_{(\ell'-\ell+2r'_{3},n_{a}-\ell,n'_{3},n'_{4},n'_{5})}^{ij|m+\ell|kl})_{F^{n_{a}-\ell_{a}}E^{r_{1}''}}^{E''^{r_{1}''}F''^{n_{b}-\ell_{b}}}[(\bar{\eta}_{2}\cdot\mathcal{S})_{E}]^{r''_{2}}(g^{E''E''})^{r''_{0}}\\
 & \times(g_{CD})^{b_{kl}}(g_{CE''})^{b_{km}}(g_{DE''})^{b_{lm}}(\tensor{\epsilon}{_{34CDF''}})^{\epsilon_{kl}}(\tensor{\epsilon}{_{34CE''F''}})^{\epsilon_{km}}(\tensor{\epsilon}{_{34DE''F''}})^{\epsilon_{lm}}(g_{CF''})^{i_{k}}(g_{DF''})^{i_{l}}.
\end{aligned}
\]
The $F$ and $F''$ indices are contracted straightforwardly, leading to
\[
\begin{aligned} & s_{(a\vert b]}^{ij\vert m+\ell\vert kl}(0,0,0,r,\bm{r}',\bm{r}''):\alpha_{2}^{s_{2}}\alpha_{3}^{s_{3}}\alpha_{4}^{s_{4}}x_{3}^{r_{3}}x_{4}^{r_{4}}\\
 & \rightarrow(g_{AB})^{a_{ij}}(\delta_{A}^{E})^{a_{im}}(\delta_{B}^{E})^{a_{jm}}(\tensor{\epsilon}{_{12AB}^{F}})^{\epsilon_{ij}}(\tensor{\epsilon}{_{12A}^{EF}})^{\epsilon_{im}}(\tensor{\epsilon}{_{12B}^{EF}})^{\epsilon_{jm}}(g_{EE})^{r'_{0}}\\
 & \times(\tensor{\mathcal{S}}{_{E}^{E''}})^{r}[(\mathcal{S}\cdot\Bar{\eta}_{4})_{E}]^{r'_{2}}(G_{(\ell'-\ell+2r'_{3},n_{a}-\ell,n'_{3},n'_{4},n'_{5})}^{ij|m+\ell|k\ell})_{A^{i_{i}}B^{i_{j}}C^{i_{k}}D^{i_{l}}F^{\epsilon_{a}}E^{r'_{1}}}^{E''^{r''_{1}}F''^{\epsilon_{b}}}[(\bar{\eta}_{2}\cdot\mathcal{S})_{E}]^{r''_{2}}\\
 & \times(g^{E''E''})^{r''_{0}}(g_{CD})^{b_{kl}}(g_{CE''})^{b_{km}}(g_{DE''})^{b_{lm}}(\tensor{\epsilon}{_{34CDF''}})^{\epsilon_{kl}}(\tensor{\epsilon}{_{34CE''F''}})^{\epsilon_{km}}(\tensor{\epsilon}{_{34DE''F''}})^{\epsilon_{lm}}.
\end{aligned}
\]
Contraction of the $E$ and $E''$ indices, on the other hand, is more complex, requiring introduction of several indices that are summed over, along with combinatorial factors, owing to the
symmetrization of these indices. The result is
\begin{equation}
\begin{aligned} & s_{(a\vert b]}^{ij\vert m+\ell\vert kl}(0,0,0,r,\bm{r}',\bm{r}''):\alpha_{2}^{s_{2}}\alpha_{3}^{s_{3}}\alpha_{4}^{s_{4}}x_{3}^{r_{3}}x_{4}^{r_{4}}\\
 & \rightarrow\frac{1}{\binom{i_{a}}{a_{im},a_{jm},\epsilon_{im}+\epsilon_{jm}}\binom{i_{b}}{b_{km},b_{lm},\epsilon_{km}+\epsilon_{lm}}}\sum_{\substack{\epsilon_{0},\epsilon_{1},\epsilon_{2},\epsilon_{3},\epsilon_{4}\\
\epsilon'_{0},\epsilon''_{0},\epsilon'_{2},\epsilon''_{2}\\
\epsilon_{0}+\epsilon_{3}+\epsilon_{4}+\epsilon'_{0}+\epsilon'_{2}=\epsilon_{im}+\epsilon_{jm}\\
\epsilon_{0}+\epsilon_{1}+\epsilon_{2}+\epsilon''_{0}+\epsilon''_{2}=\epsilon_{km}+\epsilon_{lm}
}
}\sum_{\substack{q\\
q'_{1},q''_{1}\\
q'_{2},q''_{2}
}
}(-1)^{\epsilon'_{0}\epsilon_{jm}+\epsilon''_{0}\epsilon_{lm}}2^{r'_{0}+r''_{0}}\\
 & \times(r'_{0})^{\epsilon'_{0}}(r''_{0})^{\epsilon''_{0}}\binom{r'_{1}}{q'_{1}}\binom{r''_{1}}{q''_{1}}\binom{r'_{2}}{q'_{2},r'_{2}-q'_{2}-\epsilon'_{2},\epsilon'_{2}}\binom{r''_{2}}{q''_{2},r''_{2}-q''_{2}-\epsilon''_{2},\epsilon''_{2}}\\
 & \times\binom{r}{q,q_{1}+\epsilon_{im}\epsilon'_{0},q_{3}+\epsilon_{km}\epsilon''_{0},r-q_{1}-q_{3}-\epsilon_{im}\epsilon'_{0}-\epsilon_{km}\epsilon''_{0},\epsilon_{0},\epsilon_{1}+\epsilon_{2},\epsilon_{3}+\epsilon_{4}}\\
 & \times(g_{AB})^{a_{ij}+r'_{0}-\epsilon'_{0}}(g_{CD})^{b_{kl}+r''_{0}-\epsilon''_{0}}(\tensor{\epsilon}{_{12AB}^{F}})^{\epsilon_{ij}+\epsilon'_{0}}(\tensor{\epsilon}{_{34CDF''}})^{\epsilon_{kl}+\epsilon''_{0}}\\
 & \times(\tensor{\epsilon}{_{12A}^{EF}})^{\epsilon_{im}(1-\epsilon'_{0})}(\tensor{\epsilon}{_{12B}^{EF}})^{\epsilon_{jm}(1-\epsilon'_{0})}(\tensor{\epsilon}{_{34C}^{E''F}})^{\epsilon_{km}(1-\epsilon''_{0})}(\tensor{\epsilon}{_{34D}^{E''F}})^{\epsilon_{lm}(1-\epsilon''_{0})}\\
 & \times[(\mathcal{S}\cdot\bar{\eta}_{4})_{E}]^{\epsilon'_{2}}[(\bar{\eta}_{2}\cdot\mathcal{S})^{E''}]^{\epsilon''_{2}}(\tensor{\mathcal{S}}{_{E}^{E''}})^{\epsilon_{0}}(\tensor{\mathcal{S}}{_{A}^{E''}})^{\epsilon_{1}}(\tensor{\mathcal{S}}{_{B}^{E''}})^{\epsilon_{2}}(\mathcal{S}_{EC})^{\epsilon_{3}}(\mathcal{S}_{ED})^{\epsilon_{4}}\\
 & \times[(\mathcal{S}\cdot\bar{\eta}_{4})_{A}]^{q'_{2}}[(\mathcal{S}\cdot\bar{\eta}_{4})_{B}]^{r'_{2}-q'_{2}-\epsilon'_{2}}[(\bar{\eta}_{2}\cdot\mathcal{S})_{C}]^{q''_{2}}[(\bar{\eta}_{2}\cdot\mathcal{S})_{D}]^{r''_{2}-q''_{2}-\epsilon''_{2}}\\
 & \times(\mathcal{S}_{AC})^{q}(\mathcal{S}_{BD})^{r-q_{1}-q_{3}-\epsilon_{im}\epsilon'_{0}-\epsilon_{km}\epsilon''_{0}}(\mathcal{S}_{AD})^{q_{1}+\epsilon_{im}\epsilon'_{0}}(\mathcal{S}_{BC})^{q_{3}+\epsilon_{km}\epsilon''_{0}}\\
 & \times(G_{(\ell'-\ell+2r'_{3},n_{a}-\ell,n'_{3},n'_{4},n'_{5})}^{ij|m+\ell|kl})_{\substack{A^{i_{i}+q'_{1}}B^{i_{j}+r'_{1}-q'_{1}}C^{i_{k}+q''_{1}}D^{i_{l}+r''_{1}-q''_{1}}}
}^{F^{\epsilon_{a}+\epsilon_{b}}}.
\end{aligned}
\end{equation}
Let us explain the meaning of these new variables that are summed over.  First, $\epsilon'_{0}$ ($\epsilon''_{0}$) counts how many of the $r'_{0}$ ($r''_{0}$) $g_{EE}$'s ($g^{E''E''}$'s) are contracted with the $\epsilon_{12}$ ($\epsilon_{34}$). The remaining $g$'s are contracted with both $\delta_{A}^{E}$ ($g_{CE''}$) and $\delta_{B}^{E}$ ($g_{DE''}$), leading to $r'_{0}-\epsilon'_{0}$ ($r''_{0}-\epsilon''_{0}$) additional $g_{AB}$'s ($g_{CD}$'s). Due to the symmetry of the $G$-tensor, there cannot be any contractions of index $E$- with the antisymmetric $\epsilon$ tensors. Therefore, we only need $q'_{1}$ ($q''_{1}$) to count how many of the $r'_{1}$ ($r''_{1}$) $E$ ($E''$) indices in the $G$-tensor are contracted with $\delta_{A}^{E}$ ($g_{CE''}$), while the rest contract with $\delta_{B}^{E}$ ($g_{DE''}$). Similarly, $q'_{2}$ ($q''_{2}$) and $\epsilon'_{2}$ ($\epsilon''_{2}$) count how many of the $r'_{2}$ ($r''_{2}$) $(S\cdot\bar{\eta}_{4})_{E}$ ($(\bar{\eta}_{2}\cdot S)^{E''}$) are contracted with $\delta_{A}^{E}$ ($g_{CE''}$) and $\epsilon_{12}$ ($\epsilon_{34}$), respectively, while the remaining ones are contracted with $\delta_{B}^{E}$ ($g_{DE''}$).

Among the $r$ $\tensor{S}{_{E}^{E''}}$'s, $\epsilon_{0}$ counts how many of them are contracted to both $\epsilon_{12}$ and $\epsilon_{34}$-tensors. Meanwhile, $\epsilon_{1}$ ($\epsilon_{2}$) counts how many are contracted to $\delta_{A}^{E}$ ($\delta_{B}^{E}$) and $\epsilon_{34}$, and  $\epsilon_{3}$ ($\epsilon_{4}$) indicate how many are contracted to $g_{CE''}$ ($g_{DE''}$) and $\epsilon_{12}$. Lastly, $q$ counts how many of the remaining $\tensor{S}{_{E}^{E''}}$'s are contracted to both $\delta_{A}^{E}$ and $g_{CE''}$. All the remaining contractions are determined by the total number of indices given by the spin of each operator, which  leads to indices that do not require sums and which can be expressed in terms of the others
\begin{align}
q_{1} & =a_{im}-r'_{0}-q-q'_{1}-q'_{2}-\epsilon_{1},\\
q_{3} & =b_{km}-r''_{0}-q-q''_{1}-q''_{2}-\epsilon_{3}.
\end{align}

\subsubsection{All Fermions}

When all of the operators in the four-point function are fermions,
the substitution takes the form
\[
\begin{aligned} & s_{(a\vert b]}^{ij\vert m+\ell\vert kl}(0,0,0,r,\bm{r}',\bm{r}''):\alpha_{2}^{s_{2}}\alpha_{3}^{s_{3}}\alpha_{4}^{s_{4}}x_{3}^{r_{3}}x_{4}^{r_{4}}\\
 & \rightarrow(g_{AB})^{a_{ij}}(\delta_{A}^{E})^{a_{im}}(\delta_{B}^{E})^{a_{jm}}(\delta_{A}^{F})^{i_{i}}(\delta_{B}^{F})^{i_{j}}[(\Gamma_{12}^{E})^{\delta_{a}}(\Gamma_{12}^{F})^{|\epsilon_{a}-\delta_{a}|}C_{\Gamma}^{-1}]_{ab}\\
 & \times(g_{EE})^{r'_{0}}(\tensor{\mathcal{S}}{_{E}^{E''}})^{r}[(\mathcal{S}\cdot\bar{\eta}_{4})_{E}]^{r'_{2}}(G_{(\ell'-\ell+2r'_{3},n_{a}-\ell,n'_{3},n'_{4},n'_{5})}^{ij|m+\ell|kl})_{F^{n_{a}-\ell_{a}}E^{r'_{1}}}^{E''^{r''_{1}}F''^{n_{b}-\ell_{b}}}[(\bar{\eta}_{2}\cdot\mathcal{S})_{E}]^{r''_{2}}(g^{E''E''})^{r''_{0}}\\
 & \times(g_{CD})^{b_{kl}}(g_{CE''})^{b_{km}}(g_{DE''})^{b_{lm}}(g_{CF''})^{i_{k}}(g_{DF''})^{i_{l}}[(\Gamma_{34E''})^{\delta_{b}}(\Gamma_{34F''})^{|\epsilon_{b}-\delta_{b}|}C_{\Gamma}^{-1}]_{cd}.
\end{aligned}
\]
In a manner very similar to simplifying the full boson case, this becomes
\begin{equation}
\begin{aligned} & s_{(a\vert b]}^{ij\vert m+\ell\vert kl}(0,0,0,r,\bm{r}',\bm{r}''):\alpha_{2}^{s_{2}}\alpha_{3}^{s_{3}}\alpha_{4}^{s_{4}}x_{3}^{r_{3}}x_{4}^{r_{4}}\\
 & \rightarrow\frac{1}{\binom{i_{a}}{a_{im},a_{jm},\delta_{a}}\binom{i_{b}}{b_{km},b_{lm},\delta_{b}}}\sum_{\substack{\epsilon_{0},\epsilon_{1},\epsilon_{2},\epsilon_{3},\epsilon_{4}\\
\epsilon'_{0},\epsilon''_{0},\epsilon'_{1},\epsilon''_{1},\epsilon'_{2},\epsilon''_{2}\\
\epsilon_{0}+\epsilon_{3}+\epsilon_{4}+\epsilon'_{0}+\epsilon'_{1}+\epsilon'_{2}=\delta_{a}\\
\epsilon_{0}+\epsilon_{1}+\epsilon_{2}+\epsilon''_{0}+\epsilon''_{1}+\epsilon''_{2}=\delta_{b}
}
}^{1}\sum_{\substack{q\\
q'_{1},q''_{1}\\
q'_{2},q''_{2}
}
}2^{r'_{0}+r''_{0}}(r'_{0})^{\epsilon'_{0}}(r''_{0})^{\epsilon''_{0}}\\
 & \times\binom{r'_{1}}{q'_{1},\epsilon'_{1},r'_{1}-q'_{1}-\epsilon'_{1}}\binom{r''_{1}}{q''_{1},\epsilon''_{1},r''_{1}-q''_{1}-\epsilon''_{1}}\binom{r'_{2}}{q'_{2},\epsilon'_{2},r'_{2}-q'_{2}-\epsilon'_{2}}\binom{r''_{2}}{q''_{2},\epsilon''_{2},r''_{2}-q''_{2}-\epsilon''_{2}}\\
 & \times\binom{r}{q,q_{1}+\epsilon'_{0},q_{3}+\epsilon''_{0},r-q_{1}-q_{3}-\epsilon'_{0}-\epsilon''_{0},\epsilon_{0},\epsilon_{1}+\epsilon_{2},\epsilon_{3}+\epsilon_{4}}(g_{AB})^{a_{ij}+r'_{0}-\epsilon'_{0}}(g_{CD})^{b_{kl}+r''_{0}-\epsilon''_{0}}\\
 & \times[(\Gamma_{B})^{\epsilon'_{0}}(\Gamma_{12}^{E})^{\delta_{a}-\epsilon'_{0}}(\Gamma_{12F})^{|\epsilon_{a}-\delta_{a}|}C_{\Gamma}^{-1}]_{ab}[(\Gamma_{D})^{\epsilon''_{0}}(\Gamma_{34E''})^{\delta_{b}-\epsilon''_{0}}(\Gamma_{34F})^{|\epsilon_{b}-\delta_{b}|}C_{\Gamma}^{-1}]_{cd}\\
 & \times(\tensor{\mathcal{S}}{_{E}^{E''}})^{\epsilon_{0}}(\tensor{\mathcal{S}}{_{A}^{E''}})^{\epsilon_{1}}(\tensor{\mathcal{S}}{_{B}^{E''}})^{\epsilon_{2}}(\mathcal{S}_{EC})^{\epsilon_{3}}(\mathcal{S}_{ED})^{\epsilon_{4}}[(\mathcal{S}\cdot\bar{\eta}_{4})_{E}]^{\epsilon'_{2}}[(\bar{\eta}_{2}\cdot\mathcal{S})^{E''}]^{\epsilon''_{2}}\\
 & \times[(\mathcal{S}\cdot\bar{\eta}_{4})_{A}]^{q'_{2}}[(\mathcal{S}\cdot\bar{\eta}_{4})_{B}]^{r'_{2}-q'_{2}-\epsilon'_{2}}[(\bar{\eta}_{2}\cdot\mathcal{S})_{C}]^{q''_{2}}[(\bar{\eta}_{2}\cdot\mathcal{S})_{D}]^{r''_{2}-q''_{2}-\epsilon''_{2}}\\
 & \times(S_{AC})^{q}(S_{AD})^{q_{1}+\epsilon'_{0}}(S_{BC})^{q_{3}+\epsilon''_{0}}(S_{BD})^{r-q_{1}-q_{3}-\epsilon'_{0}-\epsilon''_{0}}\\
 & \times(G_{(\ell'-\ell+2r'_{3},n_{a}-\ell,n'_{3},n'_{4},n'_{5})}^{ij|m+\ell|kl})_{\substack{A^{i_{i}+q'_{1}}B^{i_{j}+r'_{1}-q'_{1}}C^{i_{k}+q''_{1}}D^{i_{l}+r''_{1}-q''_{1}}}
}^{F^{\epsilon_{a}+\epsilon_{b}}}.
\end{aligned}
\end{equation}
In contrast to the pure boson case, there are no $\epsilon$-tensors to be contracted to. They are replaced by $\Gamma$ matrices, and the corresponding sums run over contractions to those matrices. Moreover, whereas contracting an $\epsilon$-tensor via an $E$ index to $G$-tensor produces a vanishing result, the same is not true for $\Gamma$ matrices. Hence, we have two new summation indices, $\epsilon'_{1}$ and $\epsilon''_{1}$ that count those contractions (for $E$ and $E''$ indices respectively).

\subsubsection{Mixed Case}

If the operators at points $\bar{\eta}_{1}$ and $\bar{\eta}_{2}$ are fermionic and the other two are bosonic, the resulting family of conformal blocks is clearly worked out as a hybrid of the two previous cases. Explicitly, the substitution is
\begin{equation}
\begin{aligned} & s_{(a\vert b]}^{ij\vert m+\ell\vert kl}(0,0,0,r,\bm{r}',\bm{r}''):\alpha_{2}^{s_{2}}\alpha_{3}^{s_{3}}\alpha_{4}^{s_{4}}x_{3}^{r_{3}}x_{4}^{r_{4}}\\
 & \rightarrow\frac{1}{\binom{i_{a}}{a_{im},a_{jm},\delta_{a}}\binom{i_{b}}{b_{km},b_{lm},\epsilon_{km}+\epsilon_{lm}}}\sum_{\substack{\epsilon_{0},\epsilon_{1},\epsilon_{2},\epsilon_{3},\epsilon_{4}\\
\epsilon'_{0},\epsilon''_{0},\epsilon'_{1},\epsilon'_{2},\epsilon''_{2}\\
\epsilon_{0}+\epsilon_{3}+\epsilon_{4}+\epsilon'_{0}+\epsilon'_{1}+\epsilon'_{2}=\delta_{a}\\
\epsilon_{0}+\epsilon_{1}+\epsilon_{2}+\epsilon''_{0}+\epsilon''_{2}=\epsilon_{km}+\epsilon_{lm}
}
}^{1}\sum_{\substack{q\\
q'_{1},q''_{1}\\
q'_{2},q''_{2}
}
}(-1)^{\epsilon''_{0}\epsilon_{lm}}2^{r'_{0}+r''_{0}}(r'_{0})^{\epsilon'_{0}}(r''_{0})^{\epsilon''_{0}}\\
 & \times\binom{r'_{1}}{q'_{1},\epsilon'_{1},r'_{1}-q'_{1}-\epsilon'_{1}}\binom{r''_{1}}{q''_{1}}\binom{r'_{2}}{q'_{2},\epsilon'_{2},r'_{2}-q'_{2}-\epsilon'_{2}}\binom{r''_{2}}{q''_{2},\epsilon''_{2},r''_{2}-q''_{2}-\epsilon''_{2}}\\
 & \times\binom{r}{q,q_{1}+\epsilon'_{0},q_{3}+\epsilon_{km}\epsilon''_{0}\epsilon_{0},r-q_{1}-q_{3}-\epsilon'_{0}-\epsilon_{km}\epsilon''_{0},\epsilon_{1}+\epsilon_{2},\epsilon_{3}+\epsilon_{4}}(g_{AB})^{a_{ij}+r'_{0}-\epsilon'_{0}}(g_{CD})^{b_{kl}+r''_{0}-\epsilon''_{0}}\\
 & \times[(\Gamma_{B})^{\epsilon'_{0}}(\Gamma_{12}^{E})^{\delta_{a}-\epsilon'_{0}}(\Gamma_{12F})^{|\epsilon_{a}-\delta_{a}|}C_{\Gamma}^{-1}]_{ab}(\epsilon_{34CDF})^{\epsilon_{kl}+\epsilon''_{0}}(\epsilon_{34CE''F})^{\epsilon_{km}(1-\epsilon''_{0})}(\epsilon_{34DE''F})^{\epsilon_{lm}(1-\epsilon''_{0})}\\
 & \times(\tensor{\mathcal{S}}{_{E}^{E''}})^{\epsilon_{0}}(\tensor{\mathcal{S}}{_{A}^{E''}})^{\epsilon_{1}}(\tensor{\mathcal{S}}{_{B}^{E''}})^{\epsilon_{2}}(\mathcal{S}_{EC})^{\epsilon_{3}}(\mathcal{S}_{ED})^{\epsilon_{4}}[(\mathcal{S}\cdot\bar{\eta}_{4})_{E}]^{\epsilon'_{2}}[(\bar{\eta}_{2}\cdot\mathcal{S})^{E''}]^{\epsilon''_{2}}\\
 & \times[(\mathcal{S}\cdot\bar{\eta}_{4})_{A}]^{q'_{2}}[(\mathcal{S}\cdot\bar{\eta}_{4})_{B}]^{r'_{2}-q'_{2}-\epsilon'_{2}}[(\bar{\eta}_{2}\cdot\mathcal{S})_{C}]^{q''_{2}}[(\bar{\eta}_{2}\cdot\mathcal{S})_{D}]^{r''_{2}-q''_{2}-\epsilon''_{2}}\\
 & \times(\mathcal{S}_{AC})^{q}(\mathcal{S}_{AD})^{q_{1}+\epsilon'_{0}}(\mathcal{S}_{BC})^{q_{3}+\epsilon_{km}\epsilon''_{0}}(\mathcal{S}_{BD})^{r-q_{1}-q_{3}-\epsilon'_{0}-\epsilon_{km}\epsilon''_{0}}\\
 & \times(G_{(\ell'-\ell+2r'_{3},n_{a}-\ell,n'_{3},n'_{4},n'_{5})}^{ij|m+\ell|kl})_{\substack{A^{i_{i}+q'_{1}}B^{i_{j}+r'_{1}-q'_{1}}C^{i_{k}+q''_{1}}D^{i_{l}+r''_{1}-q''_{1}}}
}^{F^{\epsilon_{a}+\epsilon_{b}}}.
\end{aligned}
\end{equation}

For completeness, we also provide the results for the case when the operators at points $\bar{\eta}_{1}$ and $\bar{\eta}_{2}$ are bosonic while the other two are fermionic
\begin{equation}
\begin{aligned} & s_{(a\vert b]}^{ij\vert m+\ell\vert kl}(0,0,0,r,\bm{r}',\bm{r}''):\alpha_{2}^{s_{2}}\alpha_{3}^{s_{3}}\alpha_{4}^{s_{4}}x_{3}^{r_{3}}x_{4}^{r_{4}}\\
 & \rightarrow\frac{1}{\binom{i_{a}}{a_{im},a_{jm},\epsilon_{km}+\epsilon_{lm}}\binom{i_{b}}{b_{km},b_{lm},\delta_{b}}}\sum_{\substack{\epsilon_{0},\epsilon_{1},\epsilon_{2},\epsilon_{3},\epsilon_{4}\\
\epsilon'_{0},\epsilon''_{0},\epsilon'_{1},\epsilon'_{2},\epsilon''_{2}\\
\epsilon_{0}+\epsilon_{3}+\epsilon_{4}+\epsilon'_{0}+\epsilon'_{2}=\epsilon_{im}+\epsilon_{jm}\\
\epsilon_{0}+\epsilon_{1}+\epsilon_{2}+\epsilon''_{0}+\epsilon''_{1}+\epsilon''_{2}=\delta_{b}
}
}^{1}\sum_{\substack{q\\
q'_{1},q''_{1}\\
q'_{2},q''_{2}
}
}(-1)^{\epsilon'_{0}\epsilon_{jm}}2^{r'_{0}+r''_{0}}(r'_{0})^{\epsilon'_{0}}(r''_{0})^{\epsilon''_{0}}\\
 & \times\binom{r'_{1}}{q'_{1}}\binom{r''_{1}}{q''_{1},\epsilon''_{1},r''_{1}-q''_{1}-\epsilon''_{1}}\binom{r'_{2}}{q'_{2},\epsilon'_{2},r'_{2}-q'_{2}-\epsilon'_{2}}\binom{r''_{2}}{q''_{2},\epsilon''_{2},r''_{2}-q''_{2}-\epsilon''_{2}}\\
 & \times\binom{r}{q,q_{1}+\epsilon_{im}\epsilon'_{0},q_{3}+\epsilon''_{0},r-q_{1}-q_{3}-\epsilon_{im}\epsilon'_{0}-\epsilon''_{0},\epsilon_{0},\epsilon_{1}+\epsilon_{2},\epsilon_{3}+\epsilon_{4}}(g_{AB})^{a_{ij}+r'_{0}-\epsilon'_{0}}(g_{CD})^{b_{kl}+r''_{0}-\epsilon''_{0}}\\
 & \times(\tensor{\epsilon}{_{12AB}^{F}})^{\epsilon_{ij}+\epsilon'_{0}}(\tensor{\epsilon}{_{12A}^{EF}})^{\epsilon_{im}(1-\epsilon'_{0})}(\tensor{\epsilon}{_{12B}^{EF}})^{\epsilon_{km}(1-\epsilon'_{0})}[(\Gamma_{D})^{\epsilon''_{0}}(\Gamma_{34E''})^{\delta_{b}-\epsilon''_{0}}(\Gamma_{34}^{F})^{|\epsilon_{b}-\delta_{b}|}C_{\Gamma}^{-1}]_{cd}\\
 & \times(\tensor{\mathcal{S}}{_{E}^{E''}})^{\epsilon_{0}}(\tensor{\mathcal{S}}{_{A}^{E''}})^{\epsilon_{1}}(\tensor{\mathcal{S}}{_{B}^{E''}})^{\epsilon_{2}}(\mathcal{S}_{EC})^{\epsilon_{3}}(\mathcal{S}_{ED})^{\epsilon_{4}}[(\mathcal{S}\cdot\bar{\eta}_{4})_{E}]^{q'_{2}}[(\bar{\eta}_{2}\cdot\mathcal{S})^{E''}]^{q''_{2}}\\
 & \times[(\mathcal{S}\cdot\bar{\eta}_{4})_{A}]^{q'_{2}}[(\mathcal{S}\cdot\bar{\eta}_{4})_{B}]^{r'_{2}-q'_{2}-\epsilon'_{2}}[(\bar{\eta}_{2}\cdot\mathcal{S})_{C}]^{q''_{2}}[(\bar{\eta}_{2}\cdot\mathcal{S})_{D}]^{r''_{2}-q''_{2}-\epsilon''_{2}}\\
 & \times(\mathcal{S}_{AC})^{q}(\mathcal{S}_{AD})^{q_{1}-\epsilon_{im}\epsilon'_{0}}(\mathcal{S}_{BC})^{q_{3}-\epsilon''_{0}}(\mathcal{S}_{BD})^{r-q_{1}-q_{3}-\epsilon_{im}\epsilon'_{0}-\epsilon''_{0}}\\
 & \times(G_{(\ell'-\ell+2r'_{3},n_{a}-\ell,n'_{3},n'_{4},n'_{5})}^{ij|m+\ell|kl})_{\substack{A^{i_{i}+q'_{1}}B^{i_{j}+r'_{1}-q'_{1}}C^{i_{k}+q''_{1}}D^{i_{l}+r''_{1}-q''_{1}}F^{\epsilon_{a}+\epsilon_{b}}}
}.
\end{aligned}
\end{equation}

\subsection{Fermionic Exchange}

For the case of fermion exchange that is when $\xi_{m}=\frac{1}{2}$  it is easy to verify, using the fermion and boson projection operators
given in \cite{Fortin:2020ncr}, that in Eq. (3.2)  $t$ runs over $0$ and $1$. For $t=0$, $\ell_{0}=d_{0}=0$, $\mathscr{A}_{1}(\ell)=\frac{\ell+1}{2\ell+1}$, and $\hat{\mathcal{Q}}_{0}=\mathbb{I}$, while for $t=1$, $\ell_{1}=1$, $d_{1}=2$, $\mathscr{A}_{2}(\ell)=\frac{-\mathscr{K}\ell}{2\ell+1}$, and $\hat{\mathcal{Q}}_{1}=\epsilon_{13}\cdot\Gamma$. The full conformal block is hence a sum of two pieces, which we consider separately.

\subsubsection{$t=0$}

The $t=0$ part of the block takes the form
\begin{equation}
\begin{aligned}\mathscr{G}_{(a\vert b]}^{ij\vert m+\ell\vert kl}\biggr\vert_{t=0}= & \frac{\ell+1}{2\ell+1}\sum_{\substack{r+2r'_{0}+r'_{1}+r'_{2}=i_{a}\\
r+2r''_{0}+r''_{1}+r''_{2}=i_{b}\\
r'_{0}+r'_{1}+r'_{3}=r''_{0}+r''_{1}+r''_{3}
}
}(-1)^{\ell-\ell'-i_{a}+r'_{1}+r'_{2}}\frac{(-2)^{r'_{3}+r"_{3}}\ell'!}{(d'/2-1)_{\ell'}}\\
\times & \mathscr{C}_{i_{a},i_{b}}^{(3,\ell)}(r,\bm{r}',\bm{r}'')(C_{\ell'}^{(d'/2-1)}(X))_{s_{(a\vert b]}^{ij\vert m+\ell\vert kl}(0,0,0,r,\bm{r}',\bm{r}'')}.
\end{aligned}
\end{equation}
Simply plugging in out tensor basis into Eq. (3.1), the substitution becomes
\begin{equation}
\begin{aligned} & s_{(a\vert b]}^{ij\vert m+\ell\vert kl}(0,0,0,r,\bm{r}',\bm{r}''):\alpha_{2}^{s_{2}}\alpha_{3}^{s_{3}}\alpha_{4}^{s_{4}}x_{3}^{r_{3}}x_{4}^{r_{4}}\\
 & \rightarrow-(g_{AB})^{a_{ij}}(\delta_{A}^{E})^{a_{im}}(\delta_{B}^{E})^{a_{jm}}(\delta_{A}^{F})^{i_{i}}(\delta_{B}^{F})^{i_{j}}\tensor{[(\Gamma_{B^{2\xi_{i}}A^{2\xi_{j}}})^{\delta_{a}}(\Gamma_{12}^{F})^{|\epsilon_{a}-\delta_{a}|}]}{_{a^{2\xi_{i}}b^{2\xi_{j}}}^{e}} g_{EE}^{r'_{0}}(\tensor{\mathcal{S}}{_{E}^{E''}})^{r}\\
 & \times [(\mathcal{S}\cdot\bar{\eta}_{4})_{E}]^{r'_{2}}(G_{(\ell'-\ell+2r'_{3},n_{a}-\ell,n'_{3},n'_{4},n'_{5})}^{ij|m+\ell|kl})_{F^{n_{a}-\ell_{a}}E^{r'_{1}}}^{E''^{r''_{1}}F''^{2+n_{b}-\ell_{b}}}\tensor{(\Gamma_{F''}\bar{\eta}_{3}\cdot\Gamma\Gamma_{F''})}{_{e}^{e''}}[(\bar{\eta}_{2}\cdot\mathcal{S})_{E}]^{r''_{2}}(g^{E''E''})^{r''_{0}}\\
 & \times(g_{CD})^{b_{kl}}(g_{CE''})^{b_{km}}(g_{DE''})^{b_{lm}}(g_{CF''})^{i_{k}}(g_{DF''})^{i_{l}}[(\Gamma_{34F''})^{|\epsilon_{b}-\delta_{b}|}(\Gamma_{D^{2\xi_{k}}C^{2\xi_{l}}})^{\delta_{b}}C_{\Gamma}^{-1}]_{e''c^{2\xi_{k}}d^{2\xi_{l}}}.
\end{aligned}
\end{equation}

These are simplified by contracting the fermion indices $e$ and $e''$. The part of this expression carrying fermion indices is then
\begin{equation}
\begin{aligned} & [(\Gamma_{12B^{2\xi_{i}}A^{2\xi_{j}}})^{\delta_{a}}(\Gamma_{12}^{F})^{|\epsilon_{a}-\delta_{a}|}\Gamma_{F''}\bar{\eta}_{3}\cdot\Gamma\Gamma_{F''}(\Gamma_{34F''})^{|\epsilon_{b}-\delta_{b}|}(\Gamma_{34D^{2\xi_{k}}C^{2\xi_{l}}})^{\delta_{b}}C_{\Gamma}^{-1}]_{a^{2\xi_{i}}b^{2\xi_{j}}c^{2\xi_{k}}d^{2\xi_{l}}}\\
 & \simeq 2\bar{\eta}_{3F''}(\bar{\eta}_{2}^{2\xi_{i}}\bar{\eta}_{1}^{2\xi_{j}})^{F}(\bar{\eta}_{1}^{2\xi_{i}}\bar{\eta}_{2}^{2\xi_{j}})^{G}(\bar{\eta}_{4}^{2\xi_{k}}\bar{\eta}_{3}^{2\xi_{l}})_{F''}(\bar{\eta}_{3}^{2\xi_{k}}\bar{\eta}_{4}^{2\xi_{l}})^{G''}\\
 & \times[(\Gamma_{B^{2\xi_{i}}A^{2\xi_{j}}})^{\delta_{a}}(\Gamma_{G})^{|\epsilon_{a}-\delta_{a}|}\Gamma_{F''}(\Gamma_{G''})^{|\epsilon_{b}-\delta_{b}|}(\Gamma_{D^{2\xi_{k}}C^{2\xi_{l}}})^{\delta_{b}}C_{\Gamma}^{-1}]_{a^{2\xi_{i}}b^{2\xi_{j}}c^{2\xi_{k}}d^{2\xi_{l}}},
\end{aligned}
\end{equation}
where we have used the fact that the $G$-tensor is symmetric and traceless when evaluating contractions to the $\Gamma$-matrices carrying $F$ and $F''$ indices, along with the $\Gamma$ matrices' anti-commutative properties. The $E$ and $E''$ indices can then be contracted almost exactly as in the purely bosonic case (with all $\epsilon$-parameters set to $0$), and the final result is
\begin{equation}
\begin{aligned} & s_{(a\vert b]}^{ij\vert m+\ell\vert kl}(0,0,0,r,\bm{r}',\bm{r}''):\alpha_{2}^{s_{2}}\alpha_{3}^{s_{3}}\alpha_{4}^{s_{4}}x_{3}^{r_{3}}x_{4}^{r_{4}}\\
 & \rightarrow\frac{1}{\binom{i_{a}}{a_{ij}}\binom{i_{b}}{b_{kl}}}\sum_{\substack{q\\
q'_{1},q''_{1}\\
q'_{2},q''_{2}
}
}2^{r'_{0}+r''_{0}+2}\binom{r}{q,q_{1},q_{3},r-q_{1}-q_{3}}\binom{r'_{1}}{q'_{1}}\binom{r''_{1}}{q''_{1}}\binom{r'_{2}}{q'_{2}}\binom{r''_{2}}{q''_{2}}\\
 & \times[(\mathcal{S}\cdot\bar{\eta}_{4})_{A}]^{q'_{2}}[(\mathcal{S}\cdot\bar{\eta}_{4})_{B}]^{r'_{2}-q'_{2}}[(\bar{\eta}\cdot\mathcal{S})_{C}]^{q''_{2}}[(\bar{\eta}_{2}\cdot\mathcal{S})_{D}]^{r''_{2}-q''_{2}}\\
 & \times(g_{AB})^{a_{ij}+r'_{0}}(g_{CD})^{b_{kl}+r''_{0}}(\mathcal{S}_{AC})^{q}(\mathcal{S}_{AD})^{q_{1}}(\mathcal{S}_{BC})^{q_{3}}(\mathcal{S}_{BD})^{r-q_{1}-q_{3}}\\
 & \times(G_{(\ell'-\ell+2r'_{3}+4\xi_{i},n_{a}-\ell,n'_{3}-2-4(\xi_{j}+\xi_{k}+\xi_{l}),n'_{4}-2-4\xi_{l},n'_{5}+4\xi_{k})}^{ij|m+\ell|kl})_{\substack{A^{i_{i}+q'_{1}}B^{i_{j}+r'_{1}-q'_{1}}C^{i_{k}+q''_{1}}D^{i_{l}+r''_{1}-q''_{1}}}
}^{F}\\
 & \times[(\Gamma_{B^{2\xi_{i}}A^{2\xi_{j}}})^{\delta_{a}}(\bar{\eta}_{1}^{2\xi_{i}}\bar{\eta}_{2}^{2\xi_{j}}\cdot\Gamma)^{|\epsilon_{a}-\delta_{a}|}\Gamma_{F}(\bar{\eta}_{3}^{2\xi_{k}}\bar{\eta}_{4}^{2\xi_{l}}\cdot\Gamma)^{|\epsilon_{b}-\delta_{b}|}(\Gamma_{D^{2\xi_{k}}C^{2\xi_{l}}})^{\delta_{b}}C_{\Gamma}^{-1}]_{a^{2\xi_{i}}b^{2\xi_{j}}c^{2\xi_{k}}d^{2\xi_{l}}}.
\end{aligned}
\end{equation}

\subsubsection{$t=1$}

For the $t=1$ part of the conformal block, we turn again to Eq. (3.1). For compactness, we introduce the variables $k_{a}=i_{a}-j_{a}$ and
analogously $k_{b}$, which run over $(0,1)$. This leads to
\begin{equation}
\begin{aligned}\mathscr{G}_{(a\vert b]}^{ij\vert m+\ell\vert kl}\biggr\vert_{t=1}= & \frac{-\mathscr{K}}{\ell(2\ell+1)}\sum_{k_{a},k_{b}=0,1}(-i_{a})^{k_{a}}(-i_{b})^{k_{b}}(-\ell_{a})^{1-k_{a}}(-\ell_{b})^{1-k_{b}}\\
\times & \sum_{\substack{r+2r'_{0}+r'_{1}+r'2=i_{a}-k_{a}\\
r+2r''_{0}+r''_{1}+r''2=i_{b}-k_{b}\\
r'_{0}+r'_{1}+r'_{3}=r''_{0}+r''_{1}+r''_{3}
}
}(-1)^{\ell-\ell'-i_{a}+r'_{1}+r'_{2}}\frac{(-2)^{r'_{3}+r''{}_{3}}\ell'!}{(d'/2-1)_{\ell'}}\\
\times & \mathscr{C}_{i_{a}-k_{a},i_{b}-k_{b}}^{(5,\ell-1)}(r,\bm{r}',\bm{r}'')(C_{\ell'}^{(d'/2-1)}(X))_{s_{(a\vert b]}^{ij\vert m+\ell\vert kl}(1,i_{a}-k_{a},i_{b}-k_{b},r,\bm{r}',\bm{r}'')},
\end{aligned}
\end{equation}
and the conformal substitution is
\begin{equation}
\begin{aligned} & s_{(a\vert b]}^{ij\vert m+\ell\vert kl}(1,i_{a}-k_{a},i_{b}-k_{b},r,\bm{r}',\bm{r}''):\alpha_{2}^{s_{2}}\alpha_{3}^{s_{3}}\alpha_{4}^{s_{4}}x_{3}^{r_{3}}x_{4}^{r_{4}}\\
 & \rightarrow-(g_{AB})^{a_{ij}}(\delta_{A}^{E})^{a_{im}}(\delta_{B}^{E})^{a_{jm}}(\delta_{A}^{F})^{i_{i}}(\delta_{B}^{F})^{i_{j}}\tensor{[(\Gamma_{12B^{2\xi_{i}}A^{2\xi_{j}}})^{\delta_{a}}(\Gamma_{12}^{F})^{|\epsilon_{a}-\delta_{a}|}]}{_{a^{2\xi_{i}}b^{2\xi_{j}}}^{e}}\\
 & \times(g_{EE})^{r'_{0}}(\tensor{S}{_{E}^{E''}})^{r}[(S\cdot\bar{\eta}_{4})_{E}]^{r'_{2}}(G_{(\ell'-\ell+2r'_{3},n_{a}-\ell,n'_{3},n'_{4},n'_{5})}^{ij|m+\ell|kl})_{F^{n_{a}-\ell_{a}}E^{r'_{1}}}^{E''^{r''_{1}}F''^{3-k_{b}+n_{b}-\ell_{b}}}(\bar{\eta}_{2}^{E'})^{1-k_{a}}\\
 & \times\tensor{[\Gamma_{F''}\bar{\eta}_{3}\cdot\Gamma\tensor{(S\cdot\epsilon_{13}\cdot\Gamma)}{_{E'^{1-k_{a}}E^{k_{a}}}^{E''^{k_{b}}E'''^{1-k_{b}}}}\Gamma_{F''}]}{_{e}^{e''}}[(\bar{\eta}_{2}\cdot S)_{E}]^{r''_{2}}(g^{E''E''})^{r''_{0}}(\mathcal{A}_{34E'''F''})^{1-k_{b}}\\
 & \times(g_{CD})^{b_{kl}}(g_{CE''})^{b_{km}}(g_{DE''})^{b_{lm}}(g_{CF''})^{i_{k}}(g_{DF''})^{i_{l}}[(\Gamma_{34F''})^{|\epsilon_{b}-\delta_{b}|}(\Gamma_{34D^{2\xi_{k}}C^{2\xi_{l}}})^{\delta_{b}}C_{\Gamma}^{-1}]_{e''c^{2\xi_{k}}d^{2\xi_{l}}}.
\end{aligned}
\end{equation}

The fermionic part reduces as
\begin{align*}
 & \bar{\eta}_{3F}\{[(\Gamma_{12B^{2\xi_{i}}A^{2\xi_{j}}})^{\delta_{a}}(\Gamma_{12F})^{|\epsilon_{a}-\delta_{a}|}(\bar{\eta}_{2}^{1-k_{a}}\cdot(S\cdot\epsilon_{13}\cdot\Gamma\Gamma_{F}-\Gamma_{F}S\cdot\epsilon_{13}\cdot\Gamma)\cdot\bar{\eta}_{4}^{1-k_{b}}\\
 & \times(\bar{\eta}_{3F})^{1-k_{b}})_{_{A^{y_{a}}B^{k_{a}-y_{a}},C^{y_{b}}D^{k_{b}-y_{b}}}}(\Gamma_{34F})^{|\epsilon_{b}-\delta_{b}|}(\Gamma_{34D^{2\xi_{k}}C^{2\xi_{l}}})^{\delta_{b}}C_{\Gamma}^{-1}]_{a^{2\xi_{i}}b^{2\xi_{j}}c^{2\xi_{k}}d^{2\xi_{l}}}\\
 & -\delta_{k_{b},0}[(\Gamma_{12B^{2\xi_{i}}A^{2\xi_{j}}})^{\delta_{a}}(\Gamma_{12F})^{|\epsilon_{a}-\delta_{a}|}(\bar{\eta}_{2}^{1-k_{a}}\cdot(S\cdot\epsilon_{13}\cdot\Gamma\Gamma_{F}-\Gamma_{F}S\cdot\epsilon_{13}\cdot\Gamma)_{F})_{_{A^{y_{a}}B^{k_{a}-y_{a}},F}}\\
 & \times(\Gamma_{34F})^{|\epsilon_{b}-\delta_{b}|}(\Gamma_{34D^{2\xi_{k}}C^{2\xi_{l}}})^{\delta_{b}}C_{\Gamma}^{-1}]_{a^{2\xi_{i}}b^{2\xi_{j}}c^{2\xi_{k}}d^{2\xi_{l}}}\}
\end{align*}
Contracting all indices together as in the $t=0$ case, this becomes
\begin{equation}
\begin{aligned} & s_{(a\vert b]}^{ij\vert m+\ell\vert kl}(1,i_{a}-k_{a},i_{b}-k_{b},r,\bm{r}',\bm{r}''):\alpha_{2}^{s_{2}}\alpha_{3}^{s_{3}}\alpha_{4}^{s_{4}}x_{3}^{r_{3}}x_{4}^{r_{4}}\\
 & \rightarrow\frac{(-1)^{k_{b}}}{\binom{i_{a}}{a_{im}}\binom{i_{b}}{b_{km}}}\sum_{y_{a},y_{b}=0}^{k_{a},k_{b}}\sum_{\substack{q\\
q'_{1},q''_{1}\\
q'_{2},q''_{2}
}
}2^{r'_{0}+r''_{0}}\binom{r}{q,q_{1},q_{3},r-q_{1}-q_{3}}\binom{r'_{1}}{q'_{1}}\binom{r''_{1}}{q''_{1}}\binom{r'_{2}}{q'_{2}}\binom{r''_{2}}{q''_{2}}(g_{AB})^{a_{ij}+r'_{0}}(g_{CD})^{b_{kl}+r''_{0}}\\
 & \times[(\mathcal{S}\cdot\bar{\eta}_{4})_{A}]^{q'_{2}}[(\mathcal{S}\cdot\bar{\eta}_{4})_{B}]^{r'_{2}-q'_{2}}[(\bar{\eta}_{2}\cdot\mathcal{S})_{C}]^{q''_{2}}[(\bar{\eta}_{2}\cdot\mathcal{S})_{D}]^{r''_{2}-q''_{2}}(\mathcal{S}_{AC})^{q}(\mathcal{S}_{AD})^{q_{1}}(\mathcal{S}_{BC})^{q_{3}}(\mathcal{S}_{BD})^{r-q_{1}-q_{3}}\\
 & \times\{(G_{(\ell'-\ell+2r'_{3}+4\xi_{i},n_{a}-\ell,n'_{3}-4-4(\xi_{j}+\xi_{k}+\xi_{l})+2k_{b},n'_{4}-4-4\xi_{l}+2k_{b},n'_{5}+4\xi_{k})}^{ij|m+\ell|kl})_{\substack{A^{i_{i}+q'_{1}}B^{i_{j}+r'_{1}-q'_{1}}C^{i_{k}+q''_{1}}D^{i_{l}+r''_{1}-q''_{1}}}
}^{F^{1+2|\epsilon_{b}-\delta_{b}|}}\\
 & \times[(\Gamma_{B^{2\xi_{i}}A^{2\xi_{j}}})^{\delta_{a}}(\bar{\eta}_{1}^{2\xi_{i}}\bar{\eta}_{2}^{2\xi_{j}}\cdot\Gamma)^{|\epsilon_{a}-\delta_{a}|}(\bar{\eta}_{2}^{1-k_{a}}\cdot(\mathcal{S}\cdot\epsilon_{13}\cdot\Gamma\Gamma_{F}-\Gamma_{F}\mathcal{S}\cdot\epsilon_{13}\cdot\Gamma)\cdot\bar{\eta}_{4}^{1-k_{b}})_{_{A^{y_{a}}B^{k_{a}-y_{a}},C^{y_{b}}D^{k_{b}-y_{b}}}}\\
 & \times(\bar{\eta}_{3}^{2\xi_{k}}\bar{\eta}_{4}^{2\xi_{l}}\cdot\Gamma)^{|\epsilon_{b}-\delta_{b}|}(\Gamma_{D^{2\xi_{k}}C^{2\xi_{l}}})^{\delta_{b}}C_{\Gamma}^{-1}]_{a^{2\xi_{i}}b^{2\xi_{j}}c^{2\xi_{k}}d^{2\xi_{l}}}\\
 & -\delta_{k_{b},0}(G_{(\ell'-\ell+2r'_{3}+4\xi_{i},n_{a}-\ell,n'_{3}-2-4(\xi_{j}+\xi_{k}+\xi_{l}),n'_{4}-2-4\xi_{l},n'_{5}+4\xi_{k})}^{ij|m+\ell|kl})_{\substack{A^{i_{i}+q'_{1}}B^{i_{j}+r'_{1}-q'_{1}}C^{i_{k}+q''_{1}}D^{i_{l}+r''_{1}-q''_{1}}}}^{F^{2}}\\
 & \times[(\Gamma_{B^{2\xi_{i}}A^{2\xi_{j}}})^{\delta_{a}}(\bar{\eta}_{1}^{2\xi_{i}}\bar{\eta}_{2}^{2\xi_{j}}\cdot\Gamma)^{|\epsilon_{a}-\delta_{a}|}(\bar{\eta}_{2}^{1-k_{a}}\cdot(\mathcal{S}\cdot\epsilon_{13}\cdot\Gamma\Gamma_{F}-\Gamma_{F}\mathcal{S}\cdot\epsilon_{13}\cdot\Gamma))_{_{A^{y_{a}}B^{k_{a}-y_{a}},F}}\\
 & \times(\bar{\eta}_{3}^{2\xi_{k}}\bar{\eta}_{4}^{2\xi_{l}}\cdot\Gamma)^{|\epsilon_{b}-\delta_{b}|}(\Gamma_{D^{2\xi_{k}}C^{2\xi_{l}}})^{\delta_{b}}C_{\Gamma}^{-1}]_{a^{2\xi_{i}}b^{2\xi_{j}}c^{2\xi_{k}}d^{2\xi_{l}}}\}.
\end{aligned}
\end{equation}

\section{Conclusions}
\label{sec:conclusions}

We presented explicit formulas for all conformal blocks in three dimensions. Complete expressions for the blocks are contained in Section~\ref{sec:4pt} in a mixed basis of tensor structures. Transformations between the two basis we use are described in Section~\ref{sec:3pt}. It is necessary to transform the blocks to pure basis that is blocks in which the pair of operators placed at points $\eta_1$ and $\eta_2$ enters on the same footing as the pair placed at $\eta_3$ and $\eta_4$. Without converting to the pure basis implementing bootstrap would be rather confusing and would need to involve transformations matrices from Section~\ref{sec:3pt} anyway.

Even a cursory examination of the results shows that the expressions for the blocks and the transformation matrices are fairly complicated. This is partially because our results are valid for any spins of both external  and exchange operators and thus the representations can be large. The size of the representations on their own may not seem like an obvious reason for convoluted results. After all, there are only two distinct types of representations in three dimensions---either bosonic or fermionic. However, with increasing sizes of representations there are many possible tensor structures  that appear in the OPE or in three-point functions. The number of tensor structures grows, roughly speaking, quadratically with the spins of operators, as discussed in 
Section~\ref{sec:counting}. Of course, we did not need to consider every tensor structure separately. The tensor structures follow patterns, but there are still several possible patterns and those need to be treated distinctly as we did in many subsections of Sections \ref{sec:3pt} and ~\ref{sec:4pt}. 

We always considered a whole tower of exchange operators. If an exchange operator was allowed by symmetry,  for given external operators, so were exchange operators with spins increased by integer values and we considered all of such operators simultaneously. It turns out that despite the towers having arbitrarily large spins the expressions for conformal blocks and the transformation matrices do not increase with the exchange spins. All sums can be recast as finite sums that depend on the spin of the smallest representation at the bottom of the tower. 

Due to their generality, our results can be used for bootstrap of any three-dimensional CFT\@. (For software that compute 3d blocks  numerically see \cite{Erramilli:2020rlr}.) Of course, in a CFT there are always some conserved currents, at the very least the energy-momentum tensor. A treatment of conserved currents in the embedding space formalism we use was described in \cite{Fortin:2020des}. Constraints implied by current conservation can be applied directly to correlators considered in this work. 

We expect that the results presented here can be generalized to four dimensions as well. The Lorentz representations are not significantly more complicated compared to three dimensions. In broad strokes, there are twice as many types of representations in 4d compared to 3d and there are additional complications due to self-duality of representations.

\ack{We thank Wen-Jie Ma and Valentina Prilepina for numerous discussions. The work of JFF is supported by NSERC.  This work was also supported by the US Department of Energy under grants DE-FG02-04ER41338 and DE-FG02-06ER41449 (JL), as well as DE-SC00-17660 (AS and WS).}

\begin{appendix}
\section{Removing $\ell$-dependence}
\label{sec:ldep}

In this appendix we show how dependence on $\ell$ of the $\omega^{klm}$ coefficients, introduced in Section \ref{sec:3pt}, can be removed by redefining certain sums. We first isolate the $t$-dependent factors
\begin{equation}
\begin{aligned} & _{b\tilde{b}}\omega_{0}{}_{(b_{0},b_{2},b_{s};c_{3},c_{4},c_{s};d_{C},d_{D},d_{E},d_{2})}^{klm}=  \lambda_{\ell}\sum_{r_{0},r_{2},s_{0},s_{2}\geq0} 
\left(\begin{array}{c}
  i_{b}\\
  r_{0},r_{2},i_{b}-r_{0}-r_{2}
\end{array}\right)\\
\times &
\frac{
\left(\begin{array}{c}
  i_{\tilde{b}}+\tilde{b}_{kl}-b_{kl}-r_{0}+b_{0}+d_{C}+d_{D}+d_{E}\\
   \tilde{b}_{kl}-b_{kl},\tilde{b}_{km}-s_{0},\tilde{b}_{lm}-r_{0}+s_{0}+b_{0},d_{C}+d_{D}+d_{E}
\end{array}\right)
\left(\begin{array}{c}
   r_{0}\\
   s_{0},r_{0}-s_{0}-b_{0},b_{0}
\end{array}\right) }
{\left(\begin{array}{c}
  i_{b}\\
   b_{km},b_{lm},a_{m}
\end{array}\right)}  \\
\times &
 \frac{\left(\begin{array}{c}
   r_{2}\\
  s_{2},r_{2}-s_{2}-b_{2},b_{2}
\end{array}\right)
\left(\begin{array}{c}
  i_{b}-r_{0}-r_{2}\\
  b_{km}-s_{0}-s_{2},b_{lm}-r_{0}-r_{2}+s_{0}+s_{2}+b_{0}+b_{2},b_{s}
\end{array}\right)}
{
\left(\begin{array}{c}
  n_{b}+2i_{b}-2r_{0}-r_{2}-a_{m}-a_{s}+b_{0}+b_{2}+e_{1}\\
   \ell-r_{0},\ell_{k}-b_{kl}-s_{0}-s_{2}-a_{k},\ell_{l}-b_{kl}-r_{0}-r_{2}+s_{0}+s_{2}-a_{l}+b_{0}+b_{2},e_{1}
\end{array}\right)}\\
\times &
 \left(\begin{array}{c}
  n_{\tilde{b}}-\ell_{\tilde{b}}-r_{2}-a_{k}-a_{l}+b_{2}-d_{C}-d_{D}+d_{2}\\
   \ell_{k}-\tilde{b}_{km}-\tilde{b}_{kl}-s_{2}-a_{k}-d_{C},n_{\tilde{b}}-\ell_{k}-\ell_{\tilde{b}}+\tilde{b}_{km}+\tilde{b}_{kl}-r_{2}+s_{2}-a_{l}+b_{2}-d_{D},d_{2}
\end{array}\right)\\
\times & (-1)^{\ell_{b}+\ell-\tilde{b}_{km}-\tilde{b}_{lm}+r_{2}-b_{0}-d_{E}+e_{2}} \, 2^{i_{\tilde{b}}+\tilde{b}_{kl}-b_{kl}-r_{0}+b_{0}+d_{C}+d_{D}+d_{E}}\rho^{(3,\ell_{b}+c_{3}+e_{2};-h-n_{b}-2i_{b}-\ell_{b}-2\xi_{m})}\\
\times & \tilde{K}^{(3,h+2r_{0}+r_{2}+c_{4}+e_{2};p-r_{0}-t_{0};i_{\tilde{b}}+\tilde{b}_{kl}-b_{kl}-r_{0}+b_{0}+d_{C}+d_{D}+d_{E},n_{\tilde{b}}-\ell_{\tilde{b}}-r_{2}-a_{k}-a_{l}+b_{2}-d_{C}-d_{D}+d_{2},0,\ell_{\tilde{b}}-b_{0}-d_{E})}\Sigma_{t},
\end{aligned}
\end{equation}
where the sums over $t$'s are given by
\[
\begin{aligned}\Sigma_{t}= & \sum_{t_{0},t_{2}\geq0}\frac{(-\ell_{b})_{t_{0}+t_{2}}(-\ell_{\tilde{b}}+b_{0}+d_{E})_{t_{0}}\left(h+p+r_{0}+r_{2}+c_{4}+e_{2}-\frac{1}{2}\right)_{t_{0}+t_{2}}}{t_{0}!t_{2}!(-\ell+r_{0})_{t_{0}}\left(h+n_{b}+2i_{b}-c_{3}-e_{2}+2\xi_{m}+\frac{3}{2}\right)_{t_{0}+t_{2}}}\\
\times & \frac{\left(-p+r_{0}+1\right)_{t_{0}}\left(h+p+n_{\tilde{b}}+i_{\tilde{b}}+\tilde{b}_{kl}-b_{kl}-a_{k}-a_{l}+b_{2}+c_{4}+d_{2}+e_{2}\right)_{t_{2}}}{\left(h+\ell+\tilde{b}_{kl}-b_{kl}+r_{0}+r_{2}+b_{0}+b_{2}+c_{4}+e_{2}+1\right)_{t_{0}+t_{2}}}.
\end{aligned}
\]
Shifting the variables by $t_{2}\rightarrow t_{2}-t_{0}$ and expressing part of the expression in terms of hypergeometric series leads to 
\begin{align*}
 & \Sigma_{t}=\sum_{t_{0},t_{2}\geq0}\frac{(-p+r_{0}+1)_{t_{0}}(-\ell_{\tilde{b}}+b_{0}+d_{E})_{t_{0}}(-t_{2})_{t_{0}}(-\ell_{b})_{t_{2}}}{t_{0}!(-h-p-n_{\tilde{b}}-i_{\tilde{b}}-\tilde{b}_{kl}+b_{kl}+a_{k}+a_{l}-b_{2}-c_{4}-d_{2}-e_{2}-t_{2}+1)_{t_{0}}(-\ell+r_{0})_{t_{0}}}\\
\times & \frac{(h+p+r_{0}+r_{2}+c_{4}+e_{2}-\frac{1}{2})_{t_{2}}(h+p+n_{\tilde{b}}+i_{\tilde{b}}+\tilde{b}_{kl}-b_{kl}-a_{k}-a_{l}+b_{2}+c_{4}+d_{2}+e_{2})_{t_{2}}}{t_{2}!(h+n_{b}+2i_{b}-c_{3}-e_{2}+2\xi_{m}+\frac{3}{2})_{t_{2}}(h+\ell+\tilde{b}_{kl}-b_{kl}+r_{0}+r_{2}+b_{0}+b_{2}+c_{4}+e_{2}+1)_{t_{2}}}\\
 & =\sum_{t_{2}\geq0}{}_{3}F_{2}\bigg[\substack{-p+r_{0}+1,-\ell_{\tilde{b}}+b_{0}+d_{E},-t_{2}\\
-h-p-n_{\tilde{b}}-i_{\tilde{b}}-\tilde{b}_{kl}+b_{kl}+a_{k}+a_{l}-c_{4}-e_{2}-b_{2}-d_{2}-t_{2}+1,-\ell_{\tilde{b}}-i_{\tilde{b}}+r_{0}
}
;1\bigg] \frac{(-\ell_{a})_{t_{2}}}{t_{2}!}\\
\times & \frac{(h+p+r_{0}+r_{2}+c_{4}+e_{2}-\frac{1}{2})_{t_{2}}(h+p+n_{\tilde{b}}+i_{\tilde{b}}+\tilde{b}_{kl}-b_{kl}-a_{k}-a_{l}+b_{2}+c_{4}+d_{2}+e_{2})_{t_{2}}}{(h+n_{b}+2i_{b}-c_{3}-e_{2}+2\xi_{m}+\frac{3}{2})_{t_{2}}(h+\ell+\tilde{b}_{kl}-b_{kl}+r_{0}+r_{2}+b_{0}+b_{2}+c_{4}+e_{2}+1)_{t_{2}}}.
\end{align*}
Because $i_{\tilde{b}}-r_{0}\geq0$ (since $i_{\tilde{b}}-r_{0}=\ell-t_{0}-r_{0}-q_{4}-b_{0}-d_{E}$
and $q_{4}\leq\ell-t_{0}-r_{0}-b_{0}-d_{E}$) we can transform $\Sigma_{t} $ into
\begin{align*}
\Sigma_{t}  =& \sum_{t_{2}\geq0}\frac{(h+n_{\tilde{b}}+i_{\tilde{b}}+\tilde{b}{}_{kl}-b_{kl}+r_{0}-a_{k}-a_{l}++b_{2}+c_{4}+d_{2}+e_{2}+1)_{t_{2}}}{(h+p+n_{\tilde{b}}+i_{\tilde{b}}+\tilde{b}{}_{kl}-b_{kl}-a_{k}-a_{l}+b_{2}+c_{4}+d_{2}+e_{2})_{t_{2}}}\\
 & \times{}_{3}F_{2}\bigg[\substack{-p+r_{0}+1,-i_{\tilde{b}}+r_{0}-b_{0}-d_{E},-t_{2}\\
     h+n_{\tilde{b}}+i_{\tilde{b}}+\tilde{b}{}_{kl}-b_{kl}-a_{k}-a_{l}+r_{0}+b_{2}+c_{4}+d_{2}+e_{2}+1,-\ell+r_{0} } ;1\bigg] \frac{(-\ell_{b})_{t_{2}}}{t_{2}!} \\
& \times  \frac{(h+p+r_{0}+r_{2}+c_{4}+e_{2}-\frac{1}{2})_{t_{2}}(h+p+n_{\tilde{b}}+i_{\tilde{b}}+\tilde{b}_{kl}-b_{kl}-a_{k}-a_{l}+b_{2}+c_{4}+d_{2}+e_{2})_{t_{2}}}{(h+n_{b}+2i_{b}-c_{3}-e_{2}+2\xi_{m}+\frac{3}{2})_{t_{2}}(h+\ell+\tilde{b}_{kl}-b_{kl}+r_{0}+r_{2}+b_{0}+b_{2}+c_{4}+e_{2}+1)_{t_{2}}}\\
 = & \sum_{t_{0}\geq0}\frac{(-1)^{t_{0}}(-\ell_{b})_{t_{0}}(-i_{\tilde{b}}+r_{0}-b_{0}-d_{E})_{t_{0}}(-p+r_{0}+1)_{t_{0}}}{t_{0}!(-\ell+r_{0})_{t_{0}}(h+n_{b}+2i_{b}-c_{3}-e_{2}+2\xi_{m}+\frac{3}{2})_{t_{0}}}\\
 & \times \frac{(h+p+r_{0}+r_{3}+c_{4}+e_{2}-\frac{1}{2})_{t_{0}}}{(h+\ell-b_{kl}+\tilde{b}_{kl}+r_{0}+r_{2}+b_{0}+b_{2}+c_{4}+e_{2}+1)_{t_{0}}} \\
& \times \sum_{t_{2}\geq0}  \frac{(-\ell_{b}+t_{0})_{t_{2}}(h+p+r_{0}+r_{2}+c_{4}+e_{2}+t_{0}-\frac{1}{2})_{t_{2}}}{t_{2}!(h+n_{b}+2i_{b}-c_{3}-e_{2}+2\xi_{m}+t_{0}+\frac{3}{2})_{t_{2}}} \\
& \times \frac{(h+n_{\tilde{b}}+i_{\tilde{b}}+\tilde{b}{}_{kl}-b_{kl}+r_{0}-a_{k}-a_{l}+b_{2}+c_{4}+d_{2}+e_{2}+t_{0}+1)_{t_{2}}}{(h+\ell+\tilde{b}_{kl}-b_{kl}+r_{0}+r_{2}+b_{0}+b_{2}+c_{4}+e_{2}+t_{0}+1)_{t_{2}}},
\end{align*}
where we have shifted by $t_{2}\rightarrow t_{2}+t_{0}$. In terms
of hypergeometric series, we have
\begin{align*}
\Sigma_{t} & =\sum_{t_{0}\geq0}\frac{(-1)^{t_{0}}(-\ell_{b})_{t_{0}}(-i_{\tilde{b}}+r_{0}-b_{0}-d_{E})_{t_{0}}(-p+r_{0}+1)_{t_{0}}}{t_{0}!(-\ell+r_{0})_{t_{0}}(h+n_{b}+2i_{b}-c_{3}-e_{2}+2\xi_{m}+\frac{3}{2})_{t_{0}}}\\
&\times \frac{(h+p+r_{0}+r_{2}+c_{4}+e_{2}-\frac{1}{2})_{t_{0}}}{(h+\ell+\tilde{b}_{kl}-b_{kl}+r_{0}+r_{2}+b_{0}+b_{2}+c_{4}+e_{2}+1)_{t_{0}}} \\
&\times  _{3}F_{2}\bigg[\substack{h+p+r_{0}+r_{2}+c_{4}+e_{2}+t_{0}-\frac{1}{2},h+n_{\tilde{b}}+i_{\tilde{b}}+\tilde{b}{}_{kl}-b_{kl}+r_{0}-a_{k}-a_{l}+b_{2}+c_{4}+d_{2}+e_{2}+t_{0}+1,-\ell_{b}+t_{0}\\
h+n_{b}+2i_{b}-c_{3}-e_{2}+2\xi_{m}+t_{0}+\frac{3}{2},h+\ell+\tilde{b}_{kl}-b_{kl}+r_{0}+r_{2}+b_{0}+b_{2}+c_{4}+e_{2}+t_{0}+1
}
;1\bigg].
\end{align*}
Since $n_{\tilde{b}}-\ell_{\tilde{b}}-r_{2}-a_{k}-a_{l}-b_{0}+d_{2}\geq0$
and $a_{s}+b_{s}+c_{s}=e_{1}+2e_{2}$, we can transform again by
\begin{equation}
\begin{aligned}\Sigma_{t} & =\sum_{t_{0}\geq0}\frac{(-1)^{t_{0}}(-\ell_{b})_{t_{0}}(-i_{\tilde{b}}+r_{0}-b_{0}-d_{E})_{t_{0}}(-p+r_{0}+1)_{t_{0}}(h+p+r_{0}+r_{2}+c_{4}+e_{2}-\frac{1}{2})_{t_{0}}}{t_{0}!(-\ell+r_{0})_{t_{0}}(h+\ell+\tilde{b}_{kl}-b_{kl}+r_{0}+r_{2}+b_{0}+b_{2}+c_{4}+e_{2}+1)_{t_{0}}}\\
 & \times\frac{(-p+n_{b}+2i_{b}-r_{0}-r_{3}-a_{s}-b_{s}+e_{1}+2)_{\ell_{b}-t_{0}}}{(h+n_{b}+2i_{b}-c_{1}-e_{2}+2\xi_{m}+t_{0}+\frac{3}{2})_{\ell_{b}-t_{0}}(h+n_{b}+2i_{b}-c_{3}-e_{2}+2\xi_{m}+\frac{3}{2})_{t_{0}}}\\
 & \times  _{3}F_{2}\bigg[\substack{h+p+r_{0}+r_{2}+c_{4}+e_{2}+t_{0}-\frac{1}{2},-n_{\tilde{b}}-i_{\tilde{b}}+\ell+r_{2}+a_{k}+a_{l}+b_{0}-d_{2},-\ell_{b}+t_{0}\\
p-n_{b}-2i_{b}-\ell_{b}+r_{0}+r_{2}+t_{0}+a_{s}+b_{s}-e_{1}-1,h+\ell+\tilde{b}_{kl}-b_{kl}+r_{0}+r_{2}+b_{0}+b_{2}+c_{4}+e_{2}+t_{0}+1
}
;1\bigg]\\
&= \frac{(-p+n_{b}+2i_{b}-r_{0}-r_{2}-a_{s}-b_{s}+e_{1}+2)_{\ell_{b}}}{(h+n_{b}+2i_{b}-c_{3}-e_{2}+2\xi_{m}+\frac{3}{2})_{\ell_{b}}} \\
&  \times  \sum_{t_{0},t_{2}\geq0}
      \frac{(-i_{\tilde{b}}+r_{0}-b_{0}-d_{E})_{t_{0}}(-n_{\tilde{b}}+\ell_{\tilde{b}}+r_{2}+a_{k}+a_{l}+b_{0}-d_{2})_{t_{2}}}{(-1)^{t_{0}+t_{2}}t_{0}!t_{2}!(-\ell+r_{0})_{t_{0}} (h+\ell+\tilde{b}_{kl}-b_{kl}+r_{0}+r_{2}+b_{0}+b_{2}+c_{4}+e_{2}+1)_{t_{0}+t_{2}}}\\
& \times  \frac{(-p+r_{0}+1)_{t_{0}}(-\ell_{b})_{t_{0}+t_{2}}(h+p+r_{0}+r_{2}+c_{4}+e_{2}-\frac{1}{2})_{t_{0}+t_{2}}}{(-p+n_{b}+2i_{b}+\ell_{b}-r_{0}-r_{2}-a_{s}-b_{s}+e_{1}+2)_{t_{0}+t_{2}}},
\end{aligned}
\end{equation}
where now the $t$-sums are bounded by the $\ell$-independent
$i_{\tilde{b}}-r_{0}+b_{0}+d_{E}$ and $n_{\tilde{b}}-\ell_{\tilde{b}}-r_{2}-a_{k}-a_{l}-b_{0}+d_{2}$.

\end{appendix}


\bibliography{Bibliography}

\end{document}